\documentclass[%
 reprint,
 amsmath,amssymb,
 aps,
 prx,
]{revtex4-2}

\usepackage{graphicx}% Include figure files
\usepackage{dcolumn}% Align table columns on decimal point
\usepackage{bm}% bold math
\usepackage{MnSymbol}
\usepackage{xcolor}
\usepackage{mathrsfs}
\usepackage{nicematrix}

\begin{document}

\preprint{APS/Angular}

\title{Preparing angular momentum eigenstates using engineered quantum walks}% Force line breaks with \\

\author{Yuan Shi}
\email{Yuan.Shi@colorado.edu}
\affiliation{%
Department of Physics, Center for Integrated Plasma Studies, University of Colorado Boulder, Boulder, CO 80309, USA
}%
\altaffiliation[Was at ]{Lawrence Livermore National Laboratory, CA, 94550, USA}%Lines break automatically or can be forced with \\

\author{Kristin M. Beck}% 
\affiliation{Lawrence Livermore National Laboratory, CA, 94550, USA}%

\author{Veronika Anneliese Kruse}
\affiliation{Lawrence Livermore National Laboratory, CA, 94550, USA}

\author{Stephen B. Libby}
\affiliation{Lawrence Livermore National Laboratory, CA, 94550, USA}%

\date{\today}% It is always \today, today,

\begin{abstract}
Coupled angular momentum eigenstates are widely
used in atomic and nuclear physics calculations, and are building blocks for spin networks and the Schur transform.
To combine two angular momenta $\mathbf{J}_1$ and $\mathbf{J}_2$, forming eigenstates of their total angular momentum $\mathbf{J}=\mathbf{J}_1+\mathbf{J}_2$, we develop a quantum-walk scheme that does not require inputting $O(j^3)$ nonzero Clebsch–Gordan (CG) coefficients classically. In fact, our scheme may be regarded as a unitary method for computing CG coefficients on quantum computers with a typical complexity of $O(j)$ and a worst-case complexity of $O(j^3)$. Equivalently, our scheme provides decompositions of the dense CG unitary into sparser unitary operations. 
Our scheme prepares angular momentum eigenstates using a sequence of Hamiltonians to move an initial state deterministically to desired final states, which are usually highly entangled states in the computational basis. 
In contrast to usual quantum walks, whose Hamiltonians are prescribed, we engineer the Hamiltonians in $\mathfrak{su}(2)\times \mathfrak{su}(2)$, which are inspired by, but different from, Hamiltonians that govern magnetic resonances and dipole interactions.
To achieve a deterministic preparation of both ket and bra states, we use projection and destructive interference to double pinch the quantum walks, such that each step is a unit-probability population transfer within a two-level system. 
We test our state preparation scheme on classical computers, reproducing CG coefficients. We also implement small test problems on current quantum hardware. 
%\begin{description}
%use the optional argument of the \verb+\item+ command to give the category of each %item. 
%\end{description}
\end{abstract}

\maketitle
\setlength{\parskip}{0pt}
%\tableofcontents
%%%%%%%%%%%%%%%%%%%%%%%%%%%%%%%%%%%%%%%%%%%%%
\section{\label{sec:intro}Introduction}
Angular momentum is a fundamental quantity in quantum mechanics. When solving the Schr{\"o}dinger equation of a single particle in a central potential, the wave function is separable into a radial part and an angular part. The angular wave function, sometimes expanded in terms of spherical harmonics, are eigenfunctions of the orbital angular momentum operator $\mathbf{L}$. 
Orbital angular momentum, whose quantum numbers are integers, can be extended to general angular momentum, whose quantum numbers are half integers. For an angular momentum with quantum number $j$, 
%$j=0, \frac{1}{2}, 1, \frac{3}{2}\dots$, 
its Hilbert space dimension is $D=2j+1$. The spin-1/2 case, for which $D=2$, is the simplest nontrivial case, where the angular momentum operators are represented by Pauli matrices.

The concept of angular momentum becomes even more crucial when solving many-body wave functions. For example, when solving atomic and nuclear shell models \cite{lindgren2012atomic, suhonen2007nucleons}, many-body wave functions can be constructed from single-particle wave functions using a Slater determinant. The angular part of the many-body wave function is formed by combining multiple angular momenta. 
In the simplest case of combining two angular momenta $\mathbf{J}_1$ and $\mathbf{J}_2$, eigenstates of their total angular momentum $\mathbf{J}=\mathbf{J}_1 + \mathbf{J}_2$ can be expanded in terms of eigenstates of $\mathbf{J}_1$ and $\mathbf{J}_2$. 
The expansion coefficients are known as the Clebsch–Gordan (CG) coefficients. The CG coefficients, or in their symmetrized forms known as the Wigner $3j$ symbols, which subsequently contract into the Wigner $6j$ and $9j$ symbols, are widely used in atomic and nuclear physics calculations \cite{bauche1988transition,bar1989super,Caurier05}. 
For example, to compute transition probabilities, one needs to evaluate matrix elements of the form $\langle\psi_f|H_{\text{int}}|\psi_i\rangle$, where the initial and final states involve angular momentum eigenstates. 
To perform calculations on quantum computers, one needs to prepare the bra state, and project to the ket state, for angular momentum eigenstates, which are often highly entangled states in the computational basis. 

%\vspace{-2pt}
In a broader context, preparing angular momentum eigenstates involves a unitary change of basis, sometimes called the Clebsch-Gordan transform.
When extending from two to $n$ angular momenta, the CG transform is a fundamental building block of spin networks and the Schur transform. 
Spin network is a computational framework \cite{marzuoli2005computing, jordan2010permutational, aquilanti2008quantum, east2023all} that has natural connections to quantum gravity \cite{Ponzano69, rovelli1995spin} and topological quantum field theories \cite{kauffman2006spin}. 
Closely related is the Schur transform \cite{bacon2006efficient}, which is a change of basis $|e_1\rangle\otimes\dots\otimes |e_n\rangle\rightarrow |p_\lambda\rangle\otimes |q_\lambda\rangle \otimes|\lambda\rangle$. 
On the left-hand side (LHS), each $|e\rangle$ is a $D$-dimensional quantum system called a qudit, which reduces to a usual qubit when $D=2$. 
On the right-hand side (RHS), $|p_\lambda\rangle$ is an eigenstate of a permutation operator, namely, an element of the symmetric group, which swaps the labels of $n$ qudits. $|q_\lambda\rangle$ is an eigenstate of a unitary operator, which acts as the same $D\times D$ matrix $U$ on all qudits. The label $|\lambda\rangle$ provides a padding such that the dimensions match on both sides of the Schur transform.
Using the Wigner-Eckart theorem for the unitary group and subgroup-adapted basis, the Schur transform can be implemented efficiently as a cascade of CG transforms \cite{bacon2007quantum, kirby2018practical, wills2023generalised}. 
Performing the Schur transform this way requires performing CG transforms repeatedly. 
Alternatively, using representation theory of the symmetric group, the Schur transform can be implemented more efficiently using quantum Fourier transform and generalized phase estimation \cite{krovi2019efficient}. 
%\newpage

Much like the quantum Fourier transform, where a state that is local in one basis becomes more global in another basis, the Schur transform is useful in many quantum information protocols. 
%One may regard the LHS of the Schur transform as copies of quantum states, and view the RHS as a more convenient basis for some applications. 
%
For example, the Schur transform can be used to encode and decode information into decoherence-free subsystems \cite{Kempe01}. In this application, one assumes that the environment couples identically to $n$ qudits. Because the $|p_\lambda\rangle$ basis is insensitive to $U$, which involves unknown details of system-environment interactions, one can transform into the $|p_\lambda\rangle$ basis to store and protect quantum information. 
As another example, the Schur transform allows two parties to communicate without a shared reference frame \cite{Bartlett03}. In quantum communication, the information sender prepares a stream of $|0\rangle$ and $|1\rangle$ states along the $z$ axis. The information receiver may not know the axis and thus measures along a different $z'$ axis. Rather than trying to align their axes, if both the sender and the receiver use the Schur basis $|p_\lambda\rangle$, then they are free to pick their own axis, which only affects the form of unitary operations they perform but does not affect the information that is being transmitted.

As a fundamental building block of the aforementioned applications, we investigate how to perform CG transforms on quantum computers.  
One method is to compute matrix elements classically, and then decompose the 
unitary transformation into elementary gates. 
In the simplest example of combining two spin-1/2, the computational basis is $\mathbf{e}=({|\!\uparrow \uparrow\rangle}, {|\!\uparrow \downarrow\rangle}, {|\!\downarrow \uparrow\rangle}, {|\!\downarrow \downarrow\rangle})$. Eigenstates of their total angular momentum are the triplet and singlet states, which forms another basis $\mathbf{q}=(|1,1\rangle, |1,0\rangle, |1,-1\rangle, |0,0\rangle)$. The change of basis $\mathbf{q} = \mathbf{e} U$ is mediated by 
%\vspace{-5pt}
\begin{equation}
    \nonumber
    U = \left( \begin{array}{cccc}
    1 & 0 & 0 & 0 \\
    0 & \frac{1}{\sqrt{2}} & 0 & \frac{1}{\sqrt{2}} \\
    0 & \frac{1}{\sqrt{2}} & 0 & -\frac{1}{\sqrt{2}} \\
    0 & 0 & 1 & 0
    \end{array} \right).
\end{equation}
%\vspace{-5pt}
In more general cases, the dimension of the unitary matrix is $D^2\times D^2$, and elements of the unitary matrix are the CG coefficients. 
For the SU(2) group, CG coefficients are given in closed analytical form by the Racah formula \cite{Racah42,wigner2012group}. 
However, the formula is expressed as alternating sum of ratios of large factorials, which is cumbersome to use in practice. 
Given the usefulness of CG coefficients, many alternative classical algorithms have been developed \cite{wei1999unified, rasch2004efficient, johansson2016fast}, which efficiently produces $O(D^3)$ nonzero CG coefficients in $O(D^3)$ steps. 
Efficient numerical algorithms are also known for the more general SU($N$) group \cite{alex2011numerical}.
After inputting $O(D^3)$ matrix elements, the unitary matrix can be decomposed into $O(D^3)$ two-qubit gates and additional single-qubit gates \cite{barenco1995elementary, shende2005synthesis, Plesch11, krol2022efficient}. 
As a simpler task, preparing a specific angular momentum ket state, which involves $O(D)$ nonzero CG coefficients, only takes $O(D\log(D))$ two-qubit gates \cite{Jordan09,malvetti2021quantum,de2022double}. Alternatively, one can use a circuit of depth $O(\log(D))$ at the expense of $O(D\log(D))$ ancillary qubits \cite{zhang2022quantum, luo2024circuit}.

In this paper, we develop a state preparation scheme that does not require pre-computing the CG coefficients classically and then loading the data to quantum computers. In fact, our scheme may be regarded as a quantum method for computing the CG coefficients. 
The query complexity of our scheme, which uses quantum Hamiltonian simulations, is $O(D)$ in a typical case, and $O(D^3)$ in the worst case.
%To prepare the hardest states, the complexity of our scheme is $O(D^3)$, whereas the complexity in a typical case is $O(D)$, with 
Moreover, our scheme does not require additional ancillary qubits beyond what is needed for quantum Hamiltonian simulations.
%, giving a polynomial speed up.
Not requiring additional ancilla makes preparing bra states as easy as preparing ket states.

Our state preparation scheme uses a modified version of quantum walks, which we call {\it engineered quantum walks}. In the usual study of quantum walks \cite{kempe2003quantum,childs2010relationship, venegas2012quantum}, one asks the question of how an initial state spreads across the Hilbert space for a given Hamiltonian. The fact that quantum random walks spread faster than classical random walks have been exploited to provide speedups for stochastic algorithms \cite{childs2003exponential, childs2004spatial}. 
In the case of engineered quantum walks, the question we ask is different: For a desired quantum trajectory, namely, a sequence of movements of quantum states, how does one design local Hamiltonians $\{ H_k\}$ to achieve the movements?
Using Hamiltonians that are local, each movement is achieved by a unitary transformation that is more sparse than the overall transformation. In other words, suppose $U$ is the desired transformation, then engineered quantum walks provide a decomposition $U=\prod_k U_k$, where $U_k$ is a simpler unitary transformation generated by a local Hamiltonian over a specific evolution time.  
Although sharing a similar spirit with adiabatic quantum state generation \cite{farhi2000quantum, aharonov2003adiabatic}, our framework of state preparation is different because our steps are discrete and we do not require states to be close to the ground state.

%\vspace{-2pt}
The key to engineering the quantum walks is twofold. 
First, we needs to understand the Lie algebra of our specific problem. In our case, the algebra of combining two angular momenta is $\mathfrak{su}(2)\times \mathfrak{su}(2)$. Understanding the algebra provides a list of candidate Hamiltonians, whose behaviors are well understood and matrix representations are sparse in both the computational basis and the problem basis. 
Second, to achieve deterministic movements along a quantum trajectory, rather than a probabilistic spread of quantum states, we introduce a technique called {\it double pinch}, where each step of the engineered quantum walk is a unit-probability population transfer within a two-level system. 
The basic idea is the following: Suppose we manage to find two independent operators $A$ and $B$ that are sparse in both the computational basis and the problem basis. In the problem basis, denotes the matrix elements $A|i\rangle = \alpha_-^i |i-1\rangle + \alpha_+^i |i+1\rangle$ and 
$B|i\rangle = \beta_-^i |i-1\rangle + \beta_+^i |i+1\rangle$. 
Then, the operator $C=\beta_-^iA-\alpha_-^iB$ blocks the transition $|i\rangle \rightarrow |i-1\rangle$ (moving left) due to destructive interference.
After proper symmetrization and projection, the Hermitian operator $H_C$ induces a unidirectional quantum walk $|i\rangle \rightarrow |i+1\rangle$ (moving right). We call this a single-pinched walk, because the state may still leak out from the right. 
With the help of another independent local operator, we can use a similar idea to also block the transition $|i+1\rangle \rightarrow |i+2\rangle$, thereby pinching the quantum walk on both left and right. The double-pinched quantum walk isolates two states $|i\rangle \leftrightarrow |i+1\rangle$, so that the relevant dynamics is confined within a two-level system in the problem basis.
On the other hand, in the computational basis, double-pinched walks are allowed to involve more states. Because the operators are local also in the computational basis, the Hamiltonian matrix is sparse and amenable to efficient quantum Hamiltonian simulations \cite{berry2015hamiltonian, low2019hamiltonian}. 
%Notice that for double-pinched quantum walks, the dynamics may be nontrivial outside the two-level system of interest. However, if the initial state is within the two-level system, then the final state will also be within this system, so the rest of the Hilbert space is irrelevant. 
%\newpage

To identify local interactions that are useful for preparing angular momentum eigenstates, we draw inspiration from experimental considerations. %Suppose we want to change the angular momentum of an atom, we can perform the following experiments. 
First, to change $J_z$, while keeping $j$ fixed, we can put an atom in a rotating magnetic field $\mathbf{B}$. When the magnetic field rotates at resonant frequencies, the spin is completely flipped, thereby sweeping its $z$ component through all allowed values. The Hamiltonian in this magnetic resonance experiment is proportional to $\mathbf{J}\cdot\mathbf{B}$, which is directly implementable in the computational basis comprised of eigenstates of $\mathbf{J}_1$ and $\mathbf{J}_2$.
Second, to change the value of $j$, one can interrogate an atom with laser light. By absorbing or emitting a single photon, the angular momentum of the atom changes by $1$. Depending on the laser polarization, the $J_z$ quantum number may remain fixed or change by $\pm 1$ due to selection rules of dipole interactions. The Hamiltonian in this experiment is proportional to the position operator $\mathbf{r}$, and the selection rules are consequences of the commutation relation $[\mathbf{L}^2, [\mathbf{L}^2, \mathbf{r}]]=2(\mathbf{r}\mathbf{L}^2 + \mathbf{L}^2\mathbf{r})$. 
Because the radial part of the wave function is separated from the angular part, we cannot directly express $\mathbf{r}$ in the computational basis. Nevertheless, we discover that $\mathbf{A}=\mathbf{J}_1\times \mathbf{J}_2$, which we call the cross-pole operator, serves an identical purpose. %The operator $\mathbf{A}$ is a key building block for preparing angular momentum eigenstates. 

The paper is organized as follows: In Sec.~\ref{sec:Review}, we review basic facts about angular momentum and introduce the notation used in this paper. In Sec.~\ref{sec:algebra}, we investigate properties of the $\mathfrak{su}(2)\times \mathfrak{su}(2)$ algebra, and derive matrix elements of local operators in both the computational and the problem bases. In Sec.~\ref{sec:walk}, we develop the central idea of this paper, namely how to use double-pinched quantum walks to move quantum states along a desired path deterministically to achieve state preparation. In Sec.~\ref{sec:test}, we validate our state preparation protocol on classical computers, and perform small test problems on quantum devices. Discussions and conclusions are made in Sec.~\ref{sec:conclusion}. Details that are not essential for understanding the paper are provided in the Appendix.

%%%%%%%%%%%%%%%%%%%%%%%%%%%%%%%%%%%%%%%%%%%%%
\section{\label{sec:Review}Review of angular momentum}
To introduce notation, we give a brief review of angular momentum, starting from a single angular momentum to combining two angular momenta. Most material in this section can be found in standard textbooks \cite{edmonds1996angular, johnson2006lectures, drake2007springer}. We also introduce a picture of the state space, which will be useful when we conduct quantum walks. % for state preparation.  

In quantum mechanics, the orbital angular momentum operator $\mathbf{L}$ about the origin is defined as $\mathbf{L}=\mathbf{r}\times\mathbf{p}$ from the position operator $\mathbf{r}$ and the linear momentum operator $\mathbf{p}$. Unlike in classical mechanics, because $[r_a, p_b]=i\delta_{ab}$, where $\delta_{ab}$ is the Kronecker delta, 
the kinematic variables do not always commute. We use a natural unit $\hbar=1$ and Cartesian coordinates with indices $a,b=1,2,3$, for $x, y, z$ components, respectively. Consequently, the three components of the vector $\mathbf{L}$ satisfy commutation relations $[L_a, L_b]=i(r_ap_b-r_bp_a)=i\epsilon_{abc}L_c$, where $\epsilon_{abc}$ is the three-dimensional Levi-Civita symbol and summation over repeated indices is assumed. The orbital angular momentum is a canonical example of the Lie algebra $\mathfrak{su}(2)$, which is a three-dimensional vector space over the complex field $\mathbb{C}$, equipped with a Lie bracket $[\cdot ,\cdot ]$ that defines multiplications between vectors.

More generally, the Lie algebra $\mathfrak{su}(2)$ is represented by angular momentum $\mathbf{J}$ acting on a Hilbert space. The angular momentum can, for example, correspond to an orbital or a spin angular momentum. Components of $\mathbf{J}$ satisfy the defining multiplication table
\begin{equation}
    \label{eq:J_table}
    [J_a, J_b] = i \epsilon_{abc} J_c.
\end{equation}
%$[\textcolor{red}{J}_a, \textcolor{red}{J}_b] = i \epsilon_{abc} \textcolor{red}{J}_c$ %%%%%%
Using property of the Levi-Civita symbol that $\epsilon_{abc}\epsilon_{abd}=2\delta_{cd}$, the above is equivalent to $iJ_c=\epsilon_{cab}J_a J_b$, or $\mathbf{J}\times \mathbf{J} =i\mathbf{J}$, where $\times$ denotes cross products between vectors in three-dimensional spaces. 
%$\textcolor{red}{\mathbf{J}}\times \textcolor{red}{\mathbf{J}} =i\textcolor{red}{\mathbf{J}}$ %%%%%%
Notice that the cross product of a classical vector with itself is always zero. However, as a quantum operator, $\mathbf{J}\times \mathbf{J}$ is nonzero.

The Hilbert spaces that $\mathbf{J}$ acts on can be parameterized by eigenvalues of $\mathbf{J}^2=J_aJ_a$. Because $[\mathbf{J}^2, J_a]=0$, there exist simultaneous eigenstates of $\mathbf{J}^2$ and $J_z$. 
Suppose the eigenstate of $J_z$ is 
\begin{equation}
    J_z |j,m\rangle = m |j,m\rangle,
\end{equation}
where classically $j$ represents the length of the vector $\mathbf{J}$, so its $z$ component $|m|\le j$.
Introducing the rising and lowering operators $J_\pm = J_x \pm i J_y$, which satisfy $[J_z, J_\pm]=\pm J_\pm$,  $J_\pm|j,m\rangle$ is another eigenstate of $J_z$, but with the eigenvalue $m\pm1$. 
Because $|m|\le j$ the rising and lowering operators must terminate at top and bottom states, namely, $J_+|j,j\rangle=0$ and $J_-|j,-j\rangle=0$. 
Since $\mathbf{J}^2=\frac{1}{2}(J_+J_- + J_-J_+)+J_z^2$ and $[J_+, J_-]=2J_z$,
acting $\mathbf{J}^2=J_-J_++J_z+J_z^2$ on the top state $|j,j\rangle$, or equivalent, acting $\mathbf{J}^2=J_+J_--J_z+J_z^2$ on the bottom state $|j,-j\rangle$, gives $\mathbf{J}^2 |j,m\rangle = \mathscr{J}_j |j,m\rangle$, where the eigenvalue of the quantum operator $\mathbf{J}^2$ is
\begin{eqnarray}
    \label{eq:J2}
    \mathscr{J}_j=j(j+1).
\end{eqnarray}
Because $J_\pm$ changes the value of $m$ by $\pm 1$, the allowable values of $m$ are $m=j, j-1,\dots, -j$. Due to the $m\rightarrow -m$ symmetry, the dimension of the Hilbert space is $D_j=2j+1$. Because $D_j$ must be an integer, $j$ must be a half integer. In other words, the allowable values of $j$ are $j=0,\frac{1}{2}, 1,\frac{3}{2}, \dots$
For a given $j$, states in the Hilbert space are connected by rising operator $J_+|j,m\rangle=\mathcal{J}_j^+(m) |j,m+1\rangle$ and lowering operator $J_-|j,m\rangle=\mathcal{J}_j^-(m) |j,m-1\rangle$, where the matrix elements are
\begin{equation}
    \label{eq:J_elements}
    \mathcal{J}_j^+(m) =\mathcal{J}_j^-(-m) = \sqrt{(j+1+m)(j-m)}.
\end{equation}
The matrix elements can be derived, for example, using $\mathcal{J}_j^+(m)^2=\langle j, m|J_-J_+|j,m\rangle=\langle j, m|J^2-J_z-J_z^2|j,m\rangle$. Here, we have used $J_+^\dagger = J_-$, where $\dagger$ denotes the Hermitian adjoint.
The matrix elements satisfy $\mathcal{J}_j^+(m) =\mathcal{J}_j^-(m+1)$ and $\mathcal{J}_j^-(m) =\mathcal{J}_j^+(m-1)$.
The Hilbert space spanned by $|j,m\rangle$ gives an irreducible representation of $\mathfrak{su}(2)$.

\begin{figure}[t]
\includegraphics[width=0.45\textwidth]{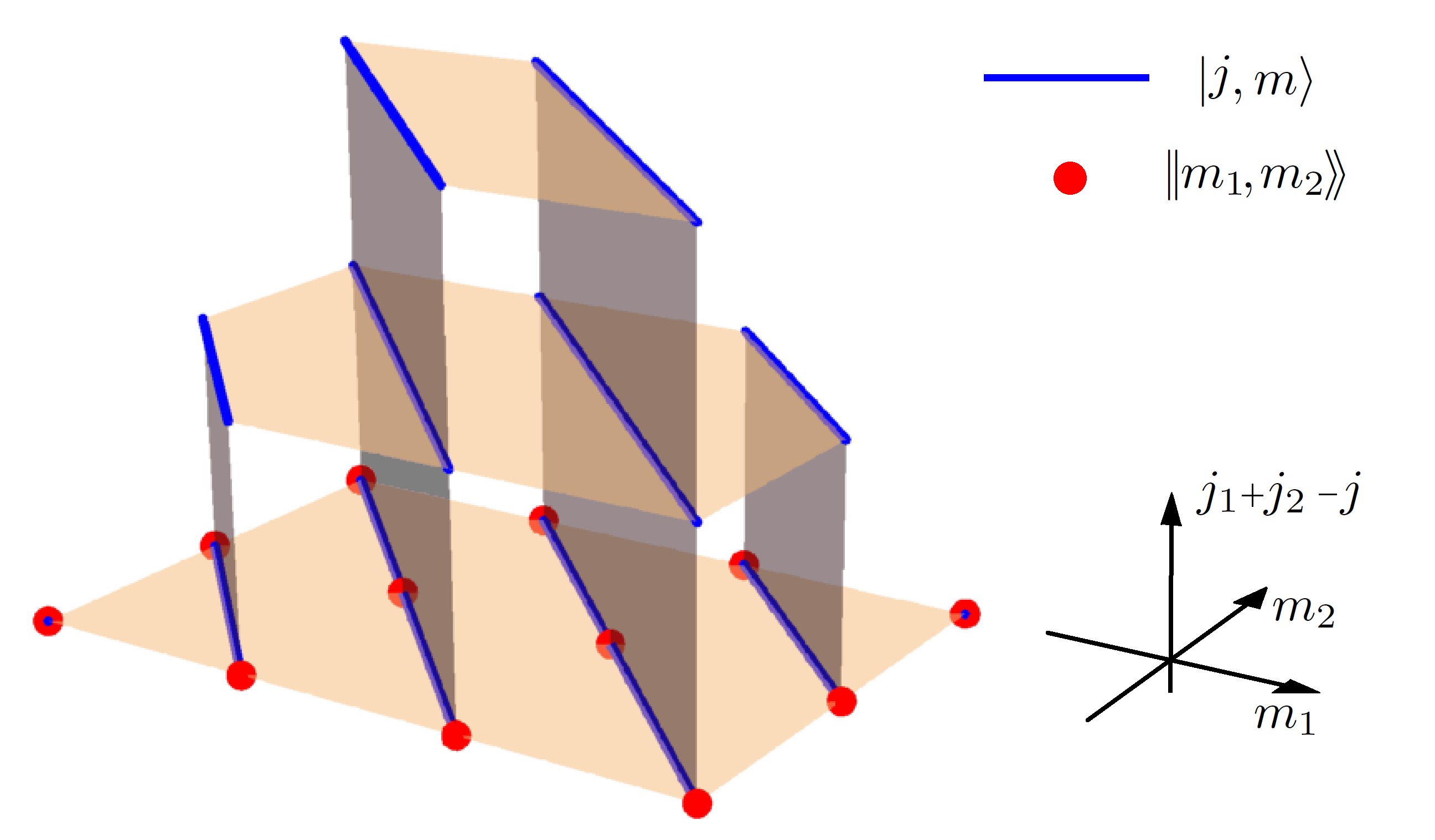}
\caption{\label{fig:domain} The Hilbert space of $\mathbf{J}=\mathbf{J}_1+\mathbf{J}_2$ can be represented as the direct product $\| m_1, m_2 \rrangle$ (red dots), or equivalently as the direct sum $|j,m\rangle$ (blue lines). Each horizontal plane (orange) is at a constant $j$, while each vertical plane (grey) is at a constant $m=m_1+m_2$. The Clebsch–Gordan coefficients give a unitary transformation from red dots to blue lines within a grey plane. 
We place $j_{\max}$ at the bottom and $j_{\min}$ at the top, so the state space is an upright pyramid. Without loss of generality, we always assume $j_1\ge j_2$. In this example, $j_1=\frac{3}{2}$ and $j_2=1$. }
\end{figure}

When there are two angular momenta $\mathbf{J}_1$ and $\mathbf{J}_2$, their total angular momentum is the vector sum 
\begin{equation}
    \label{eq:J}
    \mathbf{J}=\mathbf{J}_1+\mathbf{J}_2.
\end{equation}
%$\textcolor{red}{\mathbf{J}}=\mathbf{J}_1+\mathbf{J}_2$ %%%%%%
%$\textcolor{red}{\mathbf{J}}^2=\mathbf{J}_1^2+\mathbf{J}_2^2+2\textcolor{olive}{\Lambda}$ %%%%%%
Indices of bold symbols denote the two angular momenta, whose components are denoted by the second index of the non-bolded $J_{1a}$ and $J_{2a}$. Suppose eigenstates of $\mathbf{J}_1$ are $|j_1,m_1\rangle$ in Hilbert space $\mathcal{H}_1$, and eigenstates of $\mathbf{J}_2$ are $|j_2,m_2\rangle$ in Hilbert space $\mathcal{H}_2$. Then, eigenstates of $\mathbf{J}$ are spanned by 
\begin{equation}
    \label{eq:CG_definition}
    |j,m\rangle = \sum_{m_1+m_2=m} C^{j_1, j_2, j}_{m_1, m_2, m} |j_1,m_1\rangle \otimes |j_2,m_2\rangle,
\end{equation}
where $C^{j_1, j_2, j}_{m_1, m_2, m}$ are the CG coefficients.
The Hilbert space that $\mathbf{J}$ acts on is the tensor product $\mathcal{H}_1\otimes \mathcal{H}_2$. More rigorously, one should write $\mathbf{J}=\mathbf{J}_1\otimes\mathbb{I}_2+\mathbb{I}_1\otimes\mathbf{J}_2$, where $\mathbb{I}$ is the identity operator. We use abbreviated notation like $\mathbf{J}=\mathbf{J}_1+\mathbf{J}_2$, because it is often clear which Hilbert space the operators are acting on. Since they act on different Hilbert spaces, $\mathbf{J}_1$ and $\mathbf{J}_2$ commute in abbreviated notation.
Moreover, we use abbreviated notation $\| m_1, m_2 \rrangle := |j_1,m_1\rangle \otimes |j_2,m_2\rangle$ and $C^j_{m_1, m_2}:=C^{j_1, j_2, j}_{m_1, m_2, m}$, when values of $j_1$ and $j_2$ are clear from the context. 
We suppress $m$ in the notation of CG coefficients, because they are zero unless $m=m_1+m_2$, which is a consequence of $J_z=J_{1z}+J_{2z}$. 
As illustrated by the grey planes in Fig.~\ref{fig:domain}, CG coefficients relate states within a grey plane with a common $m=m_1+m_2$.
Using our abbreviated notation, Eq.~(\ref{eq:CG_definition}) can be written as $|j,m\rangle = C^{j}_{m_1, m_2} \|m_1,m_2\rrangle$, so $C^{j}_{m_1, m_2}=\llangle m_1, m_2\|j,m\rangle$. We see Eq.~(\ref{eq:CG_definition}) is a consequence of the completeness relation $\sum_{m_1+m_2=m} \|m_1, m_2\rrangle\llangle m_1, m_2\|=\mathbb{I}_m$, where $\mathbb{I}_m$ is the identity operator within the $m$-th grey plane. 
Another completeness relation is $\sum_j |j,m\rangle \langle j,m|=\mathbb{I}_m$, which gives $\|m_1,m_2\rrangle = \sum_j |j,m\rangle \langle j,m\|m_1,m_2\rrangle = \bar{C}^j_{m_1,m_2} |j,m\rangle$, where bar denotes complex conjugation and summation over repeated indices is again assumed. %As we shall see in the next paragraph, all CG coefficients have the same complex phase, so they can all be chosen to be real, in which case the complex conjugation is insignificant. 
As illustrated in Fig.~\ref{fig:domain}, the RHS of Eq.~(\ref{eq:CG_definition}) corresponds to a direct product of two irreducible representations of $\mathfrak{su}(2)$ (red dots), whose dimension is $D=D_{j_1}D_{j_2}$. The LHS of Eq.~(\ref{eq:CG_definition}) corresponds to a direct sum of irreducible representations of $\mathfrak{su}(2)$ (blue lines), where $j=j_{\max}, \dots, j_{\min}$, such that the total dimension $\sum_{j=j_{\min}}^{j_{\max}} D_j=D$.
Because the top $m$ value is attained when $m_1=j_1$ and $m_2=j_2$, the top state is $|j_1+j_2, j_1+j_2\rangle = \| j_1, j_2 \rrangle$, so $j_{\max}=j_1+j_2$. By dimension counting, $j_{\min}=j_1 - j_2$, where we have assumed $j_1\ge j_2$ without loss of generality. 
Geometrically, if $j_1$ and $j_2$ as two edges of a triangle, then $j$ is the third edge.
The CG coefficients give unitary transformations from one representation of $\mathfrak{su}(2)\times \mathfrak{su}(2)$, with the basis $\| m_1, m_2 \rrangle$, to another representation of $\mathfrak{su}(2)$, with the basis $|j,m\rangle$.

To compute CG coefficients, an efficient method is to consider the scalar operator
\begin{equation}
    \label{eq:Lambda}
    \Lambda = \mathbf{J}_1\cdot\mathbf{J}_2,
\end{equation}
%$ \textcolor{olive}{\Lambda} = \mathbf{J}_1\cdot\mathbf{J}_2$ %%%%%%
where $\cdot$ denotes vector inner products. The $\Lambda$ operator has two equivalent expressions. 
First, $2\Lambda = \mathbf{J}^2-\mathbf{J}_1^2-\mathbf{J}_2^2$. This expression is useful in the $|j,m\rangle$ basis because it implies that $\Lambda |j,m\rangle = \lambda_j|j,m\rangle$ and 
\begin{equation}
    \label{eq:lambda_j}
    %\lambda_j=\frac{1}{2}[j(j+1)-j_1(j_1+1)-j_2(j_2+1)].
    \lambda_j=\frac{1}{2}(\mathscr{J}_j-\mathscr{J}_{j_1}-\mathscr{J}_{j_2}),
\end{equation}
where $\mathscr{J}$ is defined in Eq.~(\ref{eq:J2}).
Second, $2\Lambda = 2J_{1z}J_{2z}+J_{1+}J_{2-}+J_{1-}J_{2+}$. This expression is useful for finding matrix elements of $\Lambda$ in the $\| m_1, m_2 \rrangle$ basis. 
The rising and lowering operators couple nearest neighbors along $m_1+m_2=m$. Because $|m|\le j$, $m_1$ and $m_2$ may not reach the full range between $\pm j_1$ and $\pm j_2$. Suppose the maximum value $m_2$ can reach is $m_u$, then we denote $|k\rangle_m=\| m - m_u+k, m_u-k \rrangle$, which counts the red dots in Fig.~\ref{fig:domain} from the top $m_2$ value for $k=0, 1, \dots, M-1$. Here, $M$ is the number of red dots, which equals to the number of blue lines, within the grey plane of a constant $m$. When $|m|\le j_{\min}$, $m_2$ can attain the full range of values, so $M=2j_2+1$; whereas when $|m|> j_{\min}$, $m_2$ can only attain a partial range, and $M=j_1+j_2+1-|m|$. 
We can compactly write $2\Lambda |k\rangle_m = \alpha_k {|k\rangle_m} + \beta_{k+1/2} {|k+1\rangle_m} + \beta_{k-1/2} {|k-1\rangle_m}$, where $\alpha_k=2m_1m_2$, $\beta_{k+1/2}=\mathcal{J}_{j_1}^+(m_1) \mathcal{J}_{j_2}^-(m_2)$, $m_1=m-m_u+k$, $m_2=m_u-k$, 
and we have used properties after Eq.~(\ref{eq:J_elements}) to relate the coefficient of $|k-1\rangle_m$ to the coefficient of $|k-1\rangle_m$.

Since we know matrix elements of $\Lambda$ in both $|j,m\rangle$ and $\| m_1, m_2 \rrangle$ bases, the $O(j^3)$ nonzero CG coefficients can be computed efficiently using recurrence relations in $O(j^3)$ steps, assuming $j_1\sim j_2\sim j$. 
%For a given $m$, the relevant states $\| m_1, m_2 \rrangle$ are those with $m_1+m_2=m$, and different combinations of them give rise to different $j$ states, living on a common grey plane in Fig.~\ref{fig:domain}. 
%
The recurrence relation is obtained by acting $2\Lambda$ on both sides of Eq.~(\ref{eq:CG_definition}), which can be rewritten more compactly as $|j,m\rangle=C^j_k |k\rangle_m$. Here, $C^j_k$ is the CG coefficient $C^j_{m_1, m_2}$ when $m_1$ and $m_2$ are evaluated at $k$. 
The resulted recurrence relation $\lambda_j C^j_k = \alpha_k C^j_k + \beta_{k+1/2} C^j_{k+1} + \beta_{k-1/2} C^j_{k-1}$ is a symmetric tridiagonal matrix equation for the vector $\mathbf{C}^j$. 
The matrix equation can be solved using forward or backward substitutions in $M=O(j)$ steps, with a final normalization step such that the 2-norm $|\mathbf{C}^j|=1$. 
Because all elements of the matrix are real valued, the phase of all CG coefficients are the same, and it is conventional to choose $C^j_{m_1, m_2}$ to be positive for top $m_1$ value. %In other words, when solving the tridiagonal matrix equation by backward substitution, it is conventional to choose the last element of $\mathbf{C}^j$ to be positive. 
This process can be repeated for different $j$, which can take $M=O(j)$ different values in the range $j_{\max} \ge j \ge \max(|m|, j_{\min})$. Finally, the process can be repeated for different $m$, which takes $2j_{\max}+1 = O(j)$ values. Therefore, it takes a total of $O(j^3)$ steps to compute all CG coefficients, which have exactly $O(j^3)$ nonzero elements. In other words, the recurrence relation gives an optimally efficient classical algorithm for computing all CG coefficients. 
With a more careful counting, the number of nonzero CG coefficients is $(2j_1+1)(2j_2+1)^2-\frac{4}{3}j_2(j_2+1)(2j_2+1)$, so more precisely, we write the scaling as $O(j_1j_2^2)$ when $j_1\ge j_2\rightarrow\infty$.
%The $O(j^3)$ scaling assumes $j\sim j_1\sim j_2$. However, when $j\sim j_1\gg j_2$, the scaling is reduced to $O(j)$. 

%%%%%%%%%%%%%%%%%%%%%%%%%%%%%%%%%%%%%%%%%%%%%
\section{\label{sec:algebra}Generators of $\mathfrak{su}(2)\times \mathfrak{su}(2)$ algebra}
Our state preparation scheme relies on moving states in the Hilbert space deterministically, using a sequence of engineered Hamiltonians. To achieve the desired movements, it is instrumental to understand properties of the $\mathfrak{su}(2)\times \mathfrak{su}(2)$ algebra, which is the goal of this section. Key results are summarized in Fig.~\ref{fig:table}, and additional details are provided in Appendix~\ref{App:generators}.
The generators of $\mathfrak{su}(2)\times \mathfrak{su}(2)$ become powerful tools after we derive their matrix elements in both $\|m_1, m_2\rrangle$ and $|j,m\rangle$ basis.

%%%%%%%%%%%%%%%%%%%%%%%%%%%%%
\subsection{Vector operators\label{sec:operators}}
With two three-dimensional vectors $\mathbf{J}_1$ and $\mathbf{J}_2$, there are $3\times3=9$ degrees of freedom (d.o.f.) in $\mathfrak{su}(2)\times \mathfrak{su}(2)$. The total angular momentum $\mathbf{J}=\mathbf{J}_1+\mathbf{J}_2$ has three d.o.f.. To account for another three d.o.f. that are orthogonal to $\mathbf{J}$, we introduce an anti-symmetric vector operator 
\begin{equation}
    \label{eq:A}
    \mathbf{A} = \mathbf{J}_1 \times \mathbf{J}_2,
\end{equation}
which we call the cross-pole operator.
Since $\mathbf{J}_1$ and $\mathbf{J}_2$ are Hermitian operators and they commute, $\mathbf{A}$ is also a Hermitian operator.
%$\textcolor{cyan}{\mathbf{A}}= \mathbf{J}_1 \times \mathbf{J}_2$ %%%%%%
If the vectors are classical, then $\mathbf{A}\cdot \mathbf{J} = \mathbf{J}\cdot \mathbf{A}=0$ due to anti-symmetry of the cross product. 
%$\textcolor{cyan}{\mathbf{A}}\cdot \textcolor{red}{\mathbf{J}} = \textcolor{red}{\mathbf{J}}\cdot \textcolor{cyan}{\mathbf{A}}=0$ %%%%%%
As quantum operators, $\mathbf{A}$ and $\mathbf{J}$ are still orthogonal, but the reason is not as trivial, as discussed in Appendix~\ref{App:generators_A}.

We compute multiplication tables between components of $\mathbf{A}$ and $\mathbf{J}$ using property of the Levi-Civita symbol.
First, for multiplication between $A_a$ and $J_b$, as derived in Appendix~\ref{App:generators_A}, we have
\begin{equation}
    \label{eq:AJ_table}
     [A_a, J_b] = i\epsilon_{abc}A_c,
\end{equation}
which is equivalent to $\mathbf{A}\times \mathbf{J} + \mathbf{J}\times \mathbf{A}=2i\mathbf{A}$.
%$[\textcolor{cyan}{A_a}, \textcolor{red}{J_b}] = i\epsilon_{abc}\textcolor{cyan}{A_c}$ %%%%%%
Second, for multiplication between $A_a$ and $A_b$, Appendix~\ref{App:generators_A} shows
\begin{equation}
    \label{eq:AA_table}
    [A_a, A_b]=i\epsilon_{abc}\Lambda J_c.
\end{equation}
%$[\textcolor{cyan}{A}_a, \textcolor{cyan}{A}_b]=i\epsilon_{abc}\textcolor{olive}{\Lambda} \textcolor{red}{J}_c$ %%%%%%
The above equals to $i\epsilon_{abc}J_c\Lambda$ because $[\Lambda, \mathbf{J}]=0$, and is equivalent to $\mathbf{A}\times \mathbf{A} = i\Lambda \mathbf{J}$. 
%$\textcolor{cyan}{\mathbf{A}} \times \textcolor{cyan}{\mathbf{A}}  = i\textcolor{olive}{\Lambda} \textcolor{red}{\mathbf{J}} $ %%%%%%
As shown in Appendix~\ref{App:generators_A}, the scalar operator $\mathbf{A}^2=A_a A_a= \mathbf{J}_1^2 \mathbf{J}_2^2 -\Lambda-\Lambda^2$,
from which it is clear that $\mathbf{A}^2$ and $\mathbf{J}$ commute. 
%$\textcolor{cyan}{\mathbf{A}}^2 = \mathbf{J}_1^2 \mathbf{J}_2^2 - \textcolor{olive}{\Lambda} - \textcolor{olive}{\Lambda}^2$ %%%%%%
Moreover, $\mathbf{A}^2$ also commutes with $\Lambda$, which means that the two scalar operators are not independent.

%\onecolumngrid
%\begin{widetext} 
\begin{figure*}[t]
\includegraphics[width=0.72\textwidth]{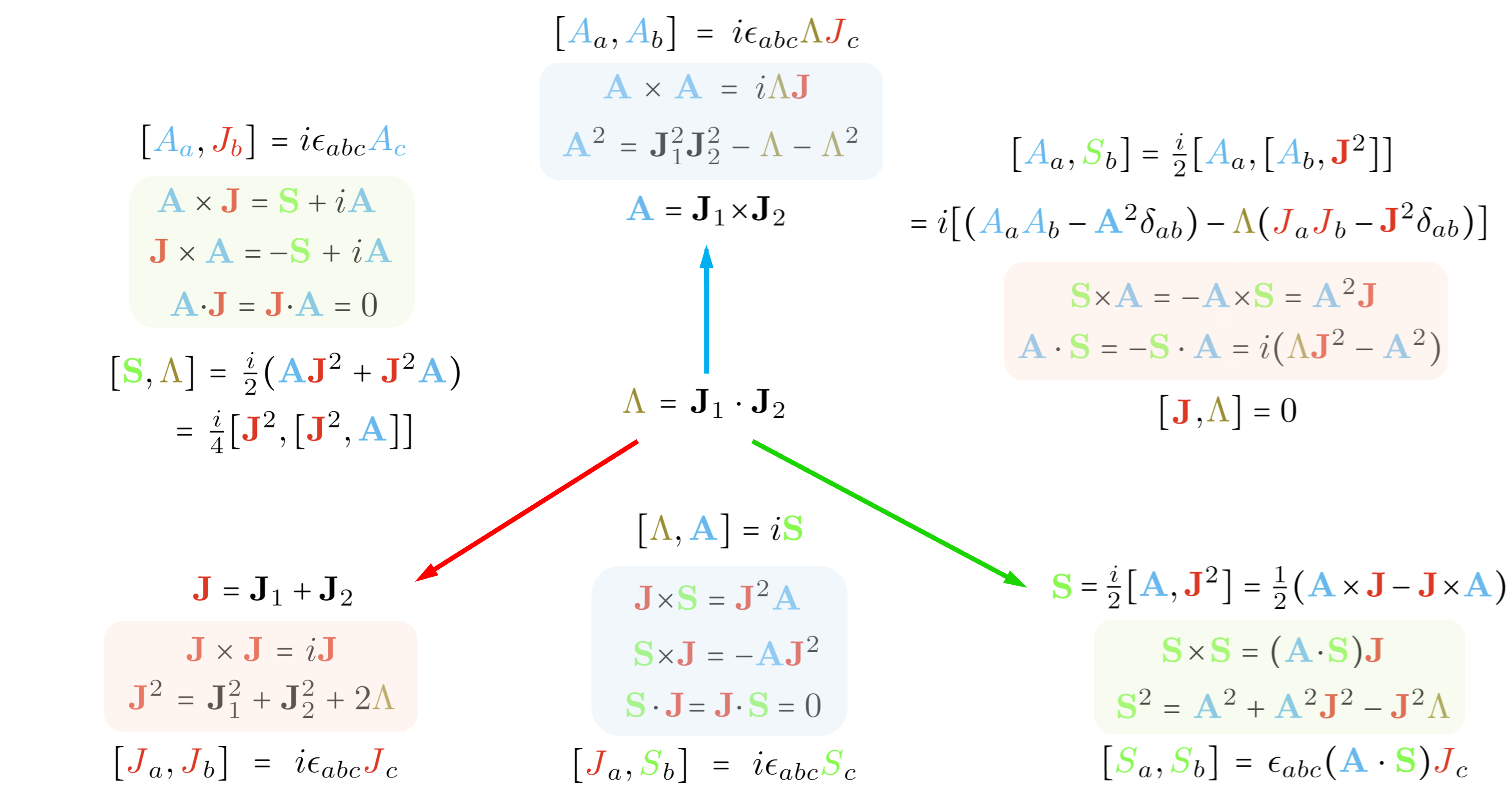}
\caption{\label{fig:table} Let $\mathbf{J}_1$ and $\mathbf{J}_2$ be generators of $\mathfrak{su}(2)$, acting on Hilbert spaces $\mathcal{H}_1$ and $\mathcal{H}_2$, respectively. Then, the $\mathfrak{su}(2)\times \mathfrak{su}(2)$ algebra, as represented by $\mathcal{H}_1\otimes \mathcal{H}_2$, is generated by vector operators $\mathbf{J}$, $\mathbf{A}$, $\mathbf{S}$, and scalar operator $\Lambda$. As classical three-vectors, the inner product $\mathbf{A}\cdot\mathbf{J}=0$, and the cross product $\mathbf{A}\times\mathbf{J}$ is along $\mathbf{S}$ direction, and so on. This figure summarizes key properties of the four quantum operators.}
\end{figure*}
%\twocolumngrid
%\end{widetext}

The remaining three d.o.f. of $\mathfrak{su}(2)\times \mathfrak{su}(2)$ are revealed by the commutator $[\Lambda, \mathbf{A}]$. 
As shown in Appendix~\ref{App:generators_S},
$[\Lambda, \mathbf{A}]=i\mathbf{A}\times\mathbf{J}+\mathbf{A}$, which can be symmetrized using the equivalent form of Eq.~(\ref{eq:AJ_table}). We introduce 
\begin{equation}
    \label{eq:S}
    \mathbf{S} = \frac{1}{2}(\mathbf{A}\times\mathbf{J} - \mathbf{J}\times\mathbf{A}) = \frac{i}{2}[\mathbf{A}, \mathbf{J}^2],
\end{equation}
%$\textcolor{green}{\mathbf{S}} = \frac{1}{2}(\textcolor{cyan}{\mathbf{A}} \times\textcolor{red}{\mathbf{J}}  - \textcolor{red}{\mathbf{J}} \times\textcolor{cyan}{\mathbf{A}} ) = \frac{i}{2}[\textcolor{cyan}{\mathbf{A}} ,  \textcolor{red}{\mathbf{J}}^2]$%%%%%%
such that $[\Lambda, \mathbf{A}]=i\mathbf{S}$. From its definition, it is clear that $\mathbf{S}$ is a Hermitian operator. 
%$[\textcolor{olive}{\Lambda},  \textcolor{cyan}{\mathbf{A}} ]=i\textcolor{green}{\mathbf{S}} $ %%%%%%
Alternative forms of $\mathbf{S}$ are $\mathbf{S}=\mathbf{A}\times\mathbf{J}-i\mathbf{A}=i\mathbf{A}-\mathbf{J}\times\mathbf{A}$. The second equality of Eq.~(\ref{eq:S}) comes from $\mathbf{J}^2=\mathbf{J}_1^2+\mathbf{J}_2^2+2\Lambda$, where $\mathbf{J}_1^2$ and $\mathbf{J}_2^2$ commute with $\mathbf{A}$. 
The vector $\mathbf{S}$ is orthogonal to both $\mathbf{J}$ and $\mathbf{A}$ at the classical level, and therefore linearly independent of $\mathbf{J}$ and $\mathbf{A}$.
At the quantum level, as shown in Appendix~\ref{App:generators_S}, 
$\mathbf{S}\cdot\mathbf{J}=\mathbf{J}\cdot\mathbf{S}=0$
%$\textcolor{green}{\mathbf{S}} \cdot\textcolor{red}{\mathbf{J}} =\textcolor{red}{\mathbf{J}} \cdot\textcolor{green}{\mathbf{S}} =0$ %%%%%%
are still orthogonal,
but $\mathbf{A}\cdot\mathbf{S}=-\mathbf{S}\cdot\mathbf{A}=i(\Lambda\mathbf{J}^2-\mathbf{A}^2)$ is nonzero. 
%$\textcolor{cyan}{\mathbf{A}} \cdot\textcolor{green}{\mathbf{S}} =-\textcolor{green}{\mathbf{S}} \cdot\textcolor{cyan}{\mathbf{A}} =i(\textcolor{olive}{\Lambda}\textcolor{red}{\mathbf{J}}^2-\textcolor{cyan}{\mathbf{A}}^2)$ %%%%%%

Another way to see the relations between the three vectors $\mathbf{J}$, $\mathbf{A}$, and $\mathbf{S}$ is to consider their cross products. 
(i) Using alternative forms of Eq.~(\ref{eq:S}), we know $\mathbf{A}\times\mathbf{J}=\mathbf{S}+i\mathbf{A}$ and  $\mathbf{J}\times\mathbf{A}=-\mathbf{S}+i\mathbf{A}$. 
%$\textcolor{cyan}{\mathbf{A}} \times \textcolor{red}{\mathbf{J}} = \textcolor{green}{\mathbf{S}} +i\textcolor{cyan}{\mathbf{A}} $ and  $ \textcolor{red}{\mathbf{J}} \times\textcolor{cyan}{\mathbf{A}} =- \textcolor{green}{\mathbf{S}} +i\textcolor{cyan}{\mathbf{A}} $. %%%%%%
Unlike classical vectors, $\mathbf{A}\times\mathbf{J}\ne-\mathbf{J}\times\mathbf{A}$ for quantum operators.
Without the quantum effect $i\mathbf{A}$, the cross product between $\mathbf{J}$ and $\mathbf{A}$ are along the $\mathbf{S}$ direction. 
(ii) Because $\mathbf{J}$ is orthogonal to both $\mathbf{A}$ and $\mathbf{S}$, the cross products of $\mathbf{A}$ and $\mathbf{S}$ are along the $\mathbf{J}$ direction even at the quantum level. 
As shown in Appendix~\ref{App:generators_S}, we have $\mathbf{S}\times\mathbf{A}=-\mathbf{A}\times\mathbf{S} = \mathbf{A}^2\mathbf{J}$.
%$ \textcolor{green}{\mathbf{S}} \times \textcolor{cyan}{\mathbf{A}}=- \textcolor{cyan}{\mathbf{A}}\times \textcolor{green}{\mathbf{S}}  =  \textcolor{cyan}{\mathbf{A}}^2\textcolor{red}{\mathbf{J}}$ %%%%%%
(iii) Finally, as shown in Appendix~\ref{App:generators_S}, the cross products between $\mathbf{J}$ and $\mathbf{S}$ satisfy
$\mathbf{J}\times\mathbf{S}=\mathbf{J}^2\mathbf{A}$ and $\mathbf{S}\times\mathbf{J}=-\mathbf{A}\mathbf{J}^2$. 
%$\textcolor{red}{\mathbf{J}}\times\textcolor{green}{\mathbf{S}}=\textcolor{red}{\mathbf{J}}^2\textcolor{cyan}{\mathbf{A}}$ %%%%%%
%$\textcolor{green}{\mathbf{S}}\times\textcolor{red}{\mathbf{J}}=-\textcolor{cyan}{\mathbf{A}}\textcolor{red}{\mathbf{J}}^2$ %%%%%%
In other words, the cross product between $\mathbf{J}$ and $\mathbf{S}$ are along the $\mathbf{A}$ direction. 
It is important to remember that although $\mathbf{A}^2$ commutes with $\mathbf{J}$, $\mathbf{J}^2$ does not commute with $\mathbf{A}$.

Since the three vectors $\mathbf{J}$, $\mathbf{A}$, and $\mathbf{S}$ already span all nine degrees of freedom of $\mathfrak{su}(2)\times \mathfrak{su}(2)$, the commutator $[\Lambda, \mathbf{S}]$ does not generate an independent vector. 
In Appendix~\ref{App:generators_S}, we show that 
\begin{equation}
    \label{eq:J2J2A}
    [\mathbf{S}, \Lambda] = \frac{i}{2}(\mathbf{A}\mathbf{J}^2 + \mathbf{J}^2\mathbf{A}) = \frac{i}{4}[\mathbf{J}^2, [\mathbf{J}^2, \mathbf{A}]].
\end{equation}
%$[\textcolor{green}{\mathbf{S}},  \textcolor{olive}{\Lambda}] = \frac{i}{2}(\textcolor{cyan}{\mathbf{A}}\textcolor{red}{\mathbf{J}}^2 + \textcolor{red}{\mathbf{J}}^2\textcolor{cyan}{\mathbf{A}}) = \frac{i}{4}[\textcolor{red}{\mathbf{J}}^2, [\textcolor{red}{\mathbf{J}}^2, \textcolor{cyan}{\mathbf{A}}]]$ %%%%%%
In other words, while $[\Lambda, \mathbf{A}]$ defines the $\mathbf{S}$ direction, $[\Lambda, \mathbf{S}]$ returns to the $\mathbf{A}$ direction.

Finally, let us compute the multiplication tables that involve $\mathbf{S}$. 
First, for multiplication between $J_a$ and $S_b$, we show in Appendix~\ref{App:generators_S} that
\begin{equation}
    \label{eq:SJ_table}
    [J_a, S_b]=i\epsilon_{abc}S_c.
\end{equation}
%$[\textcolor{red}{J}_a, \textcolor{green}{S}_b]=i\epsilon_{abc}\textcolor{green}{S}_c$ %%%%%%
The above is equivalent to $2i\mathbf{S}=\mathbf{J}\times\mathbf{S} +\mathbf{S}\times\mathbf{J}$, which is consistent with previous results. 
Second, for multiplication between $A_a$ and $S_b$, we show in Appendix~\ref{App:generators_S} that
\begin{eqnarray}
    \label{eq:AS_table}    
    [A_a, S_b] &=& \frac{i}{2}[A_a, [A_b, \mathbf{J}^2]] \\ 
    \nonumber
    %&=& i[(A_aA_b - \mathbf{A}^2\delta_{ab}) - \Lambda(J_aJ_b - \mathbf{J}^2\delta_{ab})],
    &=& i[(\Lambda \mathbf{J}^2 - \mathbf{A}^2)\delta_{ab} + (A_aA_b - \Lambda J_aJ_b)].
\end{eqnarray}
%$[\textcolor{cyan}{A}_a, \textcolor{green}{S}_b] = \frac{i}{2}[\textcolor{cyan}{A}_a, [\textcolor{cyan}{A}_b, \textcolor{red}{\mathbf{J}}^2]] = i[(\textcolor{cyan}{A}_a\textcolor{cyan}{A}_b - \textcolor{cyan}{\mathbf{A}}^2\delta_{ab}) - \textcolor{olive}{\Lambda}(\textcolor{red}{J}_a\textcolor{red}{J}_b - \textcolor{red}{\mathbf{J}}^2\delta_{ab})]$ %%%%%%
%which has the form of projection operators.
Classically, $\mathbf{v}^2\delta_{ab}-v_av_b$ is proportional to the projection operator in directions perpendicular to the vector $\mathbf{v}$. 
Moreover, using $A_aA_b-\Lambda J_aJ_b = A_bA_a - \Lambda J_b J_a$, the $a\leftrightarrow b$ symmetry is more transparent when the above is written as
$\frac{i}{2}[(A_aA_b + A_bA_a -2\mathbf{A}^2\delta_{ab}) - \Lambda(J_aJ_b + J_bJ_a -2\mathbf{J}^2\delta_{ab})]$, whose structure is reminiscent of central-difference discretization of two-dimensional Laplace operator. 
Third, for multiplication between $S_a$ and $S_b$, we show in Appendix~\ref{App:generators_S} that
\begin{equation}
    \label{eq:SS_table}
    [S_a, S_b]=\epsilon_{abc}(\mathbf{A}\cdot\mathbf{S})J_c,
\end{equation}
%$[\textcolor{green}{S}_a, \textcolor{green}{S}_b]=\epsilon_{abc}( \textcolor{cyan}{\mathbf{A}} \cdot\textcolor{green}{\mathbf{S}} )\textcolor{red}{J}_c$ %%%%%%
which is equivalent to $\mathbf{S}\times \mathbf{S} = (\mathbf{A}\cdot\mathbf{S})\mathbf{J}$, consistent with previous results. 
%$\textcolor{green}{\mathbf{S}}\times \textcolor{green}{\mathbf{S}} = (\textcolor{cyan}{\mathbf{A}}\cdot\textcolor{green}{\mathbf{S}})\textcolor{red}{\mathbf{J}}$ %%%%%%
Notice that the scalar operator $\mathbf{A}\cdot\mathbf{S}$ commutes with $\mathbf{J}$.
Lastly, the scalar operator $\mathbf{S}^2=\mathbf{A}^2+\mathbf{A}^2\mathbf{J}^2-\mathbf{J}^2\Lambda$, as shown in Appendix~\ref{App:generators_S}. We see all scalar operators are fundamentally related to $\Lambda$. 
%$\textcolor{green}{\mathbf{S}}^2=\textcolor{cyan}{\mathbf{A}}^2+\textcolor{cyan}{\mathbf{A}}^2\textcolor{red}{\mathbf{J}}^2-\textcolor{red}{\mathbf{J}}^2 \textcolor{olive}{\Lambda}$ %%%%%%

%%%%%%%%%%%%%%%%%%%%%%%%%%%%%%%%%%%%%%%%%%%%%
\subsection{\label{sec:matrix}Matrix elements}
\begin{figure}[t]
\includegraphics[width=0.45\textwidth]{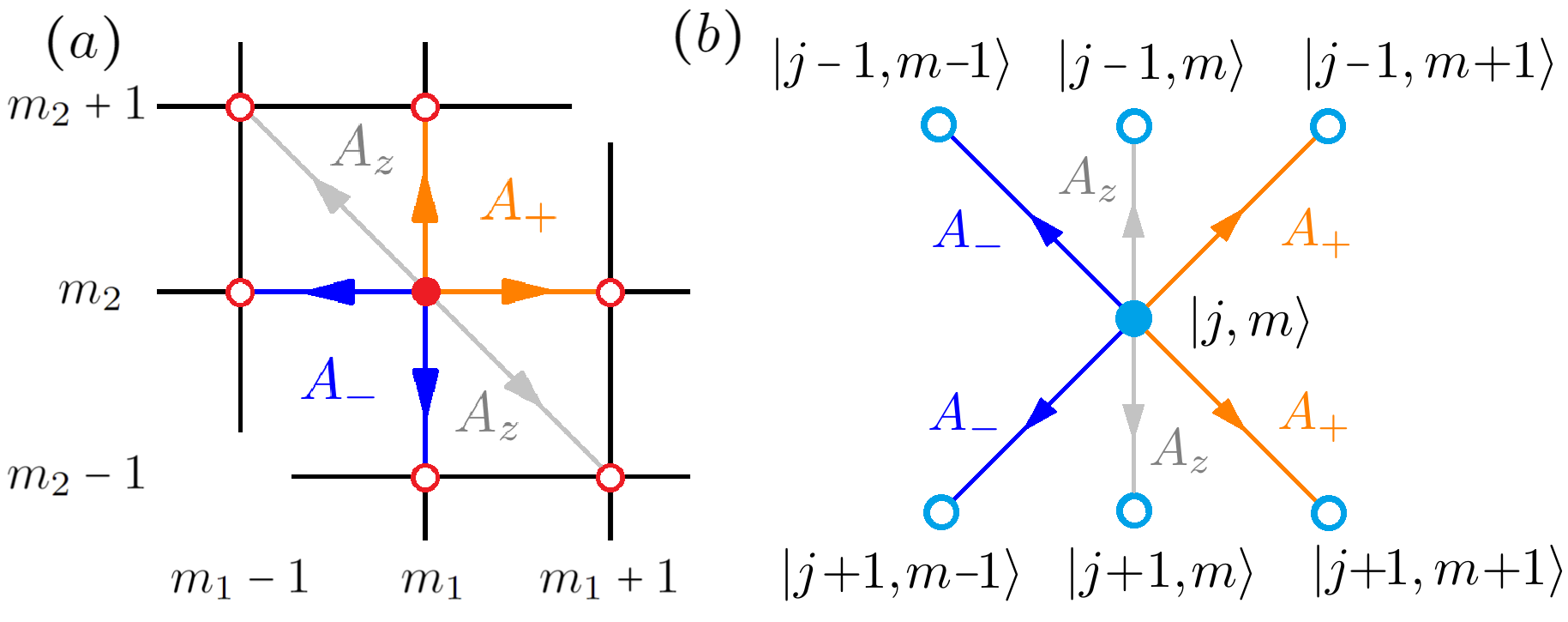}
\caption{\label{fig:coupling} Coupling patterns of the $A_z$ (grey), $A_+$ (orange), and $A_-$ (blue) operators in (a) the $\| m_1, m_2 \rrangle$ basis and in (b) the $|j,m\rangle$ basis. The operators couple a state (solid circle) to its nearest neighbors (empty circles).}
\end{figure}

Matrix elements of $\Lambda$ and $\mathbf{J}$ have already been discussed in Sec.~\ref{sec:Review}. For the vector operator $\mathbf{A}$, its matrix elements in the $\| m_1, m_2 \rrangle$ basis follow directly from its definition in Eq.~(\ref{eq:A}). Using $J_x=\frac{1}{2}(J_+ + J_-)$ and $J_y=\frac{1}{2i}(J_+ - J_-)$, $A_z=J_{1x}J_{2y}-J_{1y}J_{2x}$ is equivalent to
\begin{equation}
    \label{eq:Az}
    A_z = \frac{i}{2}(J_{1+}J_{2-} - J_{1-}J_{2+}).
\end{equation}
On a lattice of spins, this would be a hopping-type operator. 
From the above equation, %it is clear that $A_z$ is a Hermitian operator. Moreover, the 
matrix elements of $A_z$ in the $\| m_1, m_2 \rrangle$ basis can be expressed in terms of Eq.~(\ref{eq:J_elements}). 
Analogous to $J_\pm$, we introduce $A_\pm=A_x \pm i A_y$. Using Eq.~(\ref{eq:AJ_table}), we obtain
\begin{equation}
    \label{eq:Apm}
    A_\pm = \pm [ A_z, J_\pm] = \pm[J_z, A_\pm],
\end{equation}
from which matrix elements of $A_\pm$, and therefore $A_x$ and $A_y$, are readily obtained in the $\| m_1, m_2 \rrangle$ basis. A useful expression is $A_\pm = \pm i (J_{1z}J_{2\pm}-J_{1\pm}J_{2z})$. The coupling pattern of $A_z$ and $A_\pm$ is shown in Fig.~\ref{fig:coupling}(a). While $A_\pm$ have a similar stencil as $J_\pm$, the $A_z$ operator has a larger stencil than $J_z$.

To derive matrix elements of $\mathbf{A}$ in the $|j,m\rangle$ basis, an important step is to understand its selection rules. From Eq.~(\ref{eq:J2J2A}), we have $[\mathbf{J}^2, [\mathbf{J}^2, \mathbf{A}]]=2(\mathbf{A}\mathbf{J}^2 + \mathbf{J}^2\mathbf{A})$, which is identical to $[\mathbf{L}^2, [\mathbf{L}^2, \mathbf{r}]]=2(\mathbf{r}\mathbf{L}^2 + \mathbf{L}^2\mathbf{r})$ after the replacements $\mathbf{L}\rightarrow\mathbf{J}$ and $\mathbf{r}\rightarrow\mathbf{A}$. Therefore, the selection rules for $\mathbf{A}$ are exactly the same as the selection rules for an atom interacting with a plane electromagnetic wave in the electric dipole approximation. 
As shown in Appendix~\ref{App:selection}, the selection rule in $j$ is
\begin{equation}
    \label{eq:selection_rule_j}
    \langle j', m'|\mathbf{A}|j, m\rangle=0,\; \text{unless } j'=j\pm1.
\end{equation}
To obtain selection rules in $m$, we use Eq.~(\ref{eq:AJ_table}), which gives $0=\langle j', m'|[A_z, J_z]|j, m\rangle=(m-m')\langle j', m'|A_z|j, m\rangle$. Consequently,
\begin{equation}
    \label{eq:selection_rule_m}
    \langle j', m'|A_z|j, m\rangle=0,\; \text{unless } m'=m.
\end{equation}
Using this result and Eq.~(\ref{eq:Apm}), we see $\langle j', m'|A_+|j, m\rangle=0$ unless $m'=m+1$ and $\langle j', m'|A_-|j, m\rangle=0$ unless $m'=m-1$. 
The coupling pattern of $A_z$ and $A_\pm$ in the $|j,m\rangle$ basis is shown in Fig.~\ref{fig:coupling}(b). We see $\mathbf{A}$ is sparse in both the computational and the problem basis.

The selection rules require that $A_z|j,m\rangle = a_j^m|j-1, m\rangle + \bar{a}_{j+1}^{m} |j+1, m\rangle$. %, where $a_j^m=\langle j-1, m|A_z|j, m\rangle$. 
In Appendix~\ref{App:A_z}, we solve for $|a_j^m|$ using a recurrence relation induced by Eq.~(\ref{eq:AS_table}).
It is convenient to expresses $|a_j^m|=\frac{1}{2}\alpha_j \zeta_j(m)$, where
\begin{equation}
    \label{eq:zeta_j}
    %\zeta_j(m)=\sqrt{j^2-m^2},
    \zeta_j^m=\sqrt{j^2-m^2}. 
\end{equation}
%which respects the fact that all states must have $|m|\le j$. 
Because $a_j^{\pm j}$ is the coefficient of forbidden states $|j-1,m={\pm j}\rangle$, the form of $\zeta_j^m$ guarantees that $a_j^{\pm j}$ vanishes. The remaining dependence of $a_j^m$ on $j$ is captured by
\begin{equation}
    \label{eq:alpha_j}
    \alpha_j=\sqrt{\frac{[(j_1+j_2+1)^2-j^2][j^2-(j_1-j_2)^2]}{4j^2-1}}.
\end{equation}
The form of $\alpha_j$ guarantees that $a_{j_{\max}+1}^m=a_{j_{\min}}^m=0$, because $A_z$ cannot raise $j$ beyond its top value, or lower $j$ below its bottom value. 
Notice that $\alpha_j$ is well behaved when $j=0$ and $j=\frac{1}{2}$. 
This is because in order for $j=0$, we must have $j_1=j_2$, in which case $\alpha_0=0$. Likewise, in order for $j=\frac{1}{2}$, we must have $j_1-j_2=\frac{1}{2}$, in which case the denominator cancels with the second term in the numerator. 
Since $j_{\min}\le j \le j_{\max}$, $\alpha_j$ is always real and nonnegative. 
As shown in Appendix~\ref{App:A_z}, the correct phase of $a_j^m$ is such that
\begin{equation}
    \label{eq:Az_elements}
    A_z|j,m\rangle =\frac{i}{2}\Big[\zeta_j^m\alpha_j |j-1, m\rangle - \zeta_{j+1}^m\alpha_{j+1} |j+1, m\rangle\Big].
\end{equation}
The phase of $a_j^m$ is determined using two other recurrence relations induced by Eqs.~(\ref{eq:Az}) and (\ref{eq:Apm}).

%(a), (b), m_2, m_2+1, m_2-1, m_1, m_1+1, m_1-1$

From matrix elements of $A_z$, we can readily obtain matrix elements of $A_\pm$. 
Analogous to Eq.~(\ref{eq:J_elements}), we introduce
\begin{equation}
    \label{eq:J0_element}
    \mathcal{J}_j^0(m)=\sqrt{(j+1+m)(j+m)},
\end{equation}
which satisfies $\mathcal{J}_{j+l}^0(m)=\mathcal{J}_j^0(m+l)$.
Using Eq.~(\ref{eq:Apm}), we show in Appendix~\ref{App:Apm-S} that
\begin{eqnarray}
    \label{eq:Ap_element} 
    \nonumber
    A_+|j,m\rangle &=& \frac{i}{2}\Big[\alpha_j\mathcal{J}_{j-1}^0(-m)  |j-1,m+1\rangle \\
    && + \ \alpha_{j+1}\mathcal{J}_{j+1}^0(m) |j+1,m+1\rangle \Big], \\
    \label{eq:Am_element} 
    \nonumber
    A_-|j,m\rangle &=& -\frac{i}{2}\Big[\alpha_j\mathcal{J}_{j-1}^0(m)  |j-1,m-1\rangle \\
    && + \, \alpha_{j+1}\mathcal{J}_{j+1}^0(-m)  |j+1,m-1\rangle \Big].    
\end{eqnarray}
Following similar steps, we obtain matrix elements of $\mathbf{S}$. 
As a quantum operator, $\mathbf{S}$ is not orthogonal to $\mathbf{A}$, so $\mathbf{S}$ does not introduce distinctively new matrix elements.  %For our state preparation scheme, $\mathbf{S}$ turns out to be not useful. 
For completeness, we list matrix elements of $\mathbf{S}$ in Appendix~\ref{App:Apm-S}.
%Because matrix elements of $\mathbf{S}$ follow from $\mathbf{A}$, 
The coupling patterns of $\mathbf{S}$ are identical to $\mathbf{A}$ in the $|j,m\rangle$ basis, which are shown in Fig.~\ref{fig:coupling}(b). 
However, because $\mathbf{S}=i[\mathbf{A}, \Lambda]$ is a higher order operator, where $\Lambda$ couples $\| m_1, m_2 \rrangle$ with %itself as well as $\| m_1+1, m_2-1 \rrangle$ and $\| m_1-1, m_2+1 \rrangle$, 
$\| m_1\pm 1, m_2 \mp 1 \rrangle$,
the coupling patterns of $\mathbf{S}$ in the computational basis have a larger stencil. %, which means that $\mathbf{S}$ is less sparse. % in the computational basis. 

%%%%%%%%%%%%%%%%%%%%%%%%%%%%%%%%%%%%%%%%%%%%%
\section{\label{sec:walk}Double-pinched quantum walks}
Having understood the state spaces and properties of the $\mathfrak{su}(2)\times \mathfrak{su}(2)$ generators, we are ready to introduce the core concept of this paper: double-pinched quantum walks. 
A usual quantum walk is specified by a Hamiltonian acting on a Hilbert space, and quantum evolution spreads an initial state over the part of the Hilbert space that is connected by the Hamiltonian.  
Spreading is desirable in search-like applications, which aim to use quantum evolution to rapidly explore the state space. 
However, spreading is undesirable for state preparation, because instead of producing a specific state, spreading produces a superposition of states, which need to be post selected. A large spreading means a small success probability during post selection, which results in inefficiency for state preparation.

To maximize the success probability of state preparation, the key is to reduce spreading. In particular, when quantum evolution is reduced to a two-level system, a unit success probability can be attained for transferring population from one level to the other. Attaining a unit success probability does not sacrifice the ability to move across the entire Hilbert state, if we change the Hamiltonian. 
Suppose we want to evolve quantum states deterministically along a chain $|\psi_0\rangle\rightarrow |\psi_1\rangle\rightarrow \dots \rightarrow |\psi_{N}\rangle$, then what we need is a sequence of Hamiltonians $H_k$ and accompanying time steps $t_k$, such that $|\psi_{k}\rangle=\exp(-iH_{k} t_{k})|\psi_{k-1}\rangle$. 
By {\it engineered quantum walks} we mean designing the sequence $\{(H_k, t_k)\}_{k=1}^{N}$ to achieve population transfers along a chain of states.

At each step of the engineered quantum walk, we need double pinched $H_k$ to isolate the two states of interest from the rest of the Hilbert space. Because the dynamics is effectively two dimensional, the evolution time $t_k$ corresponds to the duration of a $\pi$ pulse in the two level system. With $|\psi_k\rangle$ as the initial state, evolving it under $H_k$ for a time $t_k$ results in a deterministic population transfer $|\psi_{k-1}\rangle\rightarrow |\psi_k\rangle$. 
Errors in the unitary evolution reduce the success probability of the population transfer, and cause state leakage in subsequent steps of the engineered quantum walk. Therefore, it is crucial to control errors at each step to maximize the overall success probability of state preparation. 

%%%%%%%%%%%%%%%%%%%%%%
\subsection{\label{sec:MWalk} Walks in $m$ direction}
To achieve double pinched walk along the $m$ direction, using projection operators is sufficient. Simple projections work because $m=m_1+m_2$ is already highly constrained. The projection to the $m$ subspace can be written explicitly in both bases as
% \begin{subequations}
%     \label{eq:projection_all}
%     \begin{eqnarray}    
%         \label{eq:projection_jm}
%         P_{m-1/2} &=& \sum_j \Big(|j,m-1\rangle\langle j, m-1| + |j,m\rangle\langle j, m|\Big) \\
%         \label{eq:projection_m1m2}
%         &=& \Big(\sum_{\substack{m_1+m_2\\=m-1}} + \sum_{\substack{m_1+m_2\\=m}} \Big) \| m_1, m_2 \rrangle \llangle m_1, m_2\|. \quad       
%     \end{eqnarray}
% \end{subequations}
\begin{equation}    
    \label{eq:projection}
    P_m = \sum_j |j,m\rangle\langle j, m|=\! \sum_{m_1+m_2=m}\! \| m_1, m_2 \rrangle \llangle m_1, m_2\|,   
\end{equation}
% \begin{equation}    
%     \label{eq:projection}
%     P_m = \sum_j |j,m\rangle\langle j, m|= \sum_{\substack{m_1+m_2\\=m}} \| m_1, m_2 \rrangle \llangle m_1, m_2\|,   
% \end{equation}
where the coefficients in both bases are simply one.
The two expressions give the same projection because all states with $m=m_1+m_2$ live on the same grey plane in Fig.~\ref{fig:domain} regardless of the basis. One can understand $P_m$ as the projection to the $m$-th grey plane. %, and is therefore basis independent.
For later convenience, let us also introduce
\begin{equation}
    \label{eq:project2}
    P_{m-1/2} = P_{m-1} + P_m, 
\end{equation}
which projects to two adjacent grey planes in Fig.~\ref{fig:domain}. The projection operator $P_{m-1/2}$ blocks transitions that leave the $m$ and $m-1$ subspace, and achieves double pinch in the $m$ direction.

\begin{figure}[t]
\includegraphics[width=0.45\textwidth]{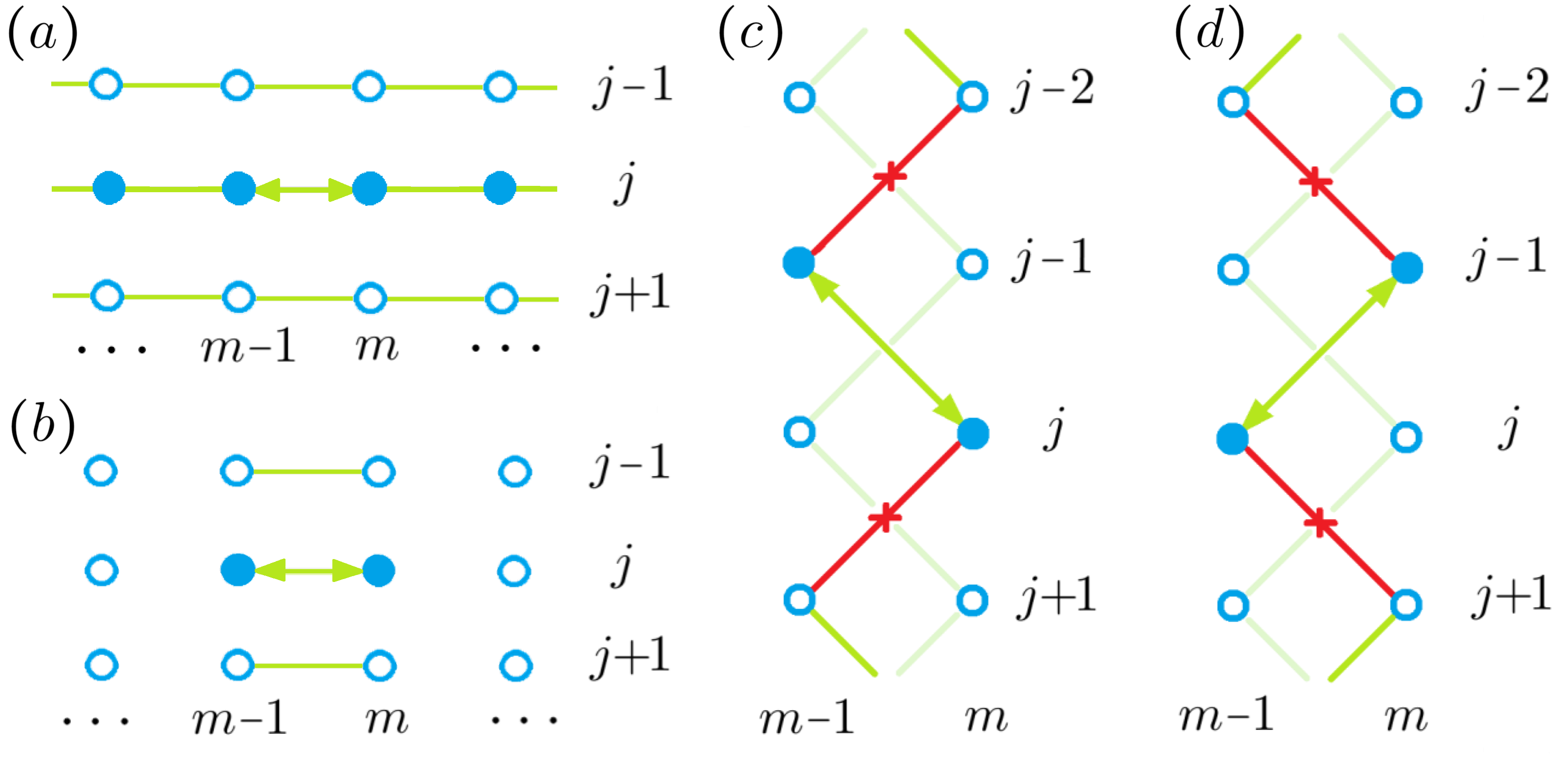}
\caption{\label{fig:pinch} Coupling patterns in the $|j,m\rangle$ basis for a step of the quantum walks. Active states with nonzero occupations are marked by solid blue dots, and inactive states are marked by empty circles. Allowed transitions are indicated by green lines, and desired transitions are indicated by green arrows. Red lines with crosses indicate transitions that are deliberately blocked. 
(a) Usual magnetic resonance couples states with the same $j$, while $m$ is allowed to spread across the entire lattice. 
(b) Using projection operators, the Hamiltonian $M_{m-1/2}$ couples $m$ and $m-1$ states for a given $j$ value. 
(c) Additionally, using destructive interference to block transitions, the Hamiltonian $L_{m-1/2}^{j-1/2}$ isolates states $|j,m\rangle$ and $|j-1, m-1\rangle$ into a two-level system. 
(d) Similarly, the Hamiltonian $R_{m-1/2}^{j-1/2}$ isolates two states $|j,m-1\rangle$ and $|j-1, m\rangle$. %, achieving double pinch using both projections and destructive interference.
}
\end{figure}

To walk between $|j,m\rangle$ and $|j, m-1\rangle$, we change $m$ using the $J_\pm$ operator. This is analogous to magnetic resonance, where the $j$ value is fixed while the spin rotates around the Bloch sphere. 
The coupling pattern of usual magnetic resonance is illustrated in Fig.~\ref{fig:pinch}(a), where $m$ continues to spread to its nearest neighbors. 
For $j>\frac{1}{2}$, we reduce the dynamics to two-level systems using the double-pinched Hamiltonian
\begin{equation}
    \label{eq:M_Hamiltonian}
    M_{m-1/2} = P_{m-1/2}(J_+ + J_-)P_{m-1/2},
\end{equation}
which is clearly a Hermitian operator. 
The projection operators on the right select to act on states with quantum numbers $m$ or $m-1$. The $J_+ + J_-=2J_x$ operator in the middle rotates the spin around the $x$ axis of the Bloch sphere. Similar effects can be achieved using $2J_y =i(J_- - J_+)$, or other combinations of $J_x$ and $J_y$. Finally, the projection operators on the left discard transitions out of the $m$ and $m-1$ subspace, thereby achieving the desired double-pinch effect.

In the $|j,m\rangle$ basis, the coupling pattern of $M_{m-1/2}$ is shown in Fig.~\ref{fig:pinch}(b), and the matrix elements are
\begin{equation}
    \label{eq:M_jm}
    M_{m-1/2} |j,m\rangle = \mathcal{J}_j^-(m) |j, m-1\rangle.
\end{equation}
In other words, $M_{m-1/2}$ acts on all $j$ states in parallel, but with a different matrix element for each $j$. Because $J_\pm$ does not change $j$, the dynamics is reduced to a collection of decoupled two-level systems. 
Using $\mathcal{J}_j^+(m-1)=\mathcal{J}_j^-(m)$, %we see $M_{m-1/2} |j,m-1\rangle = \mathcal{J}_j^-(m) |j, m\rangle$. % has the same matrix element. 
%Therefore, 
when restricted to the two-dimensional Hilbert space spanned by $|j,m\rangle$ and $|j,m-1\rangle$, the Hamiltonian matrix is $M_{m-1/2}=\mathcal{J}_j^-(m)\sigma_x$, where $\sigma_x$ is the Pauli $x$ matrix. 
In this two-level system, the unitary evolution operator is $U_{M_{m-1/2}}(t)=\exp(-iM_{m-1/2} t) = \cos[\mathcal{J}_j^-(m) t]-i\sin[\mathcal{J}_j^-(m) t]\sigma_x$. The $\pi$ pulse time when the population is completely flipped is 
\begin{equation}
    \label{eq:M_tpi}
    t_M=\frac{\pi}{2\mathcal{J}_j^-(m)}.
\end{equation}
In other words, $iU_{M_{m-1/2}}(t_M)$ is the unitary that achieves the swap
$|j,m\rangle\leftrightarrow |j,m-1\rangle$.
Notice that $t_M$ is specific to $j$, which means that a $\pi$ pulse for $j$ is not a $\pi$ pulse for $j'\ne j$. 
Therefore, when using $M_{m-1/2}$ to walk $|j,m\rangle\leftrightarrow |j,m-1\rangle$, one must target specific $j$ and $m$ values by ensuring that the initial state is one of the two levels. 
For example, to prepare the state $|j_{\max}, j_{\max}-2\rangle$, one can start from the top state $|j_{\max}, j_{\max}\rangle=\|j_1, j_2\rrangle$. The first step of the quantum walk uses $M_{j_{\max}-1/2}$ to evolve for a time $t_M$ where $j=m=j_{\max}$, which achieves $|j_{\max}, j_{\max}\rangle\rightarrow|j_{\max}, j_{\max}-1\rangle$. The second step uses a different $M_{j_{\max}-3/2}$ for a different time $t_M$ where $j=j_{\max}$ and $m=j_{\max}-1$ to achieve $|j_{\max}, j_{\max}-1\rangle\rightarrow|j_{\max}, j_{\max}-2\rangle$, completing the state preparation.
The behavior of this quantum walk is more complicated if the initial state is different.

%$(a), (b), (c), (d)$

In the computational basis $\|m_1,m_2\rrangle$, the matrix representation of $M_{m-1/2}$ is 2-sparse.
Due to the projection operators, the matrix elements are zero outside the $m$ and $m-1$ subspace. Within the subspace, because $\mathbf{J}=\mathbf{J}_1 + \mathbf{J}_2$, the matrix elements of $M_{m-1/2}$ when $m_1+m_2=m$ are 
\begin{eqnarray}
    \label{eq:M_m1m2}
    \nonumber
    M_{m-1/2} \|m_1,m_2\rrangle &=& \mathcal{J}_{j_1}^-(m_1) \|m_1-1,m_2\rrangle \\
    &+& \mathcal{J}_{j_2}^-(m_2) \|m_1,m_2-1\rrangle.
\end{eqnarray}
The matrix elements when $m_1+m_2=m-1$ are related. % by $\mathcal{J}_{j}^-(m) = \mathcal{J}_{j}^+(m-1)$. 
Notice that the matrix has at most two nonzero elements in each row and column. 
Because $M_{m-1/2}$ couples two neighboring grey planes in Fig.~\ref{fig:domain}, the number of nonzero matrix elements is $O(j_2)$, whereas the dimension of the computational basis is $O(j_1j_2)$. %Moreover, Eq.~(\ref{eq:M_m1m2}) only couples a subset of nearest neighbors in the computational basis, where 

The quantum walk induced by $M$ can be implemented using quantum Hamiltonian simulation algorithms.
For example, using a qubitization algorithm \cite{low2019hamiltonian}, the query complexity, namely, the number of terms one needs to keep in the Jacobi–Anger expansion for a given precision $\epsilon\ll 1$, is $O[\tau+\log(1/\epsilon)/\log\log(1/\epsilon)]$, where $\tau=t\|H\|_{\max}$ is the normalized simulation time. 
When $j_1\ge j_2\rightarrow\infty$, the matrix norm is $\|M\|_{\max}\simeq \sqrt{j_1^2-m_1^2}$. In the best case $|m_1|\rightarrow j_1$, we have $\|M\|_{\max}=O(1)$, and in the worst case $m_1=O(1)$, we have $\|M\|_{\max}=O(j_1)$.
On the other hand, the $\pi$-pulse time [Eq.~(\ref{eq:M_tpi})] depends on the total $j$ and $m$, rather than the individual angular momentum. In the best case $j\sim j_{\max}=O(j_1)$ and $m=O(1)$, we have $t_M=O(1/j_1)$, and in the worst case $j\sim m= O(1)$, we have $t_M=O(1)$, as shown in Fig.~\ref{fig:scaling}(a).
To attain the worst case for $t_M$, $j\ge j_{\min}$ needs to be close to zero, which requires $j_1\sim j_2$. This is why the worst complexity peaks near $j_1\sim j_2$, as shown in Fig.~\ref{fig:scaling}(d).
The first off-diagonal, which attains the minimum $j=m=1/2$, is worse than the diagonal, which attains a larger minimum $j=m=1$, by a factor of $\sqrt{2}$.
Because the best cases for $\|M\|_{\max}$ and $t_M$ cannot be attained simultaneously due to $m=m_1+m_2$, overall, $\tau=O(1)$ in the best case and $\tau\sim O(j_1)$ in the worst case, as shown in Fig.~\ref{fig:scaling}(e).  
The complexity may also be understood as the stiffness of $M$, namely, the ratio of its largest and smallest eigenvalues.  
Notice that regardless of the value of $j$, the matrix $M$ is the same [Eq.~(\ref{eq:M_Hamiltonian})], but the matrix elements [Eq.~(\ref{eq:M_jm})] differ. 
These matrix elements are the absolute values of eigenvalues of $M$.
For two-level systems with a larger coupling, a smaller simulation time $\tau=O(1)$ is sufficient; whereas for two-level systems with a smaller coupling, a longer simulation time $\tau=O(j_1)$ is required.

%%%%%%%%%%%%%%%%%%%%%%
\subsection{Walks in $j$ direction \label{sec:DWalk}}
To achieve double-pinched walk that changes the value of $j$, we use destructive interference as an additional key ingredient.  
A na\"{i}ve generalization of Eq.~(\ref{eq:M_Hamiltonian}) would invoke projection operators like $P^{j}=\sum_m |j,m\rangle\langle j,m|$, which projects to an orange plane in Fig.~\ref{fig:domain}. Although $P^{j}$ has a simple expression in the $|j,m\rangle$ basis, its expression in the computational basis involves CG coefficients $P^{j}=\sum_m C^j_{m_1, m_2} \| m_1, m_2 \rrangle \llangle m'_1, m'_2\| C^j_{m'_1, m'_2}$, where the summation is over all $m_1+m_2=m'_1+m'_2=m$. Because the CG coefficients are dense, the expression of $P^{j}$ in the computational basis is complicated, which means that a direct analogy of Eq.~(\ref{eq:M_Hamiltonian}) is a dense Hamiltonian that can not be simulated efficiently. 
To circumvent this difficulty, we utilize $\mathbf{J}$, $\mathbf{A}$, and $\Lambda$, which are sparse  operators with known matrix elements in both bases, to construct sparse Hamiltonian $L_{m-1/2}^{j-1/2}$ that achieves $|j,m\rangle \leftrightarrow|j-1, m-1\rangle$, which moves left and up the state pyramid, and $R_{m-1/2}^{j-1/2}$ that achieves $|j,m-1\rangle \leftrightarrow|j-1, m\rangle$, which moves right and up the pyramid. The coupling patterns of these two operators are illustrated in Figs.~\ref{fig:pinch}(c) and \ref{fig:pinch}(d).
One could also design operators that move states vertically, namely, change the value of $j$ at fixed $m$. However, we find such operators tend to be denser in the computational basis, so we focus on the $L$ and $R$ walks. % that move diagonally on the state pyramid.
As explained in Sec.~\ref{sec:intro}, the key idea is that when multiple independent operators have similar coupling patterns, they can be superimposed to achieve destructive interference, such that undesired transitions are blocked.

To construct a Hamiltonian that achieves a specific move in the $|j,m\rangle$ basis, it is helpful to think in this basis. 
From Sec.~\ref{sec:algebra}, we know $\mathbf{S}$ does not provide independent matrix elements. Therefore, it is sufficient to use $\mathbf{J}$, $\mathbf{A}$, and $\Lambda$. 
The operators that move $|j,m\rangle$ diagonally are $J_+A_z$, $A_zJ_+$, $\Lambda A_+$, $A_+\Lambda$, and their Hermitian conjugates. 
The four operators have similar stencils and complexities in the computational basis.
Notice that $A_+$ is not independent of these four operators because of Eq.~(\ref{eq:Apm}). 
%
%To design destructive interference, it is necessary to know matrix elements of the four operators. In Appendix~\ref{App:Apm-S}, 
When computing matrix elements, we find it convenient to abbreviate %where similar combinations of states and coefficients appear. For compactness, we abbreviate
% \begin{subequations}
%     \label{eq:abbreviate}
%     \begin{eqnarray}    
%         |+,+\rangle &=& \alpha_{j+1}\mathcal{J}^0_{j+1}(m)|j+1, m+1\rangle, \\
%         |+,-\rangle &=& \alpha_{j+1}\mathcal{J}^0_{j+1}(-m)|j+1, m-1\rangle, \\
%         |-,+\rangle &=& \alpha_j\mathcal{J}^0_{j-1}(-m)|j-1, m+1\rangle,\\        
%         |-,-\rangle &=& \alpha_j\mathcal{J}^0_{j-1}(m)|j-1, m-1\rangle,
%     \end{eqnarray}
% \end{subequations}
$|+,+\rangle := \alpha_{j+1}\mathcal{J}^0_{j+1}(m)|j+1, m+1\rangle$,
$|+,-\rangle := \alpha_{j+1}\mathcal{J}^0_{j+1}(-m)|j+1, m-1\rangle$,
$|-,+\rangle := \alpha_j\mathcal{J}^0_{j-1}(-m)|j-1, m+1\rangle$,
and
$|-,-\rangle := \alpha_j\mathcal{J}^0_{j-1}(m)|j-1, m-1\rangle$. 
%where $\alpha_j$ is given by Eq.~(\ref{eq:alpha_j}). 
%These $|\pm,\pm\rangle$ states make reference to, and are centered around, the $|j,m\rangle$ state. 
%Because $|\pm,\pm\rangle$ states absorb extra coefficients into their definitions, these states are not normalized. What makes these states useful is that they appear repeatedly in matrix elements. Notably, t
The abbreviated notation greatly simplify matrix elements of $A_+$.
%. Using the abbreviations, matrix elements of the four independent operators are
% \begin{subequations}
%     \label{eq:plus_operators}
%     \begin{eqnarray}   
%         J_+A_z|j,m\rangle &=& \frac{i}{2}\Big[(j+m)|-,+\rangle - (j+1-m)|+,+\rangle \Big], \quad \\
%         A_zJ_+|j,m\rangle &=& \frac{i}{2}\Big[(j+1+m)|-,+\rangle - (j-m)|+,+\rangle \Big], \quad \\
%         \Lambda A_+|j,m\rangle &=& \frac{i}{2}\Big(\lambda_{j-1}|-,+\rangle + \lambda_{j+1}|+,+\rangle \Big), \quad \\
%         A_+\Lambda|j,m\rangle &=& \frac{i}{2}\lambda_j\Big(|-,+\rangle + |+,+\rangle \Big).
%     \end{eqnarray}
% \end{subequations}
Moreover, 
$J_+A_z|j,m\rangle =\frac{i}{2}[(j+m)|-,+\rangle - (j+1-m)|+,+\rangle ]$, 
$A_zJ_+|j,m\rangle = \frac{i}{2}[(j+1+m)|-,+\rangle - (j-m)|+,+\rangle ]$,
$\Lambda A_+|j,m\rangle = \frac{i}{2}(\lambda_{j-1}|-,+\rangle + \lambda_{j+1}|+,+\rangle)$,
$A_+\Lambda|j,m\rangle = \frac{i}{2}\lambda_j(|-,+\rangle + |+,+\rangle )$, 
where $\lambda_j$ is given by Eq.~(\ref{eq:lambda_j}). 
Similarly, one can find matrix elements of Hermitian conjugates of these four operators.

Using $J_+A_z$, $A_zJ_+$, $\Lambda A_+$ and $A_+\Lambda$ as building blocks, whose matrix elements in the computational basis are listed in Appendix~\ref{App:buildingblocks_general}, 
we construct a general Hamiltonian that moves $|j,m\rangle$ diagonally in the $j$-$m$ lattice. The general Hamiltonian is of the form
\begin{equation}
    \label{eq:H_general}
    H = p J_+A_z + q A_zJ_+ + u\Lambda A_+ + v A_+\Lambda + \text{h.c.},
\end{equation}
where $p, q, u, v \in\mathbb{C}$ are coefficients to be determined and h.c. denotes Hermitian conjugates. When acting on $|j,m\rangle$, matrix elements of $H$ are
\begin{widetext}
\begin{eqnarray}
    \label{eq:H_general_elements}   
    \nonumber
     H|j,m\rangle = \frac{i}{2}\Big \{
    [p(j+m)+q(j+1+m)+u\lambda_{j-1}+v\lambda_j] |-,+\rangle 
    &-& [p(j+1-m)+q(j-m)-u\lambda_{j+1}-v\lambda_j] |+,+\rangle  
    \\
    - [\bar{p}(j+m)+\bar{q}(j+1+m)+\bar{u}\lambda_{j}+\bar{v}\lambda_{j+1}] |+,-\rangle
    &+& [\bar{p}(j+1-m)+\bar{q}(j-m)-\bar{u}\lambda_{j}-\bar{v}\lambda_{j-1}] |-,-\rangle \Big\}.
\end{eqnarray}
\end{widetext}
% \begin{eqnarray}
%     \label{eq:H_general_elements}
%     \nonumber
%     2iH|j,m\rangle &=& [p(j+1-m)+q(j-m)-u\lambda_{j+1}-v\lambda_j] |+,+\rangle \\ 
%     \nonumber
%     &-& [\bar{p}(j+1-m)+\bar{q}(j-m)-\bar{u}\lambda_{j}-\bar{v}\lambda_{j-1}] |-,-\rangle \\
%     \nonumber
%     &-& [p(j+m)+q(j+1+m)+u\lambda_{j-1}+v\lambda_j] |-,+\rangle \\
%     &+& [\bar{p}(j+m)+\bar{q}(j+1+m)+\bar{u}\lambda_{j}+\bar{v}\lambda_{j+1}] |+,-\rangle,
% \end{eqnarray}
To find matrix elements of $H$ acting on other $j$ and $m$ states, we just need to shift the definition of $|\pm,\pm\rangle$ states.% around the new origin. 
We engineer the Hamiltonian to achieve desired movements by choosing the four coefficients. % appropriately. 

To construct the operator $L_{m-1/2}^{j-1/2}$ that moves left and up the pyramid in the state space, we consider a Hamiltonian of the form $L=P_{m-1/2}HP_{m-1/2}$. The projection operator $P_{m-1/2}$, which is defined in Eq.~(\ref{eq:project2}), restricts the dynamics to the $m$ and $m-1$ subspace. 
As shown in Fig.~\ref{fig:pinch}(c), the goal is to block the $|j,m\rangle\leftrightarrow |j+1, m-1\rangle$ transition, which is achieved by setting the coefficient of $|+,-\rangle$ in Eq.~(\ref{eq:H_general_elements}) to zero, as well as to block the $|j-1,m-1\rangle\leftrightarrow |j-2, m\rangle$ transition, which is achieved by setting the coefficient of $|-,+\rangle$ in Eq.~(\ref{eq:H_general_elements}) to zero, after shifting $j\rightarrow j-1$ and $m\rightarrow m-1$. 
The two constraints are two linear equations for the four coefficients. We solve $p$ and $q$ in terms of $u$ and $v$, which gives the conditions for the $L_{m-1/2}^{j-1/2}$ operator as
\begin{equation}
    \label{eq:L_conditions}
    \left(\! \begin{array}{c}
    p \\ q
    \end{array} \!\right)
    =
    \left(\! \begin{array}{cc}
    \lambda_j\!-\!(k+1)(j-\frac{1}{2}) & \lambda_{j+1}\!-\!(k+1)(j+\frac{1}{2}) \\
    k(j-\frac{1}{2}) \!-\! \lambda_j & k(j+\frac{1}{2})\!-\!\lambda_{j+1} \\
    \end{array} \!\right)
    \!
    \left(\! \begin{array}{c}
    u \\ v
    \end{array} \!\right),
\end{equation}
where $k=j+m$ and we have used $\lambda_{j+1}-\lambda_{j-1}=2j+1$. 
%The solution space of Eq.~(\ref{eq:L_conditions}) is a two dimensional complex vector space. In other words, the 
The choice of coefficients is not unique, similar to Eq.~(\ref{eq:M_Hamiltonian}), where we could have chosen other linear combinations of $J_x$ and $J_y$. A viable choice for $L_{m-1/2}^{j-1/2}$ is $u=v=-1$, which gives $p=2j(k+1)-(\lambda_j+\lambda_{j+1})=j^2+2jm-1+\mathscr{J}_{j_1}+\mathscr{J}_{j_2}$ and $q=\lambda_j+\lambda_{j+1}-2jk = 2j-p$.
With this choice, we have $H=p[J_+, A_z] + 2j A_z J_+ - (\Lambda A_+ + A_+\Lambda) + $ h.c.. Using Eq.~(\ref{eq:Apm}) and denoting $\mathscr{J}=\mathscr{J}_{j_1}+\mathscr{J}_{j_2} -1$, we obtain
% \begin{equation}
%     \label{eq:L_Hamiltonian}
%     L_{m\!-\!1/2}^{j\!-\!1/2}\! =\! P_{m\!-\!\frac{1}{2}}[\!2j\!A_z\!J_+\! - (j^2\!+2jm+\!\mathscr{J}\!)A_+\! - \{\!\Lambda,\! A_+\!\} + \text{h.c.}]P_{m\!-\!\frac{1}{2}}.
% \end{equation}
\begin{eqnarray}
    \label{eq:L_Hamiltonian}
    \nonumber
    L_{m-1/2}^{j-1/2} = P_{m-1/2}[2jA_zJ_+ &-& (j^2+2jm+\mathscr{J})A_+ \\
    - (\Lambda A_+ &+& A_+\Lambda) + \text{h.c.}]P_{m-1/2}. \quad
\end{eqnarray}
By construction, the $|j,m\rangle\leftrightarrow |j+1, m-1\rangle$ and $|j-1,m-1\rangle\leftrightarrow |j-2, m\rangle$ transitions are blocked. With the additional help of the projection operator $P_{m-1/2}$, the transition of interest $|j,m\rangle\leftrightarrow |j-1, m-1\rangle$ is isolated from the rest of the Hilbert space, thereby achieving the desired double-pinch effect, provided that the coefficient of $|-,-\rangle$ is nonzero.
Let us check that with the choice $u=v=-1$, the coefficient of $|-,-\rangle$ becomes $\bar{p}(j+1-m)+\bar{q}(j-m)-\bar{u}\lambda_{j}-\bar{v}\lambda_{j-1} = p + 2j(j-m) + \lambda_j + \lambda_{j-1} = 4j^2-1$, which is indeed nonzero. 
In general, the coefficient of $|-,-\rangle$ is $(\frac{1}{2}-2j^2)(\bar{u}+\bar{v})$. If one had chosen $u=-v$, then the coefficient of $|-,-\rangle$ would be zero. In other words, while some choices block undesired transitions, they may also cut off the desired transition. 
With the additional constraint that the coefficient of $|-,-\rangle$ is nonzero, the solution space to Eq.~(\ref{eq:L_conditions}) is a two-dimensional complex vector space minus the line $u=-v$, up to an overall unimportant complex scaling factor. 
In this paper, we choose $u=v=-1$, but better choices may exist. %, but we do not explored them here.

After restoring all prefactors, the matrix element of $L_{m-1/2}^{j-1/2}$ in the two-level system $|j,m\rangle\leftrightarrow |j-1, m-1\rangle$ is
\begin{equation}
    \label{eq:L_jm}
    L_{m-1/2}^{j-1/2} |j,m\rangle = \frac{i}{2}(4j^2-1)\alpha_j\mathcal{J}^0_{j-1}(m) |j-1, m-1\rangle.
\end{equation}
Let us check the behavior of $L_{m-1/2}^{j-1/2}$ is limiting cases. First, notice that $4j^2-1$ partly cancels with the denominator of $\alpha_j$ in Eq.~(\ref{eq:alpha_j}), which means when $j=1/2$, the matrix element becomes zero. This is expected because when $j=1/2$, one cannot further lower $j\rightarrow j-1$ below zero.
Second, notice that from Eq.~(\ref{eq:J0_element}), $\mathcal{J}^0_{j-1}(m)=\sqrt{(j+m)(j+m-1)}$ becomes zero when $m=-j$ or $m=-j+1$. This is expected, because when $m=-j$ is already the bottom $m$ state, one cannot further lower $m\rightarrow m-1$. Moreover, when lowering $j$ to $j-1$, the bottom $m$ state becomes $-j+1$. Hence, starting from $m=-j+1$, one cannot move up and left on the state pyramid. 
Having checked the limiting cases, let us find the $\pi$-pulse time. Suppose we abbreviate $L| j,m\rangle = \frac{i}{2}\lambda |j-1,m-1\rangle$, then $L|j-1,m-1\rangle = -\frac{i}{2}\lambda |j,m\rangle$, because $L$ is Hermitian. In other words, in the two level system, the matrix representation is $L=\frac{\lambda}{2}\sigma_y$, where $\sigma_y$ is the Pauli $y$ matrix. 
Since the unitary $U_L(t)=\exp(-iL t) = \cos(\lambda t/2) -i\sin(\lambda t/2)\sigma_y$,  the $\pi$ pulse time when the population is completely flipped is $\lambda t=\pi$, which is given explicitly by
\begin{equation}
    \label{eq:L_tpi}
    t_L=\frac{\pi}{(4j^2-1)\alpha_j \mathcal{J}^0_{j-1}(m)}.
\end{equation}
Suppose the initial state is $|j,m\rangle$, then after evolution using $L$ for a $\pi$-pulse time, we obtain $U_L|j,m\rangle = |j-1, m-1\rangle$. Similarly, suppose the state is initially $|j-1, m-1\rangle$. 
The evolution gives $U_L|j-1,m-1\rangle = -|j, m\rangle$, with an overall minus sign.
The behavior of other initial states under $U_L$ is more complicated. In particular, if the state initially has occupations outside the two-level system, then $U_L$ spreads the state along all green lines in Fig.~\ref{fig:pinch}(c). 
For a successful state preparation, it is crucial that we initialize the state within the two level system.

\begin{figure}[t]
\includegraphics[width=0.48\textwidth]{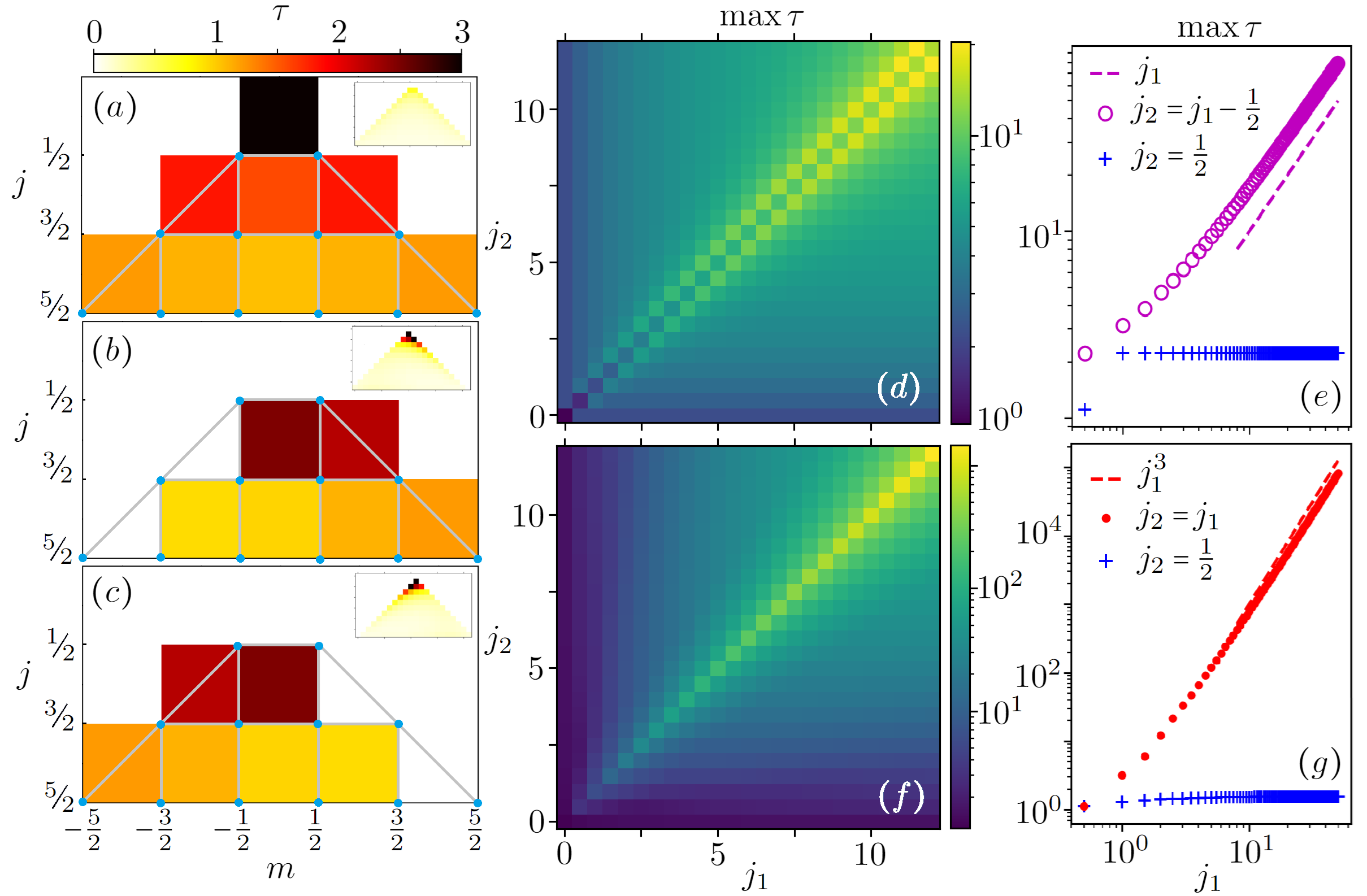}
\caption{\label{fig:scaling} Quantum walks are more expensive near the top and edge of the $|j, m\rangle$ state pyramid (blue dots).
The normalized $\pi$-pulse time $\tau=t\|H\|_{\max}$ is represented by colors in the box where the transition occurs.
For $M$ walk (a), $L$ walk (b), and $R$ walk (c), the main figures show the example $j_1=3/2$ and $j_2=1$.
The insets show the example $j_1=j_2=11/2$, whose color bar is scaled up ten times.
More generally, $\tau$ attains its maximum near $|m|\sim j \sim j_{\min}$. 
(d) For $M$ walks, $\max\tau$ sharply peaks near $j_1 = j_2\pm 1/2$. (e) Near the diagonal, $\max\tau=O(j_1)$. 
(f) For $L$ and $R$ walks, $\max\tau$ sharply peaks near $j_1 = j_2$. (g) Near the diagonal, $\max\tau=O(j_1^3)$. 
Away from the diagonall, $\max\tau$ depends weakly on $j_1\rightarrow \infty$ at a fixed $j_2$ for all walks.
}
\end{figure}
%$(a), (b), (c)$ axis: $j, m= -\frac{5}{2}, -\frac{3}{2}, -\frac{1}{2}$
%data: $0, 1, 2, 3$
%$(d), (e), (f), (g)$ axis: $j_1, j_2= 0, 10, 20, 30, 40, 50$
%data: $10^0, 10^1, 10^2, 10^3, 10^4, 10^5$
%legend: $\max{\tau}, j_2=1/2, j_2=j_1-1/2, j, j^3$

Following similar steps, we construct the $R_{m-1/2}^{j-1/2}$ operator, which moves a targeted state up and right the state pyramid. 
We consider a Hamiltonian of the form $R=P_{m-1/2}HP_{m-1/2}$, where $H$ from Eq.~(\ref{eq:H_general_elements}) is shifted to the left, such that the movements are centered around $|j, m-1\rangle$. 
As shown in Fig.~\ref{fig:pinch}(d), the goal is to block the $|j,m-1\rangle\leftrightarrow |j+1, m\rangle$ transition, which is achieved by setting the coefficient of $|+,+\rangle$ to zero, after shifting $m\rightarrow m-1$. 
Additionally, we block the $|j-1,m\rangle\leftrightarrow |j-2, m-1\rangle$ transition, which is achieved by setting the coefficient of $|-,-\rangle$ to zero, after shifting $j\rightarrow j-1$. 
Solving $p$ and $q$ in terms of $u$ and $v$ for these two linear constraints gives the conditions for the $R_{m-1/2}^{j-1/2}$ operator as
\begin{equation}
    \label{eq:R_conditions}
    \left(\! \begin{array}{c}
    p \\ q
    \end{array} \!\right)
    =
    \left(\! \begin{array}{cc}
    \lambda_{j-1}\!-\!(l-1)(j+\frac{1}{2}) & \lambda_{j}\!-\!(l+1)(j-\frac{1}{2}) \\
    l(j+\frac{1}{2}) \!-\! \lambda_{j-1} & (l+2)(j-\frac{1}{2})\!-\!\lambda_{j} \\
    \end{array} \!\right)
    \!
    \left(\! \begin{array}{c}
    u \\ v
    \end{array} \!\right),
\end{equation}
where $l=j-m$. 
The coefficient of $|-,+\rangle$ with respect to $|j,m-1\rangle$ is $(2j^2-\frac{1}{2})(u+v)$, so we again need $u\ne-v$. 
A viable choice is $u=v=1$, which gives $p=\lambda_j+\lambda_{j-1}-2jl+1=-j^2+2jm-\mathscr{J}$ and $q=2j-p$.
With this choice, $H=p[J_+, A_z] + 2j A_z J_+ + (\Lambda A_+ + A_+\Lambda) + $ h.c. is of a similar form as $L$. Combining with the projection operator, 
\begin{eqnarray}
    \label{eq:R_Hamiltonian}
    \nonumber
    R_{m-1/2}^{j-1/2} = P_{m-1/2}[2jA_zJ_+ &+& (j^2-2jm+\mathscr{J})A_+ \\
    + (\Lambda A_+ &+& A_+\Lambda) + \text{h.c.}]P_{m-1/2}, \quad
\end{eqnarray}
where $\mathscr{J}$ is the same as before Eq.~(\ref{eq:L_Hamiltonian}). 
By construction, we pinch the $|j,m-1\rangle\leftrightarrow |j+1, m\rangle$ and $|j-1,m\rangle\leftrightarrow |j-2, m-1\rangle$ transitions, such that the transition of interest $|j,m-1\rangle\leftrightarrow |j-1, m\rangle$ is isolated from the rest of the Hilbert space.

After restoring all prefactors, the matrix element of $R_{m-1/2}^{j-1/2}$ in the two-level system is
\begin{equation}
    \label{eq:R_jm}
    R_{m-1/2}^{j-1/2} |j,m-1\rangle = \frac{i}{2}(4j^2-1)\alpha_j\mathcal{J}^0_{j}(-m) |j-1, m\rangle.
\end{equation}
Similar to Eq.~(\ref{eq:L_jm}), the matrix element becomes zero when $j=1/2$, which prevents $j\rightarrow j-1$ to lower below zero.
Moreover, since $\mathcal{J}^0_{j}(-m)=\sqrt{(j-m)(j-m+1)}$, the matrix element becomes zero for the top state $m-1=j$ at the $j$-th level of the pyramid, as well as the top state $m-1=j-1$ at the $(j-1)$-th level, which prevents transitions out of the state pyramid.
To better see the behavior of $R$ in the two-level system, we abbreviate $R|j,m-1\rangle = \frac{i}{2}\lambda |j-1,m\rangle$. Then, $R|j-1,m\rangle = -\frac{i}{2}\lambda |j,m-1\rangle$, because $R$ is Hermitian. Similar to $L$, we see the matrix representation of $R$ is $R=\frac{\lambda}{2}\sigma_y$, so the unitary is $U_R(t)=\exp(-iR t) = \cos(\lambda t/2) -i\sin(\lambda t/2)\sigma_y$. The $\pi$ pulse time $t=\pi/\lambda$ is given explicitly by
\begin{equation}
    \label{eq:R_tpi}
    t_R=\frac{\pi}{(4j^2-1)\alpha_j \mathcal{J}^0_{j}(-m)}.
\end{equation}
The behavior of $U_R$ at the $\pi$-pulse time is simple only within the two-level system, where $U_R|j,m-1\rangle = |j-1, m\rangle$ and $U_R|j-1,m\rangle = -|j, m-1\rangle$. 
Outside the two level system, allowed transitions are illustrated in Fig.~\ref{fig:pinch}(d).
For a successful state preparation, it is again crucial that we initialize the state within the two level system.

Finally, let us estimate the query complexity of performing quantum Hamiltonian simulations for the $L$ and $R$ walks. 
The Hamiltonian is a linear combination of the three basic building blocks given by Eqs.~(\ref{eq:AzJ_elements_m1m2})-(\ref{eq:ApLambda_elements_m1m2}), so the Hamiltonian matrix is 4-sparse, which is amenable to efficient quantum Hamiltonian simulations. 
For example, using a qubitization algorithm \cite{low2019hamiltonian}, the query complexity is linear in $\tau=t\|H\|_{\max}$, the normalized simulation time. 
As shown in Figs.~\ref{fig:scaling}(b) and \ref{fig:scaling}(c), for given $j_1$ and $j_2$, $\tau$ is largest near $j\sim m= O(1)$, where the $\pi$ pulse is the longest. 
Moreover, when we scale up $j_1$ and $j_2$, as shown in Figs.~\ref{fig:scaling}(f) and \ref{fig:scaling}(g), the worst complexity $\max\tau$ peaks near $j_1\sim j_2$, where $j_{\min}$ is the smallest.
To estimate the worst-case complexity, consider $j_1\sim j_2\rightarrow\infty$ and $m_1+m_2\sim 0$. 
In this limit, the matrix norms $\|A_zJ_+\|_{\max}\simeq \frac{1}{2}(j_1^2-m_1^2)^{3/2}=O(j_1^3)$ maximizes when $m_1=O(1)$; 
$\|A_+\|_{\max}\simeq m_1(j_1^2-m_1^2)^{1/2}=O(j_1^2)$ maximizes when $m_1\simeq j_1/\sqrt{2}\approx0.7j_1$; and $\|\Lambda A_+ + A_+\Lambda\|_{\max}\simeq 2m_1(j_1^2-m_1^2)^{1/2}(j_1^2-3m_1^2)=O(j_1^4)$ maximizes when $m_1\approx 0.3 j_1$ or $m_1\approx 0.9 j_1$. Because the three building blocks maximize at different $m_1$, away from $|m_1|\sim j_1$, we always have $\|L\|_{\max} = \|R\|_{\max} = O(j_1^4)$ after accounting for coefficients of the three building blocks. 
For the $\pi$-pulse time Eqs.~(\ref{eq:L_tpi}) and (\ref{eq:R_tpi}), the denominators are approximately $j(j\pm m)\sqrt{(4j^2-1)(4j_1^2-j^2)}$. In the worst case $j\sim m = O(1)$, we have $t_{L/M}=O(1/j_1)$, and in the best case $j\sim |m|=O(j_1)$, we have $t_{L/M}=O(1/j_1^4)$. 
Therefore, the query complexity of L and R walks are $O(1)$ at best and $O(j_1^3)$ at worst. The complexity is largest 
near the top and edge of the state pyramid.

%%%%%%%%%%%%%%%%%%%%%%
\subsection{\label{sec:prepare}State preparation by engineered walks}
\begin{figure}[t]
\includegraphics[width=0.48\textwidth]{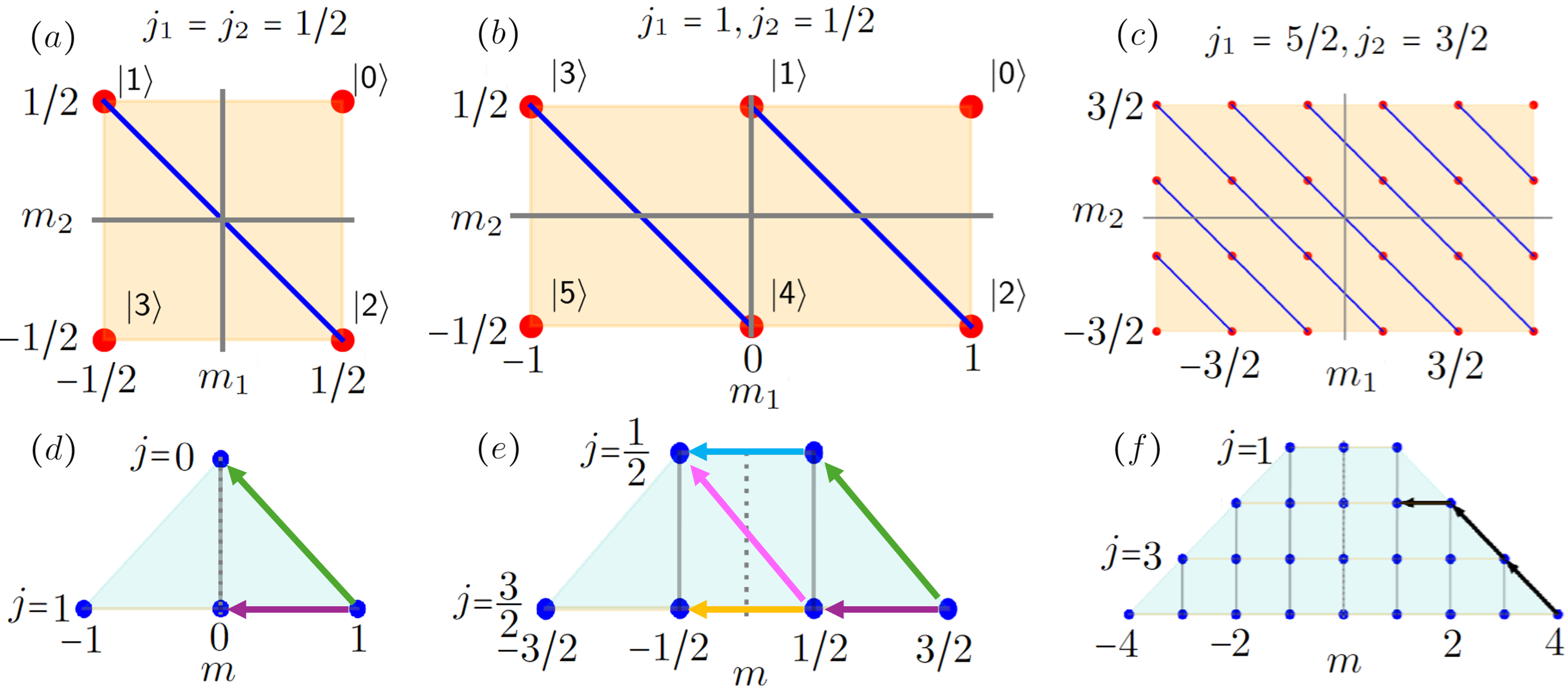}
\caption{\label{fig:examples} Projections of state pyramids in the $m_1$-$m_2$ plane (a)-(c) and the $m$-$j$ plane (d)-(f) for three examples of $j_1$ and $j_2$. The $\|m_1, m_2\rrangle$ states (red) are mapped to qubit states $|l\rangle$, and the $|j,m\rangle$ states (blue) are prepared by quantum walks from the ground state.
Colored arrows illustrate examples discussed in Secs.~\ref{sec:example1} and \ref{sec:example2}.
Black arrows show an example path for preparing the $|2,1\rangle$ state using two steps of $L$ walk followed by one step of $M$ walk. 
}
\end{figure}

With the three basic types of double-pinched walks, we can now prepare any $|j,m\rangle$ states. Recall the $M$ walk moves $|j,m\rangle\leftrightarrow{|j,m-1\rangle}$, the $L$ walk moves $|j,m\rangle\leftrightarrow{|j-1,m-1}\rangle$, and the $R$ walk moves $|j,m-1\rangle\leftrightarrow|j-1,m\rangle$. Using these basic movements, we can independently change the value of $j$ and $m$, and therefore move along a variety of paths on the state pyramid. 
Because the top angular momentum eigenstate $|j_{\max}, j_{\max}\rangle = \|j_1, j_2\rrangle$ and the bottom state $|j_{\max}, -j_{\max}\rangle = \|-j_1, -j_2\rrangle$ are particularly simple, it is convenient that we encode either of these states as the ground state of the quantum computer, and start quantum walks from there. 
Due to the $m\rightarrow -m$ symmetry, we will focus on states with $m\ge0$, and encode the top state as the ground state.
Then, to prepare any state $|j,m\rangle$, a viable path is shown in Fig.~\ref{fig:examples}. 
If $j=j_{\max}$, we use $M$ walks to reduce $m$ until it reaches the desired value.
On the other hand, if $j<j_{\max}$, we use $L$ walks to climb up the state pyramid, reducing $j$ and $m$ simultaneously. After reaching the desired $j$, we then use $M$ walks to reduce $m$ until it reaches the desired value, completing the preparation of the bra state $|j,m\rangle$.
Suppose we instead want to prepare, or project to, the ket state $\langle j,m|$, we simply reverse the steps of the quantum walk, walking the state back to the ground state of the computational basis. The quantum walk is reversible because no ancillary qubit is required beyond what is needed for quantum Hamiltonian simulations.

The query complexity of our state preparation scheme is $O(j_1)$ for a typical state, and $O(j_1^3)$ for the most expensive state. 
With the convention $j_1\ge j_2$, the width of the state pyramid is $O(j_1)$, and the height of the pyramid is $O(j_2)$. So, the path that links any two states is of length $O(j_1)$. In other words, to prepare any state takes $O(j_1)$ basic movements from the ground state. 
In a typical scenario, each step has an $O(1)$ complexity. Then, the total complexity is $O(j_1)$ after adding up all steps along the path. 
The worst case is when both angular momenta are high $j_1=j_2\rightarrow\infty$ but one wants to prepare a low angular momentum eigenstate $|0,0\rangle$. 
In this case, as one climbs up the pyramid, steps become increasingly difficult.
As shown in Fig.~\ref{fig:scaling}(b), the most costly path is climbing up along the edge of the pyramid, where the $k$-th step departs from $j=m=2j_1-k$, for $k=0, 1, \dots, 2j_1-1$. 
Because all steps have $\|L\|_{\max}=O(j_1^4)$, we can estimate the complexity by summing up the $\pi$-pulse time. For the $k$-th step, we approximate Eq.~(\ref{eq:L_tpi}) by $t_L^k\simeq\pi/[4(2j_1-k)^3\sqrt{(4j_1-k)(k+1)}]$. Approximating the discrete sum by a continuous integral gives $\sum_{k=0}^{2j_1-1} t_L^k \simeq \pi/(4j_1)=O(1/j_1)$.
Therefore, the total complexity for preparing the most costly state is $\sum_k \tau^k =O(j_1^3)$. 
%We emphasize that the typical cost of state preparation is only $O(j_1)$, unless one wants to reach the tip of the pyramid. In other words, using engineered quantum walks to prepare angular momentum eigenstates typically gives a polynomial speedup compared to a brute force gate decomposition of the unitary CG coefficients.   

%$(a), (b), (c), (d), (e), (f)$ axis: $m_1, m_2=-5/2, -3/2, -1, -1/2, 0, j, m = 0, 1, 3/2, 1/2, -1, -2, -3, -4$

The final step for implementing the double-pinched quantum walks on quantum computers is to encode the computational basis $\|m_1, m_2\rrangle$ into qubit states, which are equivalent to states $|l\rangle$ of a multi-level quantum system. In this paper, we consider perhaps the simplest, but not necessarily the most efficient, encoding scheme by ranking $\|m_1, m_2\rrangle$ states from the top right to the bottom left corners in the $m_1$-$m_2$ space, as shown in Fig.~\ref{fig:examples}. Again, we use the convention $j_1\ge j_2$.
To encode, states with a common $m=m_1+m_2$ are assigned to a line $d=j_{\max}-m$. On the $d$-th line, the total number of states is $N_d=d+1$ when $0\le d\le 2j_2$, $N_d=2j_2+1$ when $2j_2\le d \le 2j_1$, and $N_d=D_M-d$ when $2j_1\le d \le 2j_{\max}$, where $D_M=2j_{\max}+1$.
The accumulative number of states before the $d$-th line is therefore
\begin{equation}
    \label{eq:encode}
    B(d) \!=\! \left\{\! \begin{array}{ll}
    \frac{1}{2}d(d+1), & \!d\le D_{j_2}, \\
    % \!2j_2\le d \le 2j_1
    D_{j_2}(d-j_2), & \text{otherwise},\\
    D\!-\!\frac{1}{2}(\!D_M\!+\!1\!-\!d)(\!D_M\!-d), & \!d\ge D_{j_1},
    \end{array} \right.
\end{equation}
where again $D=D_{j_1} D_{j_2}$ and $D_j=2j+1$. 
On the $d$-th line, we count states from top to bottom $m_2$. The maximum value $m_2$ can reach is $m_u=j_2$ when $d\le 2j_1$, and $m_u=2j_1+j_2-d$ when $d\ge 2j_1$. 
The sub-rank of a state along the line is then $k=m_u-m_2$, and the overall rank of the state is $l=B(d)+k$. In other words, we map
\begin{equation}
    \label{eq:map}
    \|m_1, m_2\rrangle \rightleftharpoons |l\rangle = |B(d)+k\rangle,
\end{equation}
where $d$ and $k$ are piece-wise continuous functions of $m_1$ and $m_2$. To decode, namely, to find the corresponding $m_1$ and $m_2$ for a given $l$, we first identify which line the state belongs to by inverting Eq.~(\ref{eq:encode}), which gives
\begin{equation}
    \label{eq:decode}
    D(b) \!=\! \left\{\! \begin{array}{ll}
    -\frac{1}{2} +\sqrt{\frac{1}{4}+2b}, & b\le (j_2+1)D_{j_2}, \\
    j_2+b/D_{j_2}, & \textrm{otherwise},\\
    D_M\!+\!\frac{1}{2}\!-\!\sqrt{\!\frac{1}{4}\!+\!2(D\!-b)}, & b\ge(D_{j_1}\!-\!j_2)\!D_{j_2}.
    \end{array} \right.
\end{equation}
For a given $l$, we find the line by $d=\lfloor D(l) \rfloor$. Then, on the $d$-th line, we find the sub-rank of the state by $k=l-B(d)$. Finally, from the sub-rank, we obtain $m_2=m_u-k$ and $m_1=m-m_2$, where $m=j_{\max} - d$. 
Other encoding and decoding schemes from computational basis $\|m_1,m_2\rrangle$ to qubit states $|l\rangle$ are also viable.

%%%%%%%%%%%%%%%%%%%%%%
\subsubsection{\label{sec:example1} Example $j_1=j_2=1/2$}
To demonstrate our state preparation scheme, consider the simplest nontrivial example $j_1=j_2=1/2$, whose state space is four dimensional, as shown in Figs.~\ref{fig:examples}(a) and ~\ref{fig:examples}(d). 
First, let us consider $M$ walks. The Hamiltonian that moves $|1,1\rangle\leftrightarrow |1,0\rangle$, which is marked by the purple arrow in Fig.~\ref{fig:examples}(d), is represented by the matrix
\begin{equation}
    \nonumber
    M_{1/2} = \left( \begin{array}{cccc}
    0 & 1 & 1 & 0 \\
    1 & 0 & 0 & 0 \\
    1 & 0 & 0 & 0 \\
    0 & 0 & 0 & 0
    \end{array} \right),
\end{equation}
in the qubit basis $(|0\rangle, \dots, |3\rangle)$. The $\pi$-pulse time Eq.~(\ref{eq:M_tpi}) for $j=m=1$ is $t_M=\pi/(2\sqrt{2})$, and the resultant unitary $U_{M_{1/2}}^{j=1}=\exp(-iM_{1/2} t_M)$ is 
\begin{equation}
    \nonumber
    U_{M_{1/2}}^{1} = \left( \begin{array}{cccc}
    0 & -i/\sqrt{2} & -i/\sqrt{2} & 0 \\
    -i/\sqrt{2} & 1/2 & -1/2 & 0 \\
    -i/\sqrt{2} & -1/2 & 1/2 & 0 \\
    0 & 0 & 0 & 1
    \end{array} \right).
\end{equation}
When acting on the ground state, $iU_{M_{1/2}}^1 |0\rangle = \frac{1}{\sqrt{2}}(|1\rangle + |2\rangle)$, which is precisely $|1,0\rangle = \frac{1}{\sqrt{2}}(\|\frac{1}{2},-\frac{1}{2}\rrangle + \|-\frac{1}{2},\frac{1}{2}\rrangle$. 
Similarly, one can find the Hamiltonian $M_{-1/2}$, and verify that its unitary is consistent with known CG coefficients.

Next, we consider the $L$ walk indicated by the green arrow in Fig.~\ref{fig:examples}(d).
In Appendix~\ref{App:buildingblocks_example1}, we find matrix elements of the three building blocks in the qubit basis. 
We linearly superimpose these three building blocks according to Eq.~(\ref{eq:L_Hamiltonian}). 
In this case, $\mathscr{J}=1/2$, so $L_{1/2}^{1/2}=P_{1/2}(2A_zJ_+ - \frac{7}{2}A_+ - \{\Lambda, A_+\} + \textrm{h.c.})P_{1/2}$. In a matrix form,
\begin{equation}
    \nonumber
    L_{1/2}^{1/2} = \frac{3i}{2}\left( \begin{array}{cccc}
    0 & 1 & -1 & 0 \\
    -1 & 0 & 0 & 0 \\
    1 & 0 & 0 & 0 \\
    0 & 0 & 0 & 0
    \end{array} \right),
\end{equation}
where all matrix elements related to $|3\rangle$ are zeroed out by the projection operator $P_{1/2}$. 
Using $\alpha_j=j\sqrt{(4-j^2)/(4j^2-1)}$, the $\pi$-pulse time Eq.~(\ref{eq:L_tpi}) for $j=m=1$ is $t_L=\pi/(3\sqrt{2})$. The resultant unitary is 
\begin{equation}
    \nonumber
    U_{L_{1/2}^{1/2}}^{1} = \left( \begin{array}{cccc}
    0 & 1/\sqrt{2} & -1/\sqrt{2} & 0 \\
    -1/\sqrt{2} & 1/2 & 1/2 & 0 \\
    1/\sqrt{2} & 1/2 & 1/2 & 0 \\
    0 & 0 & 0 & 1
    \end{array} \right).
\end{equation}
When acting on the ground state, we obtain $U |0\rangle = \frac{1}{\sqrt{2}}(|2\rangle - |1\rangle)$, which is precisely $|0,0\rangle = \frac{1}{\sqrt{2}}(\|\frac{1}{2},-\frac{1}{2}\rrangle - \|-\frac{1}{2},\frac{1}{2}\rrangle$. 
Similarly, one can find the Hamiltonian $R^{1/2}_{-1/2}$ and compute its unitary. 
One can also verify that all possible moves of the $M$, $L$, and $R$ walks are consistent with known CG coefficients.

%%%%%%%%%%%%%%%%%%%%%%
\subsubsection{\label{sec:example2} Example $j_1=1, j_2=1/2$}
Now let us consider a slightly more complicated example, whose state space is six dimensional, as shown in Figs.~\ref{fig:examples}(b) and \ref{fig:examples}(e). For the $M$ walks, $M_1|0\rangle = J_-\|1,\frac{1}{2}\rrangle = \sqrt{2}\|0,\frac{1}{2}\rrangle + \|1,-\frac{1}{2}\rrangle = \sqrt{2}|1\rangle + |2\rangle$, which gives the lower triangular part of $M_1$. The full matrix is
% \begin{equation}
%     M_{1} = \left( \begin{array}{cccc}
%     0 & \sqrt{2} & 1 & \dots\\
%     \sqrt{2} & 0 & 0 & \dots\\
%     1 & 0 & 0 &  \dots \\
%     \vdots & \vdots & \vdots & \ddots
%     \end{array} \right),
% \end{equation}
%##############################
\begin{equation}
    \nonumber
    M_{1} = 
    \begin{pNiceArray}{c|cc|c}
        0 & \sqrt{2} & 1 & \Block{1-1}<\Large>{0}\\
        \hline
        \sqrt{2} & \Block{2-2}<\Large>{0} && \Block{2-1}<\Large>{0} \\
        1 \\
        \hline
        \Block{1-1}<\Large>{0} & \Block{1-2}<\Large>{0} && \Block{1-1}<\Large>{0}
    \end{pNiceArray},
\end{equation}
%##############################
where big zeros denote blocks of the matrix that are zero. The $\pi$-pulse time for $j=m=3/2$ is $t_M=\pi/(2\sqrt{3})$, and the resultant unitary is 
% \begin{equation}
%     U^M_1 = \left( \begin{array}{cccc}
%     0 & -i\sqrt{2/3} & -i/\sqrt{3} & \dots \\
%     -i\sqrt{2/3} & 1/3 & -\sqrt{2}/3 & \dots \\
%     -i/\sqrt{3} & -\sqrt{2}/3 & 2/3 & \dots \\
%     \vdots & \vdots & \vdots & \ddots
%     \end{array} \right),
% \end{equation}
%##############################
\begin{equation}
    \nonumber
    U_{M_1}^{3/2} = 
    \begin{pNiceArray}{c|cc|c}
        0 & -i\sqrt{2/3} & -i/\sqrt{3} & \Block{1-1}<\Large>{0}\\
        \hline
        -i\sqrt{2/3} & 1/3 & -\sqrt{2}/3  & \Block{2-1}<\Large>{0} \\
        -i/\sqrt{3} & -\sqrt{2}/3 & 2/3  \\
        \hline
        \Block{1-1}<\Large>{0} & \Block{1-2}<\Large>{0} && \Block{1-1}<\Large>{\mathbb{I}}

    \end{pNiceArray},
\end{equation}
%##############################
where $\mathbb{I}$ denotes the identity matrix.
When acting on the ground state, as indicated by the purple arrow in Fig.~\ref{fig:examples}(e), we have $iU^{3/2}_{M_1} |0\rangle = \sqrt{\frac{2}{3}}|1\rangle + \sqrt{\frac{1}{3}}|2\rangle$, which is consistent with CG coefficients of $|\frac{3}{2},\frac{1}{2}\rangle$.
Similarly, to obtain $M_0$, it is sufficient to consider $J_-|1\rangle = J_-\|0,\frac{1}{2}\rrangle=\sqrt{2}\|-1,\frac{1}{2}\rrangle + \|0,-\frac{1}{2}\rrangle = \sqrt{2}|3\rangle + |4\rangle$ and $J_-|2\rangle = J_-\|1,-\frac{1}{2}\rrangle=\sqrt{2}\|0,-\frac{1}{2}\rrangle = \sqrt{2}|4\rangle$. 
Because the projection operator $P_0$ zeroes out matrix elements outside the $m=1/2$ and $m=-1/2$ subspace, $M_0$ is nonzero only within the $(|1\rangle,\dots,|4\rangle)$ subspace, where its matrix is
%##############################
\begin{equation}
    \nonumber
    M_{0} = 
    \begin{pNiceArray}{cc|cc}
        \Block{2-2}<\Large>{0} && \sqrt{2} & 1 \\
        && 0 & \sqrt{2} \\
        \hline
        \sqrt{2} & 0 & \Block{2-2}<\Large>{0} \\
        1 & \sqrt{2} 
    \end{pNiceArray}.
\end{equation}
%##############################
% \begin{equation}
%     \nonumber
%     M_{0} = 
%     \begin{pNiceArray}{c|cc|cc|c}
%         0 & \Block{1-2}<\Large>{0} && \Block{1-2}<\Large>{0} && 0 \\
%         \hline
%         \Block{2-1}<\Large>{0} & \Block{2-2}<\Large>{0} && \sqrt{2} & 1 & \Block{2-1}<\Large>{0}\\
%         && & 0 & \sqrt{2} \\
%         \hline
%         \Block{2-1}<\Large>{0} & \sqrt{2} & 0 & \Block{2-2}<\Large>{0} && \Block{2-1}<\Large>{0} \\
%         & 1 & \sqrt{2} \\
%         \hline
%         0 & \Block{1-2}<\Large>{0} && \Block{1-2}<\Large>{0} && \Block{1-1}<\Large>{0}
%     \end{pNiceArray},
% \end{equation}
When $j=3/2$ and $m=1/2$, the $\pi$-pulse time is $t_M=\pi/4$. The resultant unitary is active only within the $(|1\rangle,\dots,|4\rangle)$ subspace, where
\begin{equation}
    \nonumber
    U_{M_{0}}^{3/2} = \frac{1}{3}\left( \begin{array}{cccc}
    1/\sqrt{2} & -1 & \!-i(\sqrt{2}\!+\!1) & i(\frac{1}{\sqrt{2}}\!-\!2) \\
    -1 & \sqrt{2} & \!i(\sqrt{2}\!-\!1) & \!-i(\sqrt{2}\!+\!1) \\
    \!-i(\sqrt{2}\!+\!1) & \!i(\sqrt{2}\!-\!1) & \sqrt{2} & -1 \\
    i(\frac{1}{\sqrt{2}}\!-\!2) & \!-i(\sqrt{2}\!+\!1) & -1 & 1/\sqrt{2}
    \end{array} \right).
\end{equation}
Outside the subspace, $U_{M_{0}}^{3/2}|0\rangle = |0\rangle$ and $U_{M_{0}}^{3/2}|5\rangle = |5\rangle$ are identity.
When acting on $|\frac{3}{2},\frac{1}{2}\rangle$, as indicated by the yellow arrow in Fig.~\ref{fig:examples}(e), we have $iU^{3/2}_{M_0} (\sqrt{\frac{2}{3}}|1\rangle + \sqrt{\frac{1}{3}}|2\rangle) = \sqrt{\frac{1}{3}}|3\rangle + \sqrt{\frac{2}{3}}|4\rangle$, which is consistent with CG coefficients of $|\frac{3}{2},-\frac{1}{2}\rangle$.
On the other hand, when $j=1/2$ and $m=1/2$, the $\pi$-pulse time $t_M=\pi/2$ is longer. Even though $M_0$ is the same matrix, the resultant unitary in the $(|1\rangle,\dots,|4\rangle)$ basis is different:
\begin{equation}
    \nonumber
    U_{M_{0}}^{1/2} = -\frac{1}{3}\left( \begin{array}{cccc}
    2 & \sqrt{2} & i\sqrt{2} & -i \\
    \sqrt{2} & 1 & -2i & i\sqrt{2} \\
    i\sqrt{2} & -2i & 1 & \sqrt{2} \\
    -i & i\sqrt{2} & \sqrt{2} & 2
    \end{array} \right).
\end{equation}
Again, the above 4-by-4 matrix omits subspaces where the unitary is inactive.
When acting on $|\frac{1}{2},\frac{1}{2}\rangle$, as indicated by the blue arrow in Fig.~\ref{fig:examples}(e), we have $iU^{1/2}_{M_0} (\sqrt{\frac{2}{3}}|2\rangle - \sqrt{\frac{1}{3}}|1\rangle) = \sqrt{\frac{1}{3}}|4\rangle - \sqrt{\frac{2}{3}}|3\rangle$, which is consistent with CG coefficients of $|\frac{1}{2},-\frac{1}{2}\rangle$.
Following similar steps, one can verify that all possible moves of $M$ walks are consistent with known CG coefficients.

Finally, we consider two examples of $L$ walks given by Eq.~(\ref{eq:L_Hamiltonian}), where $\mathscr{J}=7/4$. Matrix representations of the three building blocks are found in Appendix~\ref{App:buildingblocks_example2}.
For $j=m=3/2$, as indicated by the green arrow in Fig.~\ref{fig:examples}(e), we have $L_1^1 = P_1(3A_zJ_+ - \frac{17}{2}A_+ -\{\Lambda, A_+\} + \textrm{h.c.})P_1 = P_1(3A_zJ_+ - 8A_+  + \textrm{h.c.})P_1$. In a matrix form, because of the projection operator, all matrix elements are zero except in the $(|0\rangle, |1\rangle, |2\rangle)$ basis, where
%##############################
\begin{equation}
    \nonumber
    L_1^1 = 8i
    \begin{pNiceArray}{c|cc}
        0 & 1/\sqrt{2} & -1 \\
        \hline
        -1/\sqrt{2} & \Block{2-2}<\Large>{0} \\
        1
    \end{pNiceArray}.
\end{equation}
%##############################
Using $\alpha_j=\frac{1}{2}\sqrt{\frac{25}{4}-j^2}$, the $\pi$-pulse time Eq.~(\ref{eq:L_tpi}) for $j=m=3/2$ is $t_L=\pi/(8\sqrt{6})$. We can find the resultant unitary by exponentiating the Hamiltonian. Alternatively, it is sufficient to check that $L_1^1$ has the desired matrix elements in the $|j,m\rangle$ basis. Acting on the ground state, $L_1^1|\frac{3}{2},\frac{3}{2}\rangle = L_1^1|0\rangle = 8i(|2\rangle-\frac{1}{\sqrt{2}}|1\rangle)=\frac{i}{2}8\sqrt{6}(\sqrt{\frac{2}{3}}\|1,-\frac{1}{2}\rrangle - \sqrt{\frac{1}{3}}\|0,\frac{1}{2}\rrangle) = \frac{i}{2}8\sqrt{6} |\frac{1}{2},\frac{1}{2}\rangle$, consistent with known CG coefficients. 
Similarly, for $j=3/2$ and $m=1/2$, as indicated by the magenta arrow in Fig.~\ref{fig:examples}(e), we have $L_0^1 = P_0(3A_zJ_+ - \frac{11}{2}A_+ -\{\Lambda, A_+\} + \textrm{h.c.})P_0 = P_0(3A_zJ_+ - 5A_+  + \textrm{h.c.})P_0$. 
Due to $P_0$, matrix elements are nonzero only in the $(|1\rangle, \dots, |4\rangle)$ basis, where
%##############################
\begin{equation}
    \nonumber
    L^1_0 = i
    \begin{pNiceArray}{cc|cc}
        \Block{2-2}<\Large>{0} && 5/\sqrt{2} & -3 \\
        && 3 & -\sqrt{2} \\
        \hline
        -5/\sqrt{2} & -3 & \Block{2-2}<\Large>{0} \\
        3 & \sqrt{2}
    \end{pNiceArray}.
\end{equation}
%##############################
With $\alpha_{3/2}=1$, the $\pi$-pulse time for $m=1/2$ is $t_L=\pi/(8\sqrt{2})$. 
Again, instead of verifying the unitary, it is sufficient to check the Hamiltonian $L_0^1|\frac{3}{2},\frac{1}{2}\rangle = L_0^1 (\sqrt{\frac{2}{3}}|1\rangle + \sqrt{\frac{1}{3}}|2\rangle)
= \frac{i}{2}8\sqrt{2}(-\sqrt{\frac{2}{3}}|3\rangle + \sqrt{\frac{1}{3}}|4\rangle)
= \frac{i}{2}8\sqrt{2} |\frac{1}{2},-\frac{1}{2}\rangle$ is as desired.
Following similar steps, one can verify that all possible moves of $L$ and $R$ walks are consistent with known CG coefficients.

%%%%%%%%%%%%%%%%%%%%%%
\subsection{\label{sec:decomposition} Engineered walks as decompositions of unitaries}
More generally, our state preparation scheme using an engineered sequence of double-pinched walks may be regarded as a way to decompose unitary matrices that involve CG coefficients. 
Suppose one wants to prepare the state $|j'=j-d, m'=j-d-k\rangle$ from the state $|j, j\rangle$. A walk that serves the purpose is 
$|j, j\rangle \rightarrow |j\!-\!1, j\!-\!1\rangle \rightarrow\dots \rightarrow$ \mbox{$|j\!-\!d, j\!-\!d\rangle$} $\rightarrow |j\!-\!d, j\!-\!d\!-\!1\rangle \rightarrow\dots \rightarrow |j\!-\!d, j\!-\!d\!-\!k\rangle$, and an example is shown by the quantum walk diagram in Fig.~\ref{fig:examples}(f), where each black arrow indicates one step of the quantum walk. 
The walk is realized by a sequence of Hamiltonians 
$L_{j-1/2}^{j-1/2} \rightarrow\dots \rightarrow L_{j-d+1/2}^{j-d+1/2} \rightarrow M_{j-d-1/2} \rightarrow\dots \rightarrow M_{j-d-k+1/2}$, each used to evolve the quantum state for a corresponding $\pi$-pulse time.
Carrying out the quantum walk is effectively multiplying unitary matrices such that $|j-d, m-d-k\rangle = \prod_{q=1/2}^{k-1/2} (iU_{M_{j-d-q}}) \prod_{p=1/2}^{d-1/2} U_{L_{j-p}^{j-p}} |j,m\rangle$, where unitaries of later steps are multiplied to the left of earlier steps.

On the other hand, the state $|j',m'\rangle$ can be obtained by a unitary transformation from the $|j,m\rangle$ state using CG coefficients. Notice that when $m'\ne m$, the two states are not on the same grey plane in Fig.~\ref{fig:domain}, and hence are not directly connected by CG coefficients. The connection between the two states is mediated by some encoding and decoding scheme that maps qubit states $|l\rangle$ to $\|m_1, m_2\rrangle$ states. The encoding scheme is itself a unitary transformation $\|m_1, m_2\rrangle=E_l^{m_1,m_2} |l\rangle$, for which Eq.~(\ref{eq:map}) is an example. Again, summation over repeated indices is assumed. 
Using the completeness of the $|l\rangle$ and $\|m_1, m_2\rrangle$ basis, decoding is the unitary transformation $|l\rangle=\bar{E}_l^{m_1, m_2}\|m_1, m_2\rrangle$.
One can often choose $E$ such that all of its elements are real, so that the complex conjugation is insignificant.
After encoding, $|j',m'\rangle=C^{j'}_{m_1', m_2'}\|m_1', m_2'\rrangle = C^{j'}_{m_1', m_2'} E_{l'}^{m_1', m_2'} |l'\rangle$ is obtained as a unitary transformation of the qubits. 
We refer to the unitary $C^{j'}_{m_1', m_2'} E_{l'}^{m_1', m_2'}$ as the CG transform.
Suppose the qubits are initialized in the ground state $|0\rangle$, then the $|l'\rangle$ state can be obtained by flipping all necessary qubits to swap the two levels $|l'\rangle=S_0^{l'}|0\rangle$.
The complete unitary for preparing the state can be written as $|j',m'\rangle=U^{j'}_{m'}|0\rangle$, where 
$U^{j'}_{m'} = C^{j'}_{m_1', m_2'} E_{l'}^{m_1', m_2'} S_0^{l'}$. 
In this scheme, the ground state serves as a point of connection for all states, via which $|j',m'\rangle=U^{j'}_{m'} U^{j\dagger}_{m}|j,m\rangle$. 
In fact, using a direct swap $S_0^{l'}S^0_{l}=S^{l'}_{l}$, the ground state can be bypassed, so any two states are connected by two CG transforms up to a swap.

Comparing the two ways of relating the $|j',m'\rangle$ and $|j,m\rangle$ states, we see quantum walks give a decomposition of a dense unitary. The decomposition is not unique: By choosing a different path that connects the two states, we obtain a different decomposition. An example is
\begin{equation}
    \label{eq:decomposition}
    U^{j-d}_{m-d-k} U^{j\dagger}_{m} = \prod_{q=1/2}^{k-1/2} (iU_{M_{j-d-q}}) \prod_{p=1/2}^{d-1/2} U_{L_{j-p}^{j-p}}.
\end{equation}
In this decomposition, the LHS is a product of two unitary matrices of dimension $O(j_1j_2)\times O(j_1j_2)$, each with $O(j_1j_2^2)$ nonzero elements, assuming $j_1\ge j_2\rightarrow\infty$. Notice that the product of two such unitaries is a denser unitary in general. 
On the RHS, each unitary is of the same dimension, but with only $O(j_2^2)$ nontrivial elements. The extra degrees of freedom is now spread across $O(d+k)$ steps of the quantum walk. 
In other words, the quantum walk decomposes a unitary, with at least $O(j_1j_2^2)$ nontrivial elements, into the product of no more than $O(j_1)$ unitaries, each with only $O(j_2^2)$ nontrivial elements.

%%%%%%%%%%%%%%%%%%%%%%%%%%%%%%%%%%%%%%%%%%%%%
\section{\label{sec:test}Verification and validation}
The two examples in Secs.~\ref{sec:example1} and ~\ref{sec:example2} demonstrate how our state preparation scheme works in practice. Moreover, these examples demonstrate that known CG coefficients are recovered as coefficients of the final state of the quantum walks. 
In this section, we automate the state preparation protocol, and verify that the protocol reproduces tables of CG coefficients on classical computers, and that the protocol produces states close to desired entangled states on quantum devices.

\begin{figure}[b]
\includegraphics[width=0.49\textwidth]{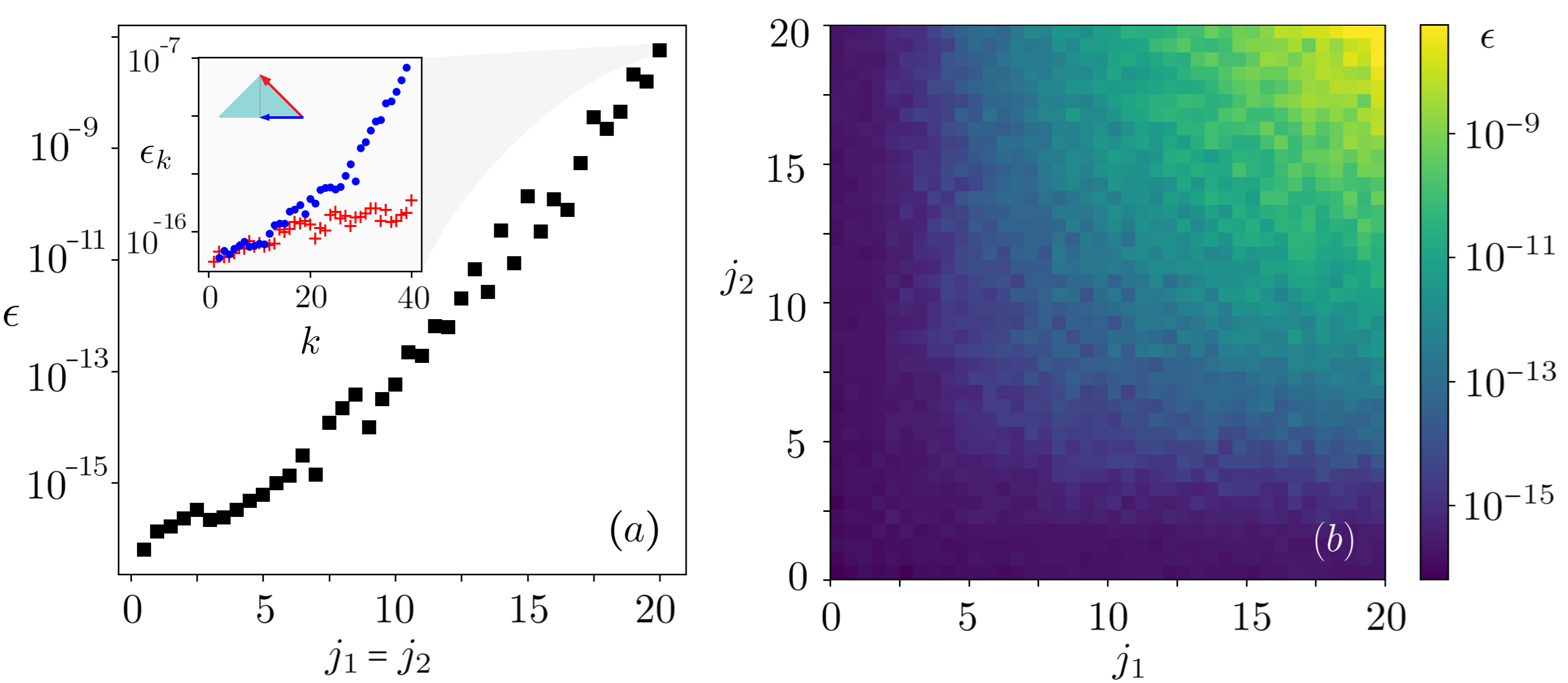}
\caption{\label{fig:verification} The difference $\epsilon$ between Clebsch–Gordan (CG) coefficients computed using quantum versus classical methods is negligible. 
(a) The minuscule $\epsilon$ increases with $j_1$ and $j_2$ due to accumulation of non-unitary rounding errors in both methods. 
The inset shows how $\epsilon_k$ increases with $k$, the path length of the quantum walk, along the right (red) and the bottom (blue) of the state pyramid for $j_1=j_2=20$. 
(b) The average separation $\epsilon$ between the full CG tables produced using the quantum and classical methods is negligible for all $j_1$ and $j_2$ values, demonstrating the equivalence of the two methods, and verifying that quantum walks produce expected angular momentum eigenstates. 
}
\end{figure}
%$(a)$ axis: $j_1, j_2, 0, 5, 10, 15, 20$, colorbar $\epsilon, 10^{-9}, 10^{-11}, 10^{-13}, 10^{-15},$
%(b)$ axis: $j_1=j_2$, inset x-axis: $l=0, 20, 40$, y-axis: $10^{-16}, 10^{-13}, 10^{-10}, 10^{-7}$

We developed an openly available Python program \cite{Shi24} to automate the state preparation protocol. 
First, for a given $j_1$ and $j_2$, the program uses the encoding and decoding scheme [Eqs.~(\ref{eq:encode})-(\ref{eq:decode})] to map qubit states $|l\rangle$ to computational basis $\|m_1, m_2\rrangle$, and vice versa.  
Second, the program prepares building blocks for the $M$ walk [Eq.~(\ref{eq:M_m1m2})] and the $L$ and $R$ walks [Eqs.~(\ref{eq:AzJ_elements_m1m2})-(\ref{eq:ApLambda_elements_m1m2})], by preloading their sparse matrix representations in the qubit basis.
Third, for a targeted $|j,m\rangle$, the program generates a path of quantum walks. The path originates from either the bottom or the top state,  and arrives at the targeted state using a sequence of $L$ walks [Eq.~(\ref{eq:L_Hamiltonian})] or $R$ walks [Eq.~(\ref{eq:R_Hamiltonian})], followed by $M$ walks [Eq.~(\ref{eq:M_Hamiltonian})].  
Finally, the path is resolved as a sequence $\{(H_k, t_k)\}$, where $H_k$ is the Hamiltonian of the $k$-th step and $t_k$ is its $\pi$-pulse time. 
For each step, $H_k$ is constructed as a linear combination of the building blocks, after projecting them to the subspace relevant for the step.

%%%%%%%%%%%%%%%%%%%%%%
\subsection{\label{sec:classical} Verification on classical computers}
On a classical computer, the resultant sequence of unitary $U_k=\exp(-iH_k t_k)$ is computed by diagonalizing and exponentiating the Hamiltonian numerically. In the qubit basis, suppose the initial state vector is $\mathbf{c}_0$. Then, after the $k$-th step of the quantum walk, the state vector becomes $\mathbf{c}_k = \prod_{p=1}^k U_p \mathbf{c}_0$. 
The program simplify calculations using the fact that $H_k$ is zero, and therefore $U_k$ is an identify, except in the subspace relevant for the $k$-th step.  
Suppose the initial state is an angular momentum eigenstate $|j_0, m_0\rangle$, then by design, the state after the $k$-th step should be another angular momentum eigenstate $|j_k, m_k\rangle$. The goal of the verification is to make sure that $\mathbf{c}_k$ indeed corresponds to $|j_k, m_k\rangle$.

We perform verification by comparing state vectors to known CG coefficients. 
We compute CG coefficients using the classical recurrence relation induced by Eq.~(\ref{eq:Lambda}). As described in Sec.~\ref{sec:Review}, for a given $j$ and $m$, the recurrence relation for $C^{j}_{m_1, m_2}$, where $m_1+m_2=m$, gives a tri-diagonal matrix equation. We solve the matrix equation numerically. 
With the encoding, % discussed in Sec.~\ref{sec:decomposition}, 
we obtain $|j,m\rangle=C^{j}_{m_1, m_2}\|m_1,m_2\rrangle = C^{j}_{m_1, m_2} E^{m_1, m_2}_l |l\rangle = C_{l}|l\rangle$. 
In other words, from CG coefficients computed using the classical method, we know what the state vector should be. 
We use capital $\mathbf{C}$ to denote state vectors from CG coefficients, and use lower-case $\mathbf{c}$ to denote state vectors from quantum walks. 
Having obtained the same state vector using two completely different methods, we compute their separation by $\epsilon_k =(\frac{1}{D}|\mathbf{C}_k - \mathbf{c}_k|^2)^{1/2}$, where $D=D_{j_1}D_{j_2}$ is the dimension of the Hilbert space and $|\mathbf{v}|^2=\mathbf{v}^\dagger \mathbf{v}$. We interpret $\epsilon_k$ as the average separation between components of the two vectors. 
The normalization by $D$ is important when comparing results across different $j_1$ and $j_2$. 
Ideally, $\mathbf{C}_k$ and $\mathbf{c}_k$ are nonzero only in a $O(j_2)$-dimensional subspace. However, due to the finite precision of floating-point arithmetic on classical computers, $\mathbf{c}_k$ may spill out of the subspace with small errors. 
We enforce states not spilling over, but then $|\mathbf{c}_k|^2$ may deviate from unity by a small but observable error. 
In the inset of Fig.~\ref{fig:verification}(a), we show $\epsilon_k$ along two paths for the example $j_1=j_2=20$. One path moves along the bottom of the state pyramid (blue), and the other path moves along the right side (red). As the path length $k$ increases, the separation $\epsilon_k$ increases due to the accumulation of round-off errors in a path-dependent way. 
A similar error behavior is also observed when comparing different classical methods for computing the CG coefficients \cite{johansson2016fast}. 
As discussed in Appendix~\ref{App:Errors}, the observed near-exponential increase of $\epsilon$ is likely due to non-unitary rounding errors in both the classical and the quantum algorithms for computing the state vectors. 
The key point of Fig.~\ref{fig:verification} is that the separation is negligible in all cases, demonstrating that state vectors produced by quantum walks are consistent with CG coefficients.

To show that the classical and quantum algorithms produce identical tables of 
CG coefficients, we remove path dependence in $\epsilon_k$ by averaging over all paths. 
Denoting $\tilde{C}^j_{m_1, m_2}$ the CG coefficients computed numerically using the classical recurrence relation, and $\tilde{c}^j_{m_1, m_2}$ the CG coefficients extracted from numerically obtained state vectors. After decoding, the final state of the quantum walks is $|j,m\rangle=\tilde{c}_{l}|l\rangle = \tilde{c}_{l}E^{m_1, m_2}_l \|m_1, m_2\rrangle$, from which we recognize $\tilde{c}^j_{m_1, m_2} = \tilde{c}_{l}E^{m_1, m_2}_l$. 
With two algorithms for computing tables of CG coefficients, we measure their distance by $\epsilon=(\frac{1}{K}\sum_{j, m_1, m_2} |\tilde{C}^j_{m_1, m_2} - \tilde{c}^j_{m_1, m_2}|^2)^{1/2}$, where $K$ is the number of nonzero CG coefficients.
We interpret $\epsilon$ as the averaged error per CG coefficient. %The normalization by $\frac{1}{K}$ is for comparing across different $j_1$ and $j_2$ values. 
In Fig.~\ref{fig:verification}(b), we show $\epsilon$ in the $j_1$-$j_2$ plan. The negligible $\epsilon$ %across the entire scanned parameter space 
demonstrates that CG tables produced by classical and quantum algorithms are identical, up to rounding errors that occur for both algorithms. We have thus verified that our state preparation protocol always %by engineered quantum walks 
produces desired angular momentum eigenstates.

%%%%%%%%%%%%%%%%%%%%%%
\subsection{\label{sec:quantum} Tests on quantum devices}
On future fault-tolerant quantum computers, unitaries resultant from sequences of quantum walks $\{(H_k, t_k)\}$ can be realized using quantum Hamiltonian simulations. However, current quantum devices are noisy, so we test our state preparation protocol on small problems by directly decomposing $U_k$ into elementary gates.
Notice that for small problems, the decomposition Eq.~(\ref{eq:decomposition}) has no advantage, because $U_{L/R}$ and $U_M$ are not yet more sparse than the total $U$.
In other words, on current hardware where only a few qubits are capable of performing high-quality gates, there is perhaps no advantage of using quantum walks for preparing eigenstates of small angular momentum: Directly decomposing the total $U$ has similar cost as decomposing the steps of the quantum walks. 
%
%However, the situation may change in the near future even before quantum Hamiltonian simulations become viable: Because the cost of decomposing a unitary into elementary gates scales with its sparsity \cite{Jordan09,malvetti2021quantum,de2022double,ramacciotti2023simple}, it is advantageous to first reduce the sparsity using Eq.~(\ref{eq:decomposition}), and then decomposing steps of quantum walks into elementary gates.   

\begin{figure}[t]
\includegraphics[width=0.47\textwidth]{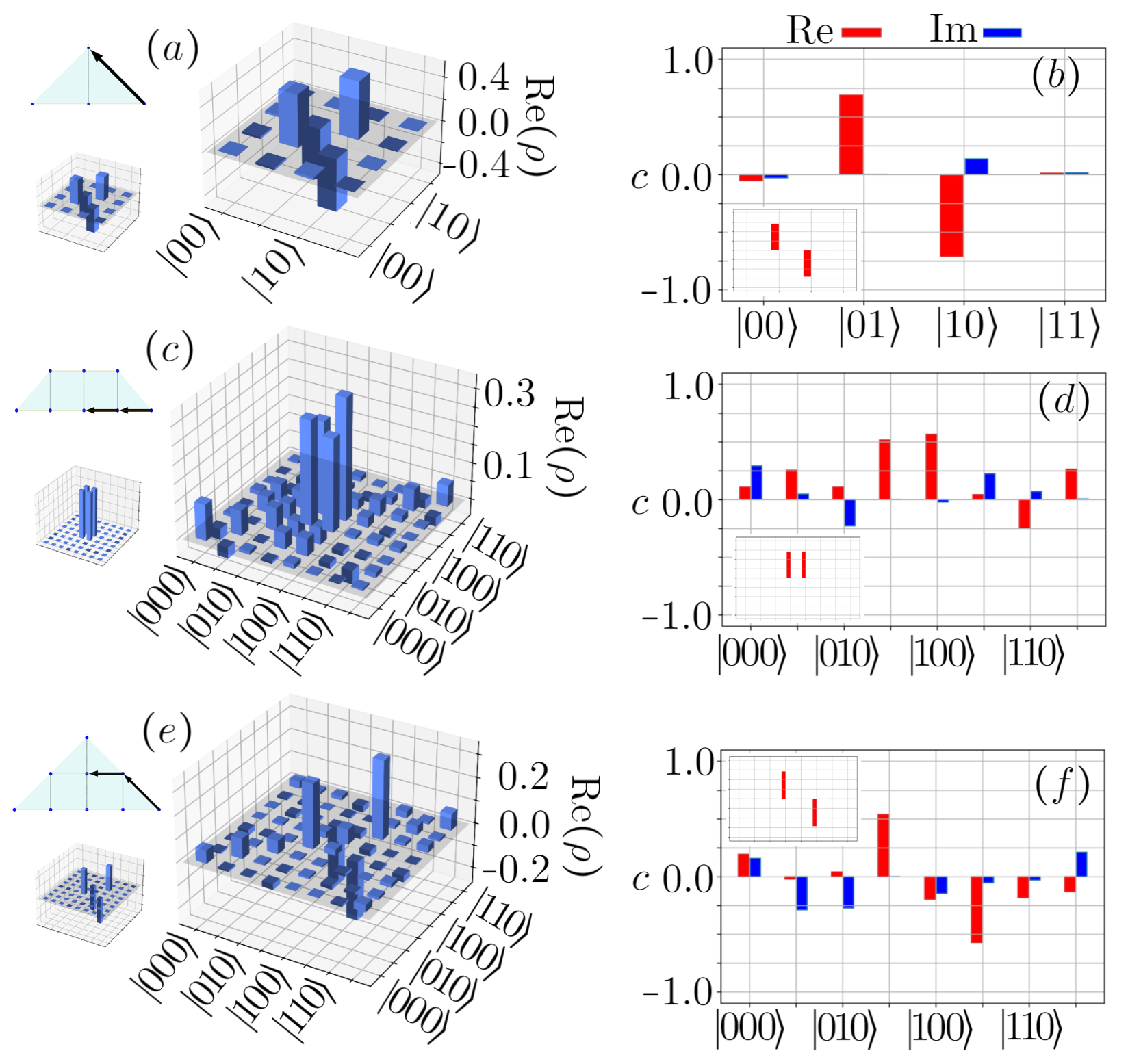}
\caption{\label{fig:validation} Engineered quantum walks are conducted on superconducting devices to perform state preparation. Full state tomography of density matrices (left) and partial tomography of state vectors (right) show that the prepared states are close to targeted angular momentum eigenstates (insets). 
(a, b) For test problem $j_1=j_2=1/2$, state $|0,0\rangle$ is prepared using one step of $L$ walk. (c, d) For $j_1=3/2$ and $j_2=1/2$, state $|2,0\rangle$ is prepared using two steps of $M$ walks. (e, f) For $j_1=j_2=1$, state $|1,0\rangle$ is prepared using one step of $L$ walk followed by one step of $M$ walk.
Results for other test problems (not shown) are similar.
}
\end{figure}
%$(a), (c), (e)$ xaxis: $|00\rangle, |01\rangle, |10\rangle, |11\rangle$; $|000\rangle, |001\rangle, |010\rangle, |011\rangle, |100\rangle, |101\rangle, |110\rangle, |111\rangle$, yaxis: $-1.0, -0.5, 0.0$
%$(b), (d), (f)$ yaxis: $\text{Re}(\rho), \text{Im}, c, -0.4, 0.0; 0.1, 0.2, 0.3, -0.2$

We perform small test problems on IBM's superconducting device Lima. At the time of our experiments, the device has 5 qubits and a reported quantum volume of 32. We use Qiskit's UnitaryGate function to convert sequences of unitaries to circuits of native gates on the quantum hardware, which are comprised of two-qubit CNOT gate and single-qubit gates. 
At the end of the quantum walk circuit $W$, we perform state tomography. A set of readout circuits $R_k$, for $k=1,\dots, K$, is needed in order to reconstruct the state of the qubits at the end of $W$. Our tomography method is described in Appendix~\ref{App:tomography}.
Appending measurement circuits $R_k$ to the state-preparation circuit $W$, we run a set of circuits $\{ W  R_k\}_{k=1}^K$ on the hardware and measure all qubits. For each circuit in the set, we perform $2^{14}$ shots to accumulate statistics of bit strings using Qiskit's Sampler. 
After finishing all shots on quantum hardware, we post process quasi probability distributions on a classical computer to infer states of the qubits. 
Example results are shown in Fig.~\ref{fig:validation}, and more results can be found in our Jupyter notebooks, which are openly available together with our Python source codes \cite{Shi24}.

For the simplest test problem $j_1=j_2=1/2$, two qubits are sufficient. As an example, we prepare the state $|0,0\rangle$, as indicated by the quantum walk diagram (top left inset) in Fig.~\ref{fig:validation}(a). 
Preparing the state requires a single step of $L$ walk, as discussed in Sec.~\ref{sec:example1}. 
The transpiled circuit typically contains two CNOT gates. Notice that such a circuit is not optimal: The state $|0,0\rangle = \frac{1}{\sqrt{2}}(|01\rangle-|10\rangle)$ is a well-known Bell state, which can be prepared using just one CNOT gate.
Notice that during tomography, we choose the global phase such that the smallest occupied $|l\rangle$ state has a real and positive coefficient $c_l$. This choice supersedes the sign convention of CG coefficients in this section.  
Although the quantum-walk circuit for preparing $|0,0\rangle$ is not optimal, the resultant density matrix is close to ideal, as shown in Fig.~\ref{fig:validation}(a). The main figure shows the real part of the reconstructed density matrix $\rho$, and the bottom left inset shows Re$(\sigma)$, where $\sigma$ denotes the density matrix of the ideal state. 
Ideally, Im$(\sigma)$ is zero. However, Im$(\rho)$ has small but nonzero components (not shown). 
The purity of the density matrix is $\text{tr}(\rho^2)\approx 0.98$, which suggest the quantum hardware is doing a decent job preparing a pure quantum state. 
The fidelity of of the quantum state is $F=\text{tr}\sqrt{\sqrt{\sigma}\rho\sqrt{\sigma}}\approx0.99$, which is computed using QuTiP. 
Approximating the state as pure, we reconstruct the state vector $\mathbf{c}$, whose components correspond to CG coefficients. As shown in Fig.~\ref{fig:validation}(b), the inferred coefficients are close to ideal (inset). However, we observe leakages to states that should not be occupied, as well as phase errors between occupied states.

For three-qubits test problems, we show two examples in Fig.~\ref{fig:validation}. First, for $j_1=3/2$ and $j_2=1/2$, the state space dimension is $D=8$, which fully utilizes all three qubits. To prepare state $|2,0\rangle$, two steps of $M$ walks are required. The two unitaries are transpiled to a total of $\sim60$ CNOT gates plus single-qubit gates. The circuits are not optimal, and the large gate depth is challenging for the quantum device. As shown in Fig.~\ref{fig:validation}(c), the realized density matrix deviates noticeably from the ideal case. The purity of the state is $0.38$, indicating that decoherence is significant. The fidelity of the state is $0.75$.
Approximating the state as pure, we infer coefficients of the prepared state, which are shown in Fig.~\ref{fig:validation}(d). The ideal state $|2,0\rangle=\frac{1}{\sqrt{2}}(|011\rangle+|100\rangle)$, as shown in the inset, could have been prepared using far fewer CNOT gates. Due to the large gate depth, the state vector prepared by our circuit deviates noticeably from the ideal state. 
Second, for $j_1=j_2=1$, the state space dimension is $D=9$. To fit into three qubits, we discard the bottom state $\|-1,-1\rrangle$, which does not participate in the quantum walk that prepares $|1,0\rangle$. This example utilizes both $M$ and $L$ walks, and the transpiled circuit also has $\sim 60$ CNOT gates. The resultant density matrix is shown in Fig.~\ref{fig:validation}(e). The purity is $0.40$ and the fidelity is $0.75$, which are comparable to the previous example because of a similar gate depth. 
The inferred state vector, as shown in Fig.~\ref{fig:validation}(f), has noticeable leakage and phase errors compared to the ideal state $|1,0\rangle = \frac{1}{\sqrt{2}}(|011\rangle-|101\rangle)$. 
In all these examples, the targeted states could have been prepared by much simpler and specialized circuits, which would have much better performance on noisy devices. Nevertheless, our state preparation protocol using engineered quantum walks provides a systematic way to scale up angular momentum on future quantum computers.

%%%%%%%%%%%%%%%%%%%%%%%%%%%%%%%%%%%%%%%%%%%%%
\section{\label{sec:conclusion}Discussions and Conclusions}
To prepare angular momentum eigenstates, we develop an approach based on engineered quantum walks. Rather than computing Clebsch–Gordan (CG) coefficients classically, and inputting them as a dense unitary matrix to quantum computers, which then implements the unitary by decomposing it into elementary gates, our approach does not require knowing CG coefficients \textit{a priori}. In fact, our approach may be regarded as a unitary method for computing CG coeffcients, which can be implemented both on classical and on quantum computers.

For large angular momentum, our approach typically offers a polynomial speedup, compared to a brute-force decomposition of the dense CG unitary. 
Notice that if one is interested in extracting every single CG coefficient from the quantum states, then our approach has a similar cost as classical methods, which are already optimally efficient. However, polynomial speedup arises if the goal is instead to prepare angular momentum eigenstates and use them for subsequent applications. In the best case, when preparing most eigenstates, the speed up is cubic. In the worst case, when preparing low angular momentum eigenstates from two large angular momenta, there is no speedup.

In a broader context, engineered quantum walks may be generalized as a framework for state preparation. While in usual studies of quantum walks an initial state spreads across the Hilbert space, we develop an approach to double pinch quantum walks such that each step is a unit-probability population transfer within a two-level system. We achieve double-pinched walks using a combination of projection and destructive interference, such that each step is induced by a Hamiltonian that is sparse in the computational basis. By choosing an appropriate sequence of Hamiltonians, and perform unitary evolution for their $\pi$-pulse time, we engineer a deterministic path for moving any initial state to any final state.

Moreover, engineered quantum walks may be regarded as a framework for decomposing  unitary matrices. In many cases, state preparation is difficult because the desired states, while simple in some basis, may be highly entangled in the computation basis. In this sense, state preparation is a change of basis, mediated by a unitary matrix that is potentially dense. While there is no efficient way to decompose a dense unitary matrix in general, if the problem has special structures, as in the case of angular momentum, then the state preparation can be implemented as a sequence of sparser unitaries, which effectively provides a decomposition of a dense unitary.

In summary, by investigating structures of the $\mathfrak{su}(2)\times \mathfrak{su}(2)$ algebra beyond their lowest-dimensional representations (Sec.~\ref{sec:operators}), we manage to determine matrix representations of a complete set of operators in both the tensor-product basis $\|m_1, m_2\rrangle$ and the direct-sum basis $|j,m\rangle$ (Sec.~\ref{sec:matrix}). 
Using these operators as building blocks, we construct Hamiltonians that isolate two adjacent $|j,m\rangle$ states from the rest of the Hilbert space (Secs.~\ref{sec:MWalk} and \ref{sec:DWalk}). 
After evolving an initial eigenstate using the Hamiltonian by its known $\pi$-pulse time, the population is completely transferred to the other eigenstate within the two level system. By repeating this process for a sequence of two level systems, an initial state is moved deterministically to any desired final state (Sec.~\ref{sec:prepare}).
We have verified our state preparation scheme on classical computers and tested it on quantum devices (Sec.~\ref{sec:test}). Although its advantage is yet to be realized on current devices, we expect our scheme to offer efficient state preparation for larger problems on future quantum computers, paving the way towards fast atomic and nuclear physics calculations, as well as applications enabled by spin networks and the Schur transform.  

\vspace{-10pt}
\begin{acknowledgments}
\vspace{-10pt}
This work was performed under the auspices of US DOE by LLNL under Contract DE-AC52-07NA27344, and reviewed and released under LLNL-JRNL-867573.
\end{acknowledgments}

\vspace{-10pt}
\section*{Code and data availability}
\vspace{-10pt}
The code and data that support the findings of this article are openly available \cite{Shi24}.

%%%%%%%%%%%%%%%%%%%%%%%%%%%%%%%%%%%%%%%%%%%%%%%
%%%%%%%%%%%%%%%%%%%%%%%%%%%%%%%%%%%%%%%%%%%%%%%
\appendix
\section{$\mathfrak{su}(2)\times \mathfrak{su}(2)$ generators\label{App:generators}}
In this appendix, we derive properties involving the vector operators $\mathbf{A}$ and $\mathbf{S}$, which are summarized in Sec.~\ref{sec:operators}. Additionally, we derive selection rules and matrix elements presented in Sec.~\ref{sec:matrix}.

%%%%%%%%%%%%%%%%%%%
\subsection{Vector operator $\mathbf{A}$ \label{App:generators_A}}
The operator $\mathbf{A}$ naturally arises when one consider the multiplication between $\Lambda$ and $\mathbf{J}_1$, because $[\Lambda, J_{1a}]=[J_{1b}, J_{1a}]J_{2b} = i\epsilon_{bac}J_{1c}J_{2b} = iA_a$. In a coordinate-independent form, $[\Lambda, \mathbf{J}_1]=[\mathbf{J}_2, \Lambda]=i\mathbf{A}$. %As a corollary, $[\Lambda, \mathbf{J}]=0$, which also follows from the first expression after Eq.~(\ref{eq:Lambda}).
%$[\textcolor{olive}{\Lambda}, \textcolor{red}{\mathbf{J}}]=0$ %%%%%%

As quantum operators, $\mathbf{A}$ is orthogonal to $\mathbf{J}$. To see this, using the multiplication table Eq.~(\ref{eq:J_table}), we have ${\mathbf{J}_1\cdot \mathbf{A}}=J_{1a}\epsilon_{abc}J_{1b}J_{2c}=iJ_{1c}J_{2c}=i\Lambda$. In other words, $\mathbf{A}$ is in fact not orthogonal to $\mathbf{J}_1$ at the quantum level. Following similar calculations, we have $\mathbf{J}_1\cdot \mathbf{A} = -\mathbf{J}_2\cdot \mathbf{A} = - \mathbf{A} \cdot \mathbf{J}_1  = \mathbf{A} \cdot \mathbf{J}_2  = i\Lambda$. Therefore, even though $\mathbf{J}_1$ and $\mathbf{J}_2$ are not separately orthogonal to $\mathbf{A}$, their sum $\mathbf{J}$ is.

To compute the multiplication between $A_a$ and $J_b$, using property of the Levi-Civita symbol that $\epsilon_{abc}\epsilon_{ade}=\delta_{bd}\delta_{ce}-\delta_{be}\delta_{cd}$.
By definition, $[A_a, J_{1b}]=\epsilon_{acd}[J_{1c}, J_{1b}]J_{2d} = i\epsilon_{acd} \epsilon_{cbe} J_{1e}J_{2d}=i(J_{1a} J_{2b} -\delta_{ab}\Lambda)$. Similarly, $[A_a, J_{2b}]=i(\delta_{ab}\Lambda - J_{1b} J_{2a})$. Using these expressions, it is easy to verify that $[\mathbf{A}, \mathbf{J}_1^2]=[\mathbf{A}, \mathbf{J}_2^2]=0$, which is expected because $\mathbf{A}$ can be measured simultaneously with $\mathbf{J}_1^2$ and $\mathbf{J}_2^2$.
Moreover, $[A_a, J_b] = i(J_{1a} J_{2b} - J_{1b} J_{2a})$, which gives Eq.~(\ref{eq:AJ_table}).

To compute multiplications between $A_a$ and $A_b$, we again expand $[A_a, A_b]=\epsilon_{acd}\epsilon_{bef}[J_{1c}J_{2d}, J_{1e}J_{2f}]$. The commutator can be decomposed as $[J_{1c}, J_{1e}]J_{2d}J_{2f} + J_{1e}J_{1c}[J_{2d}, J_{2f}]=i\epsilon_{ceg}J_{1g}J_{2d}J_{2f} + i\epsilon_{dfh}J_{1e}J_{1c} J_{2h}$. Using properties of the Levi-Civita symbol and the equivalent form of Eq.~(\ref{eq:J_table}), multiple equivalent expressions can be found, one of which is $[A_a, A_b]=\epsilon_{abc}[A_c+i(\Lambda J_{1c}+J_{2c}\Lambda)]$. The expression can be further simplified using $J_{2c}\Lambda = \Lambda J_{2c}+iA_c$, which gives Eq.~(\ref{eq:AA_table}).

Finally, the scalar operator $\mathbf{A}^2=A_a A_a$
can be expressed in terms of existing operators. By definition, $A_a A_a=\epsilon_{abc}J_{1b}J_{2c} \epsilon_{ade}J_{1d}J_{2e} = \mathbf{J}_1^2 \mathbf{J}_2^2 - J_{1a}\Lambda J_{2a}$. The second term equals to $J_{1a}([\Lambda, J_{2a}] + J_{2a}\Lambda)$. Using previous results, we obtain $\mathbf{A}^2 = \mathbf{J}_1^2 \mathbf{J}_2^2 -\Lambda-\Lambda^2$.

%%%%%%%%%%%%%%%%%%%
\subsection{Vector operator $\mathbf{S}$ \label{App:generators_S}}
The vector operator $\mathbf{S}$ arises when computing the commutation relation $[\Lambda, \mathbf{A}]$.
Using their definitions, $[\Lambda, A_a]=\epsilon_{abc}[J_{1d}J_{2d}, J_{1b}J_{2c}]$. The commutator can be decomposed as $[J_{1d}, J_{1b}] J_{2d}J_{2c} + J_{1b}J_{1d}[J_{2d}, J_{2c}]=i(A_b J_{2c}-J_{1b} A_c)$. We eliminate $\mathbf{J}_1$ and $\mathbf{J}_2$ in favor of $\mathbf{J}$, for example, by expressing $J_{2c}=J_c-J_{1c}$. Using previous results, we can simplify $\epsilon_{abc}(A_bJ_{1c} + J_{1b} A_c)=\epsilon_{abc}[A_b, J_{1c}]=iA_a$, which gives the form before Eq.~(\ref{eq:S}).

At the quantum level, $\mathbf{S}$ and $\mathbf{J}$ are orthogonal.
To see this, notice that $(\mathbf{A}\times\mathbf{J})\cdot\mathbf{J}= \mathbf{A}\cdot(\mathbf{J}\times\mathbf{J})= i\mathbf{A}\cdot\mathbf{J}=0$. Similarly, $\mathbf{J}\cdot(\mathbf{J}\times\mathbf{A})=0$. 
Moreover, $\mathbf{J}\cdot(\mathbf{A}\times\mathbf{J})=\epsilon_{abc}J_a([A_b, J_c]+J_cA_b)=i\mathbf{J}\cdot\mathbf{A}=0$, and similarly, $(\mathbf{J}\times\mathbf{A})\cdot\mathbf{J}=0$. 
Hence, $\mathbf{S}\cdot\mathbf{J}=\mathbf{J}\cdot\mathbf{S}=0$.

At the quantum level, $\mathbf{S}$ is not orthogonal to $\mathbf{A}$. Using the equivalent form of Eq.~(\ref{eq:AA_table}), $\mathbf{A}\cdot(\mathbf{A}\times\mathbf{J})=(\mathbf{A}\times\mathbf{A})\cdot\mathbf{J}=i\Lambda\mathbf{J}^2$. 
Similarly, because $\Lambda$ and $\mathbf{J}$ commute, we have $(\mathbf{J}\times\mathbf{A})\cdot\mathbf{A}=i\Lambda\mathbf{J}^2$.
Slightly more involved, $(\mathbf{A}\times\mathbf{J})\cdot\mathbf{A}=\epsilon_{abc}A_bJ_cA_a = \epsilon_{abc}A_b(A_aJ_c + [J_c,A_a])=2i\mathbf{A}^2-i\Lambda\mathbf{J}^2$, which equals to $\mathbf{A}\cdot(\mathbf{J}\times\mathbf{A})$ for similar reasons. Hence, $\mathbf{A}\cdot\mathbf{S}=-\mathbf{S}\cdot\mathbf{A}=i(\Lambda\mathbf{J}^2-\mathbf{A}^2)$ is nonzero.

The cross products of $\mathbf{A}$ and $\mathbf{S}$ are along the $\mathbf{J}$ direction. 
To see this, $[\mathbf{A}\times(\mathbf{J}\times\mathbf{A})]_a = \epsilon_{abc}\epsilon_{cde}A_bJ_dA_e=A_b(J_aA_b-J_bA_a)$. The second term is zero because $\mathbf{A}\cdot\mathbf{J}=0$. The first term equals to $([A_b, J_a] + J_a A_b)A_b=i\epsilon_{acb}A_cA_b+J_a\mathbf{A}^2=(\mathbf{A}^2-\Lambda)J_a$. Similarly, we have $(\mathbf{A}\times\mathbf{J})\times\mathbf{A}=\mathbf{A}\times(\mathbf{J}\times\mathbf{A})=(\mathbf{A}^2-\Lambda)\mathbf{J}$ and $(\mathbf{J}\times\mathbf{A})\times\mathbf{A}=\mathbf{A}\times(\mathbf{A}\times\mathbf{J})=-(\mathbf{A}^2+\Lambda)\mathbf{J}$. Combining these results, we see $\mathbf{S}\times\mathbf{A}=-\mathbf{A}\times\mathbf{S} = \mathbf{A}^2\mathbf{J}$.

The cross product between $\mathbf{J}$ and $\mathbf{S}$ are along the $\mathbf{A}$ direction. 
Consider the two terms of $\mathbf{S}$ separately, we have
$[(\mathbf{J}\times\mathbf{A})\times\mathbf{J}]_a=\epsilon_{abc}\epsilon_{bde}J_dA_eJ_c = (J_cA_a - J_aA_c)J_c$. The second term is zero, and the first term equals to 
$J_c([A_a, J_c]+J_cA_a)=i\epsilon_{acb} J_cA_b + \mathbf{J}^2A_a=(i\mathbf{J}\times\mathbf{A} + \mathbf{J}^2\mathbf{A})_a$.
Using the second equality of Eq.~(\ref{eq:S}), alternative forms can be obtained.
Similarly, $\mathbf{J}\times(\mathbf{A}\times\mathbf{J}) = i \mathbf{A}\times\mathbf{J} + \mathbf{A}\mathbf{J}^2$, $\mathbf{J}\times(\mathbf{J}\times\mathbf{A})=i \mathbf{J}\times\mathbf{A} - \mathbf{J}^2\mathbf{A}$, and $(\mathbf{A}\times\mathbf{J})\times\mathbf{J} = i \mathbf{A}\times\mathbf{J} - \mathbf{A}\mathbf{J}^2$. 
Combining these results, $\mathbf{J}\times\mathbf{S}=i\mathbf{S}+\frac{1}{2}(\mathbf{A}\mathbf{J}^2 + \mathbf{J}^2\mathbf{A})=\mathbf{J}^2\mathbf{A}$ and $\mathbf{S}\times\mathbf{J}=i\mathbf{S}-\frac{1}{2}(\mathbf{A}\mathbf{J}^2 + \mathbf{J}^2\mathbf{A})=-\mathbf{A}\mathbf{J}^2$. 
%The second equalities are obtained using $i\mathbf{S}=\frac{1}{2}[\mathbf{J}^2, \mathbf{A}]$. 

To compute $[\mathbf{S},\Lambda]$, consider the two terms of $\mathbf{S}$ separately.
We have $[(\mathbf{A}\times\mathbf{J})_a, \Lambda]=\epsilon_{abc}(A_b[J_c, \Lambda] + [A_b,\Lambda]J_c)$. The first commutator is zero, and the second equals to $-iS_b$. Hence, $[\mathbf{A}\times\mathbf{J}, \Lambda]=-i\mathbf{S}\times\mathbf{J}=i\mathbf{A}\mathbf{J}^2$, and similarly, $[\mathbf{J}\times\mathbf{A}, \Lambda]=-i\mathbf{J}^2\mathbf{A}$. Combining the two terms gives the first equality in Eq.~(\ref{eq:J2J2A}). 
To see the second equality, recall that from Eq.~(\ref{eq:S}), we have $[\mathbf{J}^2, \mathbf{A}]=2i\mathbf{S}$. Hence, it is sufficient to consider $[\mathbf{J}^2, \mathbf{S}]$. Because $\mathbf{J}_1^2$ and $\mathbf{J}_2^2$ commute with both $\mathbf{J}$ and $\mathbf{A}$, they also commute with $\mathbf{S}$. Then, $[\mathbf{J}^2, \mathbf{S}]=2[\Lambda, \mathbf{S}]$, so the second equality follows.

To compute $[J_a, S_b]$, we need $[J_a, (\mathbf{J}\times\mathbf{A})_b]=\epsilon_{bcd}([J_a, J_c]A_d + J_c[J_a, A_d])=i(J_aA_b - J_b A_a)=i\epsilon_{abc}(\mathbf{J}\times\mathbf{A})_c$. Similarly, $[J_a, (\mathbf{A}\times\mathbf{J})_b]=i\epsilon_{abc}(\mathbf{A}\times\mathbf{J})_c$. 
Combining the two terms gives the very simple result in Eq.~(\ref{eq:SJ_table}).

To compute $[A_a, S_b]$, we need $[A_a, (\mathbf{J}\times\mathbf{A})_b]=\epsilon_{bcd}([A_a, J_c]A_d + J_c[A_a, A_d]) =i[(\mathbf{A}^2\delta_{ab}-A_bA_a)+\Lambda(J_aJ_b-\mathbf{J}^2\delta_{ab})]$. Similarly, $[A_a, (\mathbf{A}\times\mathbf{J})_b]=i[(A_aA_b-\mathbf{A}^2\delta_{ab})+\Lambda(\mathbf{J}^2\delta_{ab}-J_bJ_a)]$. 
Using $A_aA_b-\Lambda J_aJ_b = A_bA_a - \Lambda J_b J_a$, we obtain Eq.~(\ref{eq:AS_table}).

To compute $[S_a, S_b]$, one approach is to express $S_b=i[A_b, \Lambda]$, which gives $[S_a, S_b]=i([[S_a, A_b],\Lambda]+[A_b, [S_a, \Lambda]])$. Using existing multiplication rules, the first double commutator equals to $i[\Lambda, A_bA_a]=-(A_bS_a+S_bA_a)$, and the second equals to $\frac{i}{2}[A_b, A_a\mathbf{J}^2+\mathbf{J}^2A_a]=A_aS_b+S_bA_a +\epsilon_{abc}\Lambda\mathbf{J}^2 J_c$. Adding the two terms gives the combination $A_aS_b-A_bS_a=\epsilon_{abc}(\mathbf{A}\times\mathbf{S})_c$. Using previous result for $\mathbf{A}\times\mathbf{S}$, we have $[S_a, S_b]=i\epsilon_{abc}(\Lambda\mathbf{J}^2-\mathbf{A}^2)J_c$. Recognizing the scalar prefactor as $\mathbf{A}\cdot\mathbf{S}$ gives Eq.~(\ref{eq:SS_table}).

Finally, to compute the scalar $\mathbf{S}^2$, replacing one occurrence of $S_a$ with $S_a=i[A_a,\Lambda]$, we obtain $S_aS_a=i(\mathbf{S}\cdot\mathbf{A}\Lambda-S_a\Lambda A_a)$. The second term equals to $([S_a,\Lambda] +\Lambda S_a)S_a = \frac{i}{2}(A_a\mathbf{J}^2A_a+\mathbf{J}^2\mathbf{A}^2)+\Lambda \mathbf{S}\cdot\mathbf{A}$, where $A_a\mathbf{J}^2A_a=[A_a,\mathbf{J}^2]A_a + \mathbf{J}^2\mathbf{A}^2$. Using Eq.~(\ref{eq:S}) and the fact that $\mathbf{S}\cdot\mathbf{A}$ and $\Lambda$ commute, these terms add to $\mathbf{S}^2=\mathbf{J}^2\mathbf{A}^2-i\mathbf{S}\cdot\mathbf{A}$, which further equals to $\mathbf{S}^2=\mathbf{A}^2+\mathbf{A}^2\mathbf{J}^2-\mathbf{J}^2\Lambda$.

%%%%%%%%%%%%%%%%%%%%%%%%%%%%%%%%%%%%%%%%%%%%%%%
\subsection{Selection rules of $\mathbf{A}$ \label{App:selection}}
To derive the selection rule in $j$, considering two expressions of $\langle \mathbf{O}\rangle :=\langle j', m'|[\mathbf{J}^2, [\mathbf{J}^2, \mathbf{A}]]|j, m\rangle=\langle j', m'|\mathbf{J}^2[\mathbf{J}^2, \mathbf{A}] - [\mathbf{J}^2, \mathbf{A}]\mathbf{J}^2|j, m\rangle$. Using Eq.~(\ref{eq:J2}) and acting $\mathbf{J}^2$ on the bra (ket) states when it appears on the left (right) gives $\langle \mathbf{O}\rangle=(\mathscr{J}_{j'}-\mathscr{J}_j)\langle j', m'|[\mathbf{J}^2, \mathbf{A}]|j, m\rangle = (\mathscr{J}_{j'}-\mathscr{J}_j)^2\langle j', m'|\mathbf{A}|j, m\rangle$.
On the other hand, $\langle \mathbf{O}\rangle =\langle j', m'|2(\mathbf{A}\mathbf{J}^2 + \mathbf{J}^2\mathbf{A})|j, m\rangle=2(\mathscr{J}_{j'}+\mathscr{J}_j) \langle j', m'|\mathbf{A}|j, m\rangle$. 
Comparing the two expressions of $\langle \mathbf{O}\rangle$, we see $[(\mathscr{J}_{j'}-\mathscr{J}_j)^2 - 2(\mathscr{J}_{j'}+\mathscr{J}_j)]\langle j', m'|\mathbf{A}|j, m\rangle=0$. The term in the square brackets can be factorized to $[(j+j'+1)^2-1][(j-j')^2-1]$. Unless $j+j'+1=\pm1$ or $j-j'=\pm1$, the matrix element is zero.  
The first case is possible only for the trivial case $j=j'=0$. For nontrivial cases where $j$ or $j'$ are positive, only the second case is possible, giving rise to the selection rule in $j$.

%%%%%%%%%%%%%%%%%%%%%%%%%%%%%%%%%%%%%%%%%%%%%%%
\subsection{Matrix elements of $A_z$ \label{App:A_z}}
To determine the absolute value $|a_j^m|$, we use the $z$ components of Eq.~(\ref{eq:AS_table}), namely, $\frac{1}{2}[A_z, [A_z, \mathbf{J}^2]]=A_z^2-\mathbf{A}^2-\Lambda (J_z^2-\mathbf{J}^2)$. The relation gives us two ways to compute $\langle O\rangle :=\frac{1}{2}\langle j, m|A_z[A_z, \mathbf{J}^2]-[A_z, \mathbf{J}^2]A_z|j, m\rangle$. 
Acting $A_z$ on the bra (ket) states when it appears on the left (right) gives
$\langle j, m|A_z[A_z, \mathbf{J}^2]|j, m\rangle = (\bar{a}_j^{m}\langle j-1, m| + a_{j+1}^m \langle j+1, m|)(A_z\mathbf{J}^2- \mathbf{J}^2 A_z)|j, m\rangle$.
Using the orthonormality of the $|j,m\rangle$ basis, the first term equals to $(|a_j^{m}|^2 + |a_{j+1}^m|^2) \mathscr{J}_j$, where $\mathscr{J}_j$ is given after Eq.~(\ref{eq:J2}), %, because the $j+2$ and $j-2$ states do not overlap with the $j$ state. 
and the second term equals to %$(\bar{a}_j^{m}\langle j-1, m| + a_{j+1}^m \langle j+1, m|)\mathbf{J}^2 (a_j^m|j-1, m\rangle + \bar{a}_{j+1}^{m} |j+1, m\rangle) = |a_j^{m}|^2 \mathscr{J}_{j-1} + |a_{j+1}^{m}|^2 \mathscr{J}_{j+1}$.
$-(|a_j^{m}|^2 \mathscr{J}_{j-1} + |a_{j+1}^{m}|^2 \mathscr{J}_{j+1})$.
Using similar methods to compute the other term, we obtain $\langle O\rangle=2[ j |a_j^{m}|^2 - (j+1)|a_{j+1}^{m}|^2]$.
On the other hand, $\langle O\rangle=\langle j, m|[A_z^2-\mathbf{A}^2-\Lambda (J_z^2-\mathbf{J}^2)]|j, m\rangle$. The first term is $\langle j, m|A_z^2|j, m\rangle=|a_j^{m}|^2 + |a_{j+1}^m|^2$, and the remaining terms are $\langle j, m|\Lambda (\mathbf{J}^2-J_z^2)-\mathbf{A}^2|j, m\rangle = \lambda_j(\mathscr{J}_j-m^2)-(\mathscr{J}_{j_1}\mathscr{J}_{j_2}-\lambda_j-\lambda_j^2)$, where $\lambda_j$ is given by Eq.~(\ref{eq:lambda_j}).
We thus obtain a recurrence relation 
\begin{eqnarray}
    \label{eq:recurrence}
    \nonumber
    &&(2j-1)|a_j^m|^2-(2j+3)|a_{j+1}^m|^2 \\
    &=&\lambda_j(\lambda_j+\mathscr{J}_j+1-m^2)-\mathscr{J}_{j_1}\mathscr{J}_{j_2}.
\end{eqnarray}

To solve the above nonlinear recurrence relation in $j$, first, we note that $a_{j_{\max}+1}^m=0$, because $A_z$ cannot raise $j$ beyond its top value $j_{\max}=j_1+j_2$. Using $\lambda_{j_1+j_2}=j_1 j_2$, the recurrence relation simplifies to $[2(j_1+j_2)-1]|a_{j_1+j_2}^m|^2=j_1j_2[(j_1+j_2)^2-m^2]$. 
Second, we note that $a_{j_1+j_2}^m=0$ when $m=\pm (j_1+j_2)$. More generally, we expect $a_j^{\pm j}=0$ for all $j$, because $a_j^{\pm j}$ is the coefficient of $|j-1,m={\pm j}\rangle$, which is not allowed. As a generalization, we postulate $a_j^m\propto \zeta_j^m$, where $\zeta_j^m$ is given by Eq.~(\ref{eq:zeta_j}). 
Third, we note that $a_{j_{\min}}^m=0$, because $A_z$ cannot lower $j$ below its bottom value $j_{\min}=j_1-j_2$. Using $\lambda_{j_1-j_2}=-j_2 (j_1+1)$, the recurrence relation simplifies to $[2(j_1-j_2)+3]|a_{j_1-j_2+1}^m|^2=j_2(j_1+1)(\zeta_{j_1-j_2+1}^m)^2$. 
To match both $a_{j_{\max}+1}^m=0$ and $a_{j_{\min}}^m=0$, we postulate $|a_j^m|=\frac{1}{2}\zeta_j^m\alpha_j$, where $\alpha_j$ is given by Eq.~(\ref{eq:alpha_j}).

Now let us prove that the recurrence relation Eq.~(\ref{eq:recurrence}) is solved by $|a_j^m|=\frac{1}{2}\zeta_j^m\alpha_j$. 
Factorizing $4(j+1)^2-1=(2j+1)(2j+3)$ and substituting Eqs.~(\ref{eq:zeta_j})-(\ref{eq:alpha_j}) into the LHS of Eq.~(\ref{eq:recurrence}), $4(2j+1)\text{LHS}=(j^2-m^2)[(j_{\max}+1)^2-j^2](j^2-j_{\min}^2) - (j\rightarrow j+1)$, where the second term means replacing all occurrences of $j$ by $j+1$ in the first term. 
The first term can be expanded as a polynomial $p(x)=-x^3+c_1 x^2 - c_2 x+c_3$, where $x=j^2$. Coefficients of the polynomial involves $(j_{\max}+1)^2 + j_{\min}^2=2(\mathscr{J}_{j_1}+\mathscr{J}_{j_2})+1$ and $(j_{\max}+1)j_{\min}=\mathscr{J}_{j_1}-\mathscr{J}_{j_2}$.
The $(j\rightarrow j+1)$ term is the same polynomial, but for $y=(j+1)^2$. 
Notice that $x+y=2\mathscr{J}_j+1$ and $xy=\mathscr{J}_j^2$.
Using $x^3-y^3=(x-y)(x^2+xy+y^2)$ and $x^2-y^2=(x-y)(x+y)$, we see $p(x)-p(y)$ has a common factor $y-x=2j+1$. Therefore, $4\text{LHS}= x^2+xy+y^2-c_1(x+y) + c_2 = 3\mathscr{J}_j^2+4\mathscr{J}_j+1-2c_1 \mathscr{J}_j+(c_2-c_1)$. 
Plugging in $c_1=2(\mathscr{J}_{j_1}+\mathscr{J}_{j_2}+1)+(m^2-1)$, $c_2-c_1=(\mathscr{J}_{j_1}-\mathscr{J}_{j_2})^2-1+2(m^2-1)(\mathscr{J}_{j_1}+\mathscr{J}_{j_2})$, and reorganizing terms into $\lambda_j$, it is a straightforward calculation to verify that $4\text{LHS}=4\text{RHS}$. In other words, $|a_j^m|=\frac{1}{2}\zeta_j^m\alpha_j$ satisfies the recurrence relation. $\blacksquare$

To prove that the complex phase of $a_j^m$ is $\pi/2$, we use two other recurrence relations.
First, to show that the phase is independent of $m$, we use Eq.~(\ref{eq:Apm}) to obtain $J_+ A_z J_- - J_-A_z J_+ = 2J_zA_z-(J_+A_-+J_-A_+)$. The second term equals to $2(J_xA_x+J_yA_y)=-2J_zA_z$ because $\mathbf{J}\cdot\mathbf{A}=0$. Taking matrix elements on the LHS of  $4J_zA_z=J_+ A_z J_- - J_-A_z J_+$ gives $\langle j-1,m|4A_zJ_z|j,m\rangle = 4ma_j^m$. % = \langle j-1,m|J_+ A_z J_- - J_-A_z J_+|j,m\rangle$ % = \zeta_j(m)[\zeta_j(m-1)a_j^{m-1} - \zeta_j(m+1)a_j^{m+1}]$, 
Taking the same matrix element on the RHS gives
\begin{equation}
    4ma_j^m = \zeta_j^m[\zeta_j^{m-1}a_j^{m-1} - \zeta_j^{m+1}a_j^{m+1}].
\end{equation}
The above recurrence relation is consistent with $|a_j^m|=\frac{1}{2}\alpha_j \zeta_j(m)$. More importantly, when $m=j-1$, because $\zeta_j(j)=0$, the recurrence relation reduces to $4(j-1)a_j^{j-1}=\zeta_j(j-1)\zeta_j(j-2)a_j^{j-2}$, from which we conclude that $a_j^{j-2}$ has the same phase as $a_j^{j-1}$. Using the recurrence relation to continue lowering the value of $m$, we see $a_j^m$ has the same phase for all $m$. 
Second, to show that the phase of $a_j^m$ is independent of $j$, we use CG coefficients to write $A_z|j,m\rangle = a_j^m|j-1, m\rangle + \bar{a}_{j+1}^{m} |j+1, m\rangle=(a_j^mC^{j-1}_{m_1,m_2}+\bar{a}_{j+1}^{m}C^{j+1}_{m_1,m_2})\| m_1, m_2 \rrangle$. %, where the summation is over $m_1+m_2=m$.
On the other hand, using Eq.~(\ref{eq:Az}), we find another expression of $A_z|j,m\rangle$ in the $\| m_1, m_2 \rrangle$ basis. Comparing the two expressions gives
\begin{eqnarray}
    && a_j^mC^{j-1}_{m_1,m_2}+\bar{a}_{j+1}^{m}C^{j+1}_{m_1,m_2} \\
    \nonumber
    &=&\! \frac{i}{2}\! \Big[\!C^j_{m_1\!-\!1, m_2\!+\!1}\!\mathcal{J}_{j_1}^-\!(\!m_1\!)\!\mathcal{J}_{j_2}^+\!(\!m_2\!) \!-\! C^j_{m_1\!+\!1,m_2-1}\!\mathcal{J}_{j_1}^+\!(\!m_1\!)\mathcal{J}_{j_2}^-\!(\!m_2\!)\!\Big].
\end{eqnarray}
%$A_z|j,m\rangle = \frac{i}{2}\sum[C^j_{m_1-1, m_2+1}\mathcal{J}_{j_1}^-(m_1)\mathcal{J}_{j_2}^+(m_2) - C^j_{m_1+1, m_2-1}\mathcal{J}_{j_1}^+(m_1)\mathcal{J}_{j_2}^-(m_2)] \| m_1, m_2 \rrangle$.
Denoting $a_j^m=e^{i\phi_j}|a_j^m|$, the above recurrence relation is of the form $p_je^{i\phi_j} + q_{j+1}e^{-i\phi_{j+1}}=ir_j$, where $p_j, q_j, r_j\in\mathbb{R}$. 
At $j=j_{\max}$, the CG coefficient $C^{j+1}_{m_1,m_2}=0$, which means $q_{j+1}=0$, so $\phi_{j_{\max}}=\pm\frac{\pi}{2}$. To determine the sign, we use the fact that $a_j^m$ has the same phase for all $m$. Hence, it is sufficient to consider the special case where $m_1=j_1$ and $m_2=j_2-1$. Using $C^{j_{\max}}_{j_1-1, j_2}=C^{j_{\max}-1}_{j_1, j_2-1}=\sqrt{j_1/(j_1+j_2)}$, we obtain $a_{j_{\max}}^{j_{\max}-1}=i\sqrt{j_1j_2}$, which means $\phi_{j_{\max}}=\frac{\pi}{2}$. 
Using the recurrence relation to continue lowering $j$, we see $\phi_{j}=\pm\frac{\pi}{2}$ for all $j$. 
Finally, to see all angles must take the plus sign, consider $j\sim j_1\sim j_2$ and
the limit $m_1\rightarrow j_1$ and $m_2\rightarrow -j_2$, such that $m\sim 0$. 
In this case, $p_j$ and $q_j$ are both $O(j^2)$, whereas $r_j=O(j)$, where we have used the fact that the CG coefficients are $O(1)$. 
The recurrence relation is satisfied only when $e^{i\phi_j}$ and $e^{-i\phi_{j+1}}$ are of opposite signs, so that the leading terms of $p_j$ and $q_j$ cancels. Since $\phi_{j_{\max}}=\frac{\pi}{2}$, we conclude that $\phi_j=\frac{\pi}{2}$ for all $j$. $\blacksquare$

%%%%%%%%%%%%%%%%%%%%%%%%%%%%%%%%%%%%%%%%%%%%%%%
\subsection{Matrix elements of $A_{\pm}$ and $\mathbf{S}$\label{App:Apm-S}}
From matrix elements of $A_z$, we determine matrix elements of $A_\pm$. Using Eq.~(\ref{eq:Apm}), $A_+|j,m\rangle = [A_z, J_+]|j,m\rangle = A_z\mathcal{J}_j^+(m)|j,m+1\rangle - J_+(a_j^m|j-1,m\rangle + \bar{a}_{j+1}^{m}|j+1,m\rangle) = \mathcal{J}_j^+(m)( a_j^{m+1}|j-1,m+1\rangle + \bar{a}_{j+1}^{m+1}|j+1,m+1\rangle) - (a_j^m\mathcal{J}_{j-1}^+(m)|j-1,m+1\rangle + \bar{a}_{j+1}^{m} \mathcal{J}_{j+1}^+(m)|j+1,m+1\rangle)$. The coefficient of $|j-1,m+1\rangle$ is $\mathcal{J}_j^+(m) a_j^{m+1} - \mathcal{J}_{j-1}^+(m)a_j^m = \frac{i}{2}\alpha_j[\mathcal{J}_j^+(m)\zeta_j^{m+1} - \mathcal{J}_{j-1}^+(m)\zeta_j^m]$. Using Eqs.~(\ref{eq:J_elements}) and (\ref{eq:zeta_j}), the terms in the square bracket simplify to $\sqrt{(j-m)(j-m-1)}$, which can be expressed in terms of Eq.~(\ref{eq:J0_element})
Following similar steps to compute the coefficient of $|j+1,m+1\rangle$ for $A_+$ and $A_-|j,m\rangle$, we obtain Eqs.~(\ref{eq:Ap_element}) and (\ref{eq:Am_element}).

We verify the property that $A_+^\dagger = A_-$ from their matrix elements. Taking Hermitian conjugate of $\langle j-1, m+1|A_+|j,m\rangle = \frac{i}{2}\alpha_j\mathcal{J}_{j-1}^0(-m)$ gives $\langle j, m|A_+^\dagger|j-1,m+1\rangle = -\frac{i}{2}\alpha_j\mathcal{J}_{j-1}^0(-m)$. On the other hand, using Eq.~(\ref{eq:Am_element}) and replacing $j\rightarrow j-1$ and $m\rightarrow m+1$, we have
$\langle j, m|A_-|j-1,m+1\rangle = -\frac{i}{2}\alpha_j\mathcal{J}_{j}^0(-m-1)$.
Because $\mathcal{J}_{j-1}^0(-m) = \mathcal{J}_{j}^0(-m-1)$, the two expressions are consistent. 
Similarly, we can verify other matrix elements of $A_\pm$ respect the property that $A_+^\dagger = A_-$.

Finally, we obtain matrix elements of $\mathbf{S}$. 
Using $S_z=i[A_z, \Lambda]$ and matrix elements of $A_z$ and $\Lambda$, we obtain
$S_z|j,m\rangle=i[A_z\lambda_j|j,m\rangle - \Lambda(a_j^m|j-1,m\rangle + \bar{a}_{j+1}^{m}|j+1,m\rangle)] = i[\lambda_j(a_j^m|j-1,m\rangle + \bar{a}_{j+1}^{m}|j+1,m\rangle) - (a_j^m\lambda_{j-1}|j-1,m\rangle + \bar{a}_{j+1}^{m}\lambda_{j+1}|j+1,m\rangle)]$. Because $\lambda_j-\lambda_{j-1}=j$, the expression simplifies to
% \begin{eqnarray}
%     \label{eq:Sz_element}
%     \nonumber
%     S_z|j,m\rangle&=&-\frac{1}{2}\Big[j\alpha_j\zeta_j^m |j-1,m\rangle \\
%     && + (j+1)\alpha_{j+1}\zeta_{j+1}^m |j+1,m\rangle\Big].
% \end{eqnarray}
\begin{eqnarray}
    \label{eq:Sz_element}
    S_z|j,m\!\rangle\!=\!-\frac{1}{2}\!\Big[\!j\alpha_j\zeta_j^m |j\!-\!1,m\!\rangle \!+\! (j\!+\!1)\alpha_{j+1}\zeta_{j+1}^m |j\!+\!1,m\!\rangle\!\Big].\quad\quad
\end{eqnarray}
To obtain matrix elements of $S_\pm=S_x\pm i S_y$, we use Eq.~(\ref{eq:SJ_table}) to obtain $[S_z, J_\pm]=\pm S_\pm$. Acting both sides on $|j,m\rangle$ and following similar steps as $A_\pm$, we obtain
\begin{eqnarray}
    \label{eq:Sp_element} 
    \nonumber
    S_+|j,m\rangle &=& -\frac{1}{2}\Big[j \alpha_j \mathcal{J}_{j-1}^0(-m)  |j-1,m+1\rangle \\
    && - (j+1)\alpha_{j+1} \mathcal{J}_{j+1}^0(m)  |j+1,m+1\rangle \Big], \\
    \label{eq:Sm_element} 
    \nonumber
    S_-|j,m\rangle &=& \frac{1}{2}\Big[j \alpha_j \mathcal{J}_{j-1}^0(m)  |j-1,m-1\rangle \\
    && -(j+1)\alpha_{j+1} \mathcal{J}_{j+1}^0(-m)  |j+1,m-1\rangle \Big].    
\end{eqnarray}
The matrix elements for $S_\pm$ can also be derived from other commutation relations, such as $S_\pm = i[A_\pm, \Lambda]$, and the results are consistent. 
One can also verify the property that that $S_+^\dagger = S_-$ from their matrix elements.

%%%%%%%%%%%%%%%%%%%%%%%%%%%%%%%%%%%%%%%%%%%%%%%
\section{Building blocks of $j$ walks\label{App:buildingblocks}}
%To implement the $L$ and $R$ walks, we need their matrix elements.
In this appendix, we list matrix elements of the common building blocks $J_+A_z$, $A_zJ_+$, $\Lambda A_+$ and $A_+\Lambda$ in the computational basis $\|m_1, m_2\rrangle$. We give the matrix elements both in the general case, and in two illustrative examples.

%%%%%%%%%%%%%%%%%%%%%%%%
\subsection{General case\label{App:buildingblocks_general}}
In the computational basis, $A_z$ is given by Eq.~(\ref{eq:Az}), $A_\pm$ is given after Eq.~(\ref{eq:Apm}), and $\Lambda$ is given after Eq.~(\ref{eq:lambda_j}).
Due to their coupling patterns, it is sufficient to focus on the $m$ and $m-1$ subspace. 
First, using the expression $A_zJ_+ = \frac{i}2{}(J_{1+}^2J_{2-} -J_{1-}J_{2+}^2 + J_{1+}J_{2-}J_{2+}-J_{1-}J_{1+}J_{2+})$, we find that
\begin{eqnarray}
    \label{eq:AzJ_elements_m1m2}
    %&&P_{m-1/2}(A_zJ_+ + \text{h.c.})P_{m-1/2} \|m_1+m_2=m-1\rrangle \\
    &&(A_zJ_+ + \text{h.c.}) \|m_1+m_2=m-1\rrangle \\
    \nonumber
    &&=\!\frac{i}{2}\!\Big\{ 
    \mathcal{J}_{j_1}^+(m_1+1)\mathcal{J}_{j_1}^+(m_1) \mathcal{J}_{j_2}^-(m_2) \|m_1+2, m_2-1\rrangle \\
    \nonumber
    &&\;-\;\,\;\,
    \mathcal{J}_{j_2}^+(m_2+1)\mathcal{J}_{j_2}^+(m_2)\mathcal{J}_{j_1}^-(m_1) \|m_1-1, m_2+2\rrangle \\
    \nonumber
    &&\;+\;\,\;\,
    \mathcal{J}_{j_1}^+(m_1) [\mathcal{J}_{j_2}^+(m_2)]^2 \|m_1+1, m_2\rrangle \\
    \nonumber
    &&\;-\;\,\;\,
    \mathcal{J}_{j_2}^+(m_2)[\mathcal{J}_{j_1}^+(m_1)]^2  \|m_1, m_2+1\rrangle
    \Big\}.
\end{eqnarray}
We see the operator is represented by a 4-sparse matrix in the computational basis.
Second, using $A_+=i(J_{1z}J_{2+}-J_{1+}J_{2z})$, we have
\begin{eqnarray}
    \label{eq:Ap_elements_m1m2}
    &&\quad(A_+ + \text{h.c.})\|m_1+m_2=m-1\rrangle \\
    \nonumber
    &&=i\Big[
    m_1 \mathcal{J}_{j_2}^+(\!m_2\!) \|m_1, m_2+1\rrangle -
    m_2 \mathcal{J}_{j_1}^+(\!m_1\!) \|m_1+1, m_2\rrangle
    \Big],
\end{eqnarray}
so the operator is a 2-sparse matrix. 
Finally, the operator $\Lambda A_+ + A_+\Lambda$ is slightly more complicated. We expand the operators and collect terms according to how they move the states.  
Terms that move $\|m_1, m_2\rrangle\rightarrow \|m_1, m_2+1\rrangle$ are $i$ times $(J_{1z}^2-\frac{1}{2}J_{1+}J_{1-})J_{2z}J_{2+} + (J_{1z}^2-\frac{1}{2}J_{1-}J_{1+})J_{2+}J_{2z}$. Using expressions of $J_+ J_-$ and $J_-J_+$ before Eq.~(\ref{eq:J2}), these terms simplify to $\frac{1}{2}[(3J_{1z}^2-\mathbf{J}_1^2)\{J_{2z}, J_{2+}\} - J_{1z}J_{2+}]$, where the curly brackets denote anti-commutators. 
From this expression, terms that move $\|m_1, m_2\rrangle\rightarrow \|m_1+1, m_2\rrangle$ are $-i$ times $(1\leftrightarrow2)$, where the two angular momenta are swapped. 
We also have terms that move $\|m_1, m_2\rrangle\rightarrow \|m_1-1, m_2+2\rrangle$, which are $i$ times $\frac{1}{2}\{J_{1z}, J_{1-}\} J_{2+}^2$, and terms that move $\|m_1, m_2\rrangle\rightarrow \|m_1+2, m_2-1\rrangle$, which are $-i$ times $(1\leftrightarrow2)$. 
Using these expressions, the matrix elements are 
\begin{eqnarray}
    \label{eq:ApLambda_elements_m1m2}
    &&(\{\Lambda, A_+\} + \text{h.c.}) \|m_1\!+\!m_2\!=\!m-1\rrangle \\
    \nonumber
    &&=\frac{i}{2}\Big\{ 
    [(3m_1^2\!-\!\mathscr{J}_{j_1})(2m_2+1)\!-\!m_1] \mathcal{J}_{j_2}^+(\!m_2) \|m_1, m_2+1\rrangle \\
    \nonumber
    &&-\;\,\;\,
    [(3m_2^2-\mathscr{J}_{j_2})(2m_1+1)\!-m_2] \mathcal{J}_{j_1}^+(m_1) \|m_1+1, m_2\rrangle \\
    \nonumber
    &&+
    (2m_1\!-\!1) \mathcal{J}_{j_1}^-(\!m_1\!)\mathcal{J}_{j_2}^+(\!m_2\!+\!1)\mathcal{J}_{j_2}^+(\!m_2\!)  \|m_1\!-\!1, m_2\!+\!2\rrangle\\
    &&-
    \nonumber
    (2m_2\!-\!1) \mathcal{J}_{j_2}^-(\!m_2\!)\mathcal{J}_{j_1}^+(\!m_1\!+\!1)\mathcal{J}_{j_1}^+(\!m_1\!)  \|m_1\!+\!2, m_2\!-\!1\rrangle
    \Big\}.
\end{eqnarray}
so the operator is a 4-sparse matrix and has the same stencil as Eq.~(\ref{eq:AzJ_elements_m1m2}).
Because the operators are Hermitian, by taking conjugate transpose, matrix elements for $\|m_1+m_2=m\rrangle$ states are readily obtained. 

%As in the general case, to find the three building blocks for $L$ and $R$ walks, it is sufficient to consider the upper triangles of their matrices, which correspond to rising the value of $m$. 

%%%%%%%%%%%%%%%%%%%%%%%%
\subsection{Example $j_1=j_2=1/2$ \label{App:buildingblocks_example1}}
%When $j_1=j_2=1/2$, we find matrix elements of the three building blocks in the qubit basis as follows. 
Using Eq.~(\ref{eq:AzJ_elements_m1m2}), we find 
%$A_zJ_+|0\rangle = A_zJ_+|1\rangle=A_zJ_+|2\rangle=0$, 
$A_zJ_+$ acting on $|0\rangle$, $|1\rangle$ and $|2\rangle$ are zero,
and $A_zJ_+|3\rangle=\frac{i}{2}(|2\rangle-|1\rangle)$, which gives the upper triangle of the matrix for $A_zJ_+ + $h.c..
Using Eq.~(\ref{eq:Ap_elements_m1m2}), we find $A_+|0\rangle =0$, $A_+|1\rangle =-\frac{i}{2}|0\rangle$, $A_+|2\rangle =\frac{i}{2}|0\rangle$, and $A_+|2\rangle =\frac{i}{2}(|2\rangle-|1\rangle)$, which gives the upper triangle of the matrix for $A_+ + $h.c..
Using Eq.~(\ref{eq:ApLambda_elements_m1m2}), we find $\{A_+,\Lambda\}|0\rangle =0$, $\{A_+,\Lambda\}|1\rangle =\frac{i}{4}|0\rangle$, $\{A_+,\Lambda\}|2\rangle =-\frac{i}{4}|0\rangle$, and $\{A_+,\Lambda\}|2\rangle =\frac{i}{4}(|1\rangle-|2\rangle)$, which gives the upper triangle of the matrix for $\{A_+,\Lambda\} + $h.c..
Notice that in this special case, $\{A_+,\Lambda\}+\textrm{h.c.}=-\frac{1}{2}(A_+ + \textrm{h.c.})$.

%%%%%%%%%%%%%%%%%%%%%%%%
\subsection{Example $j_1=1, j_2=1/2$ \label{App:buildingblocks_example2}}
%When $j_1=1$ and $j_2=1/2$, matrix elements of the three building blocks in the qubit basis are found as follows. 
Using Eq.~(\ref{eq:AzJ_elements_m1m2}), we find 
$A_zJ_+$ acting on $\|1,\frac{1}{2}\rrangle$, $\|0,\frac{1}{2}\rrangle$, and  
$\|1,-\frac{1}{2}\rrangle$ gives zero, while
$A_zJ_+\|-1,\frac{1}{2}\rrangle=i\|1,-\frac{1}{2}\rrangle$, 
$A_zJ_+\|0,-\frac{1}{2}\rrangle=i(\frac{1}{\sqrt{2}} \|1,-\frac{1}{2}\rrangle -\|0,\frac{1}{2}\rrangle)$, 
and $A_zJ_+\|-1,-\frac{1}{2}\rrangle=i(\frac{1}{\sqrt{2}} \|0,-\frac{1}{2}\rrangle -\|-1,\frac{1}{2}\rrangle)$, which gives the upper triangle of $A_zJ_+ + \textrm{h.c.}$.
The full matrix is %in the $(|0\rangle, \dots, |5\rangle)$ basis is
\begin{equation}
    %\vspace{-2pt}
    \nonumber
    A_zJ_+ + \textrm{h.c.} = \frac{i}{\sqrt{2}}
    \left(\! \begin{array}{cccccc}
    0 & 0 & 0 & 0 & 0 & 0 \\
    0 & 0 & 0 & 0 & -\sqrt{2} & 0 \\
    0 & 0 & 0 & \sqrt{2} & 1 & 0 \\
    0 & 0 & -\sqrt{2} & 0 & 0 & -\sqrt{2} \\
    0 & \sqrt{2} & -1 & 0 & 0 & 1 \\
    0 & 0 & 0 & \sqrt{2} & -1 & 0
    \end{array}\! \right).
    %\vspace{-2pt}
\end{equation}
Notice that the matrix is 3-sparse, rather than 4-sparse in the general case, because the state space is not large enough.
Using Eq.~(\ref{eq:Ap_elements_m1m2}), we find 
$A_+\|1,\frac{1}{2}\rrangle = 0$,
$A_+\|0,\frac{1}{2}\rrangle=-\frac{i}{\sqrt{2}} \|1,\frac{1}{2}\rrangle$, 
$A_+\|1,-\frac{1}{2}\rrangle=i \|1,\frac{1}{2}\rrangle$, 
$A_+\|-1,\frac{1}{2}\rrangle=-\frac{i}{\sqrt{2}} \|0,\frac{1}{2}\rrangle$,
$A_+\|0,-\frac{1}{2}\rrangle=\frac{i}{\sqrt{2}} \|1,-\frac{1}{2}\rrangle$, 
and $A_+\|-1,-\frac{1}{2}\rrangle=i(\frac{1}{\sqrt{2}} \|0,-\frac{1}{2}\rrangle -\|-1,\frac{1}{2}\rrangle)$.
From its upper triangle, the full matrix is
%\vspace{-2pt}
\begin{equation}
    \nonumber
    A_+ + \textrm{h.c.} = \frac{i}{\sqrt{2}}
    \left(\! \begin{array}{cccccc}
    0 & -1 & \sqrt{2} & 0 & 0 & 0 \\
    1 & 0 & 0 & -1 & 0 & 0 \\
    -\sqrt{2} & 0 & 0 & 0 & 1 & 0 \\
    0 & 1 & 0 & 0 & 0 & -\sqrt{2} \\
    0 & 0 & -1 & 0 & 0 & 1 \\
    0 & 0 & 0 & \sqrt{2} & -1 & 0
    \end{array}\! \right),
\end{equation}
%\vspace{-2pt}
which is only 2-sparse. %this matrix looks very similar to that of $A_zJ_+ + \textrm{h.c.}$, but is only 2-sparse.
Finally, using Eq.~(\ref{eq:ApLambda_elements_m1m2}), we find the upper triangle 
$\{A_+,\Lambda\}\|1,\frac{1}{2}\rrangle = 0$,
$\{A_+,\Lambda\}\|0,\frac{1}{2}\rrangle=\frac{i}{2\sqrt{2}} \|1,\frac{1}{2}\rrangle$, 
$\{A_+,\Lambda\}\|1,-\frac{1}{2}\rrangle=-\frac{i}{2}\|1,\frac{1}{2}\rrangle$, 
$\{A_+,\Lambda\}\|-1,\frac{1}{2}\rrangle=\frac{i}{2\sqrt{2}} \|0,\frac{1}{2}\rrangle$,
$\{A_+,\Lambda\}\|0,-\frac{1}{2}\rrangle=-\frac{i}{2\sqrt{2}} \|1,-\frac{1}{2}\rrangle$, 
and $\{A_+,\Lambda\}\|-1,-\frac{1}{2}\rrangle=\frac{i}{2}(\|-1,\frac{1}{2}\rrangle -\frac{1}{\sqrt{2}} \|0,-\frac{1}{2}\rrangle)$, which shows
\begin{equation}
    \nonumber
    \{A_+,\Lambda\} + \textrm{h.c.} = -\frac{1}{2}(A_+ + \textrm{h.c.}).
\end{equation}
This relation is only true in special cases where there are multiple cancellations. For example, because $\mathscr{J}_{j_2}=3/4$, it always cancels with $3m_2^2$. Similarly, $2m_2\pm 1$ is always zero when $m_2=\mp1/2$. Moreover, $m_1=0$ removes additional terms that would be present in general cases.

%%%%%%%%%%%%%%%%%%%%%%%%%%%%%%%%%%%%%%%%%%%%%%%
\section{Error analysis of classical verification\label{App:Errors}}
The observed near-exponential separation between state vectors is likely due to non-unitary rounding errors present in both the classical and the quantum algorithms. 
For the classical algorithm, the exponential increase of error can be suppressed using multiword integer arithmetic \cite{johansson2016fast}.
For the quantum-walk algorithm, we expect errors to accumulate only polynomially with the path length, provided that the errors are purely unitary. The reason is as follows.
Suppose the exact state vector is $\mathbf{c}_l=V_l\mathbf{c}_0$, where $V_l=\prod_{k=1}^lU_k$, whereas the realized state vector is $\tilde{\mathbf{c}}_l=\tilde{V}_l\mathbf{c}_0$, where $\tilde{V}_l=\prod_{k=1}^l\tilde{U}_k$.
As long as both $V_l$ and $\tilde{V}_l$ are unitary, $|\mathbf{c}_l-\tilde{\mathbf{c}}_l|^2=2-(\mathbf{c}_0^\dagger V_l^\dagger \tilde{V}_l \mathbf{c}_0 + \text{c.c.})$, because the initial state vector is normalized. Notice that $|\mathbf{c}_l-\tilde{\mathbf{c}}_l|^2\in[0,4]$ is bounded. 
To obtain a tighter upper bound, suppose $\tilde{U}_k=U_k\exp(i\Delta_k)$. Although slightly different from $U_k$, if the realized $\tilde{U}_k$ is nevertheless still a unitary matrix, then $\Delta_k$ is a Hermitian matrix and hence can be diagonalized. We denote the eigenvalues of $\Delta_k$ by $\lambda_{kp}$, where $p=1, \dots, D$, and assume eigenvalues are nondegenerate, so their eigenvectors $\bm{\xi}_{kp}$ form an orthonormal basis.
We expand $\mathbf{c}_0=\sum_p \gamma_{0p}\bm{\xi}_{1p}$, then $\mathbf{c}_1^\dagger \tilde{\mathbf{c}}_1 = \mathbf{c}_0^\dagger V_1^\dagger \tilde{V}_1 \mathbf{c}_0 = \mathbf{c}_0^\dagger \exp(i\Delta_1) \mathbf{c}_0 =\sum_pe^{i\lambda_{1p}}|\gamma_{0p}|^2$. 
Therefore, $\mathbf{c}_1^\dagger \tilde{\mathbf{c}}_1 + \text{c.c.}=2 \sum_p |\gamma_{0p}|^2 \cos\lambda_{1p}\ge 2\cos|\lambda_1|$, where $|\lambda_k|=\max_{p}|\lambda_{kp}|$ and we have used the fact that $\bm{\gamma}_0$ is a normalized vector.
Then, an upper bound is $|\mathbf{c}_1-\tilde{\mathbf{c}}_1|^2\le 2(1-\cos|\lambda_1|)$, 
Next, when computing $|\mathbf{c}_2-\tilde{\mathbf{c}}_2|^2$, we need $\mathbf{c}_0^\dagger V_2^\dagger \tilde{V}_2 \mathbf{c}_0 = \mathbf{c}_1^\dagger \exp(i\Delta_2)\tilde{\mathbf{c}}_1$. 
To estimate this term, we expand $\mathbf{c}_1=\sum_p \gamma_{1p}\bm{\xi}_{2p}$ and $\tilde{\mathbf{c}}_1=\sum_p \tilde{\gamma}_{1p}\bm{\xi}_{2p}$, then
$\mathbf{c}_1^\dagger \tilde{\mathbf{c}}_1=\sum_p \bar{\gamma}_{1p}\tilde{\gamma}_{1p}$, where bar denotes complex conjugation.
Moreover, $\mathbf{c}_1^\dagger \exp(i\Delta_2)\tilde{\mathbf{c}}_1 + \text{c.c.}=\sum_p[\cos\lambda_{2p}(\bar{\gamma}_{1p}\tilde{\gamma}_{1p} + \text{c.c.}) + i\sin\lambda_{2p}(\bar{\gamma}_{1p}\tilde{\gamma}_{1p} - \text{c.c.})]$.
Suppose errors are small, namely, $|\lambda_{kp}|\ll 1$, then the $\sin$ terms can be ignored. Moreover, since $\bm{\gamma}_{1}$ is close to  $\bm{\tilde{\gamma}}_{1}$, coefficients of the $\cos$ terms are positive. 
Then, $\mathbf{c}_1^\dagger \exp(i\Delta_2)\tilde{\mathbf{c}}_1 + \text{c.c.}\gtrsim \cos|\lambda_2|\sum_p (\bar{\gamma}_{1p}\tilde{\gamma}_{1p} + \text{c.c.}) = \cos|\lambda_2| (\mathbf{c}_1^\dagger \tilde{\mathbf{c}}_1 + \text{c.c.})\ge 2\cos|\lambda_1|\cos|\lambda_2|$. 
By induction, we find an upper bound in the small-error limit to be
\begin{equation}
    \label{eq:error_bound}
    |\mathbf{c}_l-\tilde{\mathbf{c}}_l|^2 \lesssim 2\Big(1-\prod_{k=1}^l \cos|\lambda_k|\Big),
\end{equation}
where $|\lambda_k|\ll1 $ is again the largest eigenvalue of the error matrix $\Delta_k$. We may interpret $\lambda_k$ as the leading angle error for the $k$-th unitary rotation.
To see a possible dependence of $\lambda_k$ on $k$, consider the two example paths shown in the inset of Fig.~\ref{fig:verification}(a). Along these two paths, the Hamiltonian matrices, and therefore the error matrices, become increasingly dense as $k$ increases. The number of nonzero elements is $O(k^2)$. Suppose a typical nonzero element is $O(\delta)$ for some $\delta\ll1$, then the characteristic polynomial of $\Delta_k$ is $\lambda^{D-k}[\lambda^k - (\sum_{p=1}^k\delta_p)\lambda^{k-1}+\dots]$. Because the sum of all roots is $\sum_p\lambda_{kp} = \sum_{p=1}^k\delta_p = O(k\delta)$, the largest eigenvalue is bounded from below by $|\lambda_k|=O(k\delta)$.  
In more general cases, $|\lambda_k|=Q_k(\delta)$ is a polynomial. As long as $\delta\ll1$, we can expand $\cos|\lambda_k|\simeq 1-\frac{1}{2}Q_k^2(\delta)$, then $2(1-\prod_{k=1}^l \cos|\lambda_k|)\simeq \sum_k Q_k^2(\delta)$ is also a polynomial.
In other words, as long as the error per step is unitary, the accumulated error along any paths only increases polynomially with the path length.

In contrast, when testing on classical computers, the rounding errors are often non-unitary. Suppose the norm error per step is $O(\delta)$, namely, $|\tilde{\mathbf{c}}_{l}|\simeq (1+\delta_l)|\tilde{\mathbf{c}}_{l-1}|$, where $\delta_l$ may be viewed as a random variable with $\langle\delta_l\rangle=0$ and $\langle (1+\delta_l)(1+\delta_{l'})\rangle = 1+\langle \delta_l \delta_{l'}\rangle$, where $\langle \delta_l \delta_{l'}\rangle=O(\delta^2)$ is the error correlation. The correlation is nonzero because errors are path dependent.
For the quantum algorithm, we may estimate $|\tilde{\mathbf{c}}_{l}-\mathbf{c}_{l}|^2\simeq 1+(1+\delta^2)^{l}-2(1+\delta^2)^{l/2}\cos\theta_l$, where $\theta_l$ is the leading angle error. 
Similarly, the classical algorithm also accumulates errors exponentially, so $|\tilde{\mathbf{c}}_{l}-\tilde{\mathbf{C}}_{l}|^2$ depends on $l$ exponentially. 
In other words, Fig.~\ref{fig:verification} does not imply that errors of the quantum algorithm scales exponentially with the problem size, as long as we ensure that each step of the quantum walk is a unitary evolution.

%%%%%%%%%%%%%%%%%%%%%%%%%%%%%%%%%%%%%%%%%%%%%%%
\section{Quantum state tomography\label{App:tomography}}
Multiple methods have been investigated for estimating mixed states \cite{James01, paris2004quantum, cramer2010efficient, Rambach21, gebhart2023learning} and pure states \cite{Hayashi05, Lee18, zambrano2020estimation, bantysh2020comparison}. 
For completeness, this appendix describes the tomography method we use for Sec.~\ref{sec:quantum}.
%Symbols used here are separate from, and should not be confused with, those used in other sections.
Since errors are dominated by two-qubit gates during state preparation, instead of using more advanced tomography schemes, we simply assume that single-qubit gates used in measurement circuits are error-free. 
%For $n$ qubits, whose Hilbert space dimension is $N=2^n$, we first describe how we infer full density matrices, which cost $3^n$ circuits, and then describe how we infer pure states, which cost $(2n+1)$ circuits. 

%%%%%%%%%%%%%%%%%%%%%%%%%%%
\subsection{Mixed states\label{sec:mixed}}
A mixed state is described by its density matrix $\rho$, which is a $N\times N$ Hermitian matrix and can thus be spanned by Pauli basis. The number of real degrees of freedom (d.o.f.) along the diagonal of $\rho$ is $N$. Since off-diagonal elements are complex, the d.o.f. there is $\sum_{k=1}^{N-1}2k=N(N-1)$. Constrained by tr$(\rho)=1$, the total d.o.f. is $N+N(N-1)-1=N^2-1=4^n-1$, where $n$ is the number of qubits.
To account for these d.o.f., we expand $\rho$ in Pauli basis as
\begin{equation}
    \label{eq:rho_pauli}
    \rho=\frac{1}{2^n}\mathbb{I}+\sum_{k=1}^{4^n-1} R_k \sigma^n_{k},
\end{equation}
where $\mathbb{I}$ is the identity matrix, and $\sigma^n_k$ is $n$-qubit Pauli basis. For example, when $n=3$ and $k=13=(031)_4$, where $(031)_4=0\times 4^2 + 3\times 4^1 + 1\times 4^0$ is the base-$4$ representation of the decimal number $13$, the pauli basis is $\sigma^3_{13} = \sigma^3_{(031)_4} := \mathbb{I}_1\otimes Z_2 \otimes X_3$, where $Z=\sigma_z$ and so on are Pauli matrices and the subscripts indicate the qubits.
Notice that we count qubits from left to right, starting from $1$, with the first qubit being the most significant bit. 
Consistent with this ordering, we use the convention for tensor products that $(A\otimes B)_{(i_1 i_2)_2, (j_1 j_2)_2}=A_{i_1,j_1}B_{i_2, j_2}$, where $A$ and $B$ are $2\times 2$ matrices so their tensor product $A\otimes B$ is a $4\times 4$ matrix. %Tensor products of higher-dimensional matrices use a similar convention.
Notice that we count indices of vectors and matrices starting from $0$.
More generally, for $k=(\mu_1 \dots \mu_{n})_4$, where $\mu=0,\dots,3$, the Pauli basis is $\sigma^n_k :=\bigotimes_{l=1}^{n}\sigma_{\mu_l}$, where $\sigma_{\mu}=(\mathbb{I}, X, Y, Z)$ is the Pauli 4-vector.
The goal of full state tomography is to measure $R_k$, for $k=1,\dots, 4^n-1$.
Notice that a physical density matrix must be positive semi-definite, so the magnitude of these coefficients are constrained.

The diagonal elements of $\rho$ are measurable probabilities. For example, when $n=2$, the element $\rho_{1,1}$ is the probability of measuring the state $|1\rangle = |01\rangle$. More generally, the probability of measuring the state $|q_1\dots q_{n}\rangle = |q_1\rangle\otimes\dots\otimes|q_n\rangle$ is $p_l=\rho_{l,l}$, where $l=(q_1 \dots q_{n})_2$ and $q=0, 1$.  
To see how to map $p_l$ to $R_k$ in Eq.~(\ref{eq:rho_pauli}), consider the example $n=2$, where the four matrices $\mathbb{I}_1\otimes\mathbb{I}_2$, $\mathbb{I}_1\otimes Z_2$, $Z_1 \otimes\mathbb{I}_2$, and $Z_1\otimes Z_2$ form a basis for diagonal matrices.
By enumerating elements of the four matrices, we see the $(\dots q_i\dots q_j\dots)_2$-th diagonal element of the matrix $(\dots Z_i\otimes\dots\otimes\mathbb{I}_j\dots)$ is $\dots(-1)^{q_i} \dots  (+1)^{q_j} \dots$, 
% \begin{equation}
%     \nonumber
%     (\!\dots Z_i\otimes\dots\otimes\mathbb{I}_j\dots\!)_{(\!\dots b_i\dots b_j\dots\!)_2} \!=\! \dots\! (-1)^{b_i}\!\dots\! (+1)^{b_j}\!\dots,
% \end{equation}
which can be proven by induction by adding the next qubit to the left. 
Therefore, focusing on diagonal elements of Eq.~(\ref{eq:rho_pauli}), we have
\begin{eqnarray}
    \label{eq:rho_diag}
    \nonumber
    p_{(q_1\dots q_n)_2} = \frac{1}{2^n} &+& R_{(3\dots 3)_4}(-1)^{q_1}\dots(-1)^{q_n} + \dots \\
    \nonumber
    &+& \sum_{i<j} R_{(0\dots 3_i\dots3_j\dots 0)_4}(-1)^{q_i} (-1)^{q_j}\\
    &+& \sum_{i} R_{(0\dots 3_i\dots0)_4}(-1)^{q_i}.
\end{eqnarray}
On both sides, there are $2^n-1=N-1$ independent variables. Therefore, from measured probabilities, the coefficients can be solved. 
To simplify the RHS, we map a binary number $(b_1\dots b_n)_2$ to a base-4 index $(\mu_1\dots\mu_n)_4$ by setting $\mu_i=0$ when $b_i=0$, and $\mu_i=3$ when $b_i=1$.
We denote the mapping by $\mu(b_1\dots b_n)_2=(\mu_1\dots\mu_n)_4$.
Then, Eq.~(\ref{eq:rho_diag}) can be written as a matrix equation. 
First, we introduce a vector $\mathbf{p}$, whose $l$-th element is the probability $p_{(q_1\dots q_n)_2}$ when $l=(q_1\dots q_n)_2=0,1,\dots,N-2$. The last probability $p_{N-1}$ is not independent because $\sum_{l=0}^{N-1}p_l=1$.
Second, we introduce a vector $\mathbf{r}$, whose $k$-th element is the coefficient $R_{\mu(b_1\dots b_n)_2}$ when $k=(b_1\dots b_n)_2-1=0,1,\dots,N-2$. 
At least one $b_i$ is nonzero because the index of $R_k$ starts from $1$.
Finally, we introduce a matrix $G$, which allows us to map coefficients to probabilities and vice versa. The matrix elements are
\begin{equation}
    \label{eq:tomography_matrix}
    G_{(q_1\dots q_n)_2, (b_1\dots b_n)_2-1} = \prod_{i=1}^n (-1)^{b_i q_i},
\end{equation}
whereby Eq.~(\ref{eq:rho_diag}) can be written as $\mathbf{p}=2^{-n} + G\mathbf{r}$.
Because rows of $G$ are linearly independent, the matrix is full rank and invertible. We infer $\mathbf{r}=G^{-1}(\mathbf{p}-2^{-n})$ from the probabilities when all qubits are measured in the $Z$ basis. %Notice that elements of the matrix $G$ are $\pm 1$ deterministically, and the inverse $G^{-1}$ is well behaved.  

The remaining Pauli coefficients of $\rho$ can be measured using a similar method after rotating the basis of measurements to $X$ and $Y$. 
Under a unitary transformation $|\psi\rangle\rightarrow U|\psi\rangle$, the density matrix is transformed by $\rho\rightarrow U\rho U^\dagger$. 
Therefore, when performing right-handed rotation of a qubit around the $X$ axis of its Bloch sphere by angle $\pi/2$, such that the $Y$ axis is turned into the $Z$ axis, the unitary $U_X=\exp(-i\frac{\pi}{4}X)$ transforms the $Y$ basis in $\rho$ to the $Z$ basis, because $\exp(-i\frac{\pi}{4}X) Y \exp(+i\frac{\pi}{4}X)=Z$.
Similarly, when rotating the qubit by $U_Y=\exp(+i\frac{\pi}{4}Y)$, the $X$ basis is turned into the $Z$ basis, because $\exp(i\frac{\pi}{4}Y) X \exp(-i\frac{\pi}{4}Y)=Z$.
After single-qubit rotations, we can now measure the remaining Pauli coefficients.
For example, when $n=2$, to measure coefficients including $R_{(13)_4}$, we apply single-qubit gates $U=U_{Y_1}\otimes\mathbb{I}_2$, after which the density matrix becomes $U\rho U^\dagger = 2^{-2}\mathbb{I} + R_{(13)_4} Z_1\otimes Z_2 + R_{(10)_4} Z_1\otimes \mathbb{I}_2 + R_{(03)_4} \mathbb{I}_1\otimes Z_2 + \dots$, where omitted terms have zero diagonal elements. % and hence do not contribute to probabilities. 
While $R_{(03)_4}$ can already be extracted by measuring the two qubits in the $(Z_1, Z_2)$ basis, now we can extract additional coefficients $R_{(13)_4}$ and $R_{(10)_4}$ by measuring in the $(X_1, Z_2)$ basis. 
More generally, to measure along a given axis $\mathbf{a}=(a_1,\dots ,a_{n})$, where $a_i=1,2,3$ corresponds to $X_i, Y_i, Z_i$ axes for qubit $i$, we perform single-qubit gates $U=\bigotimes_{i=1}^{n} U_i$, where $U_i=U_{Y_i}$ when $a_i=1$, $U_i=U_{X_i}$ when $a_i=2$, and $U_i=\mathbb{I}$ when $a_i=3$. After single-qubit rotations, the probability of measuring the qubits in the state $|q_1, q_2,\dots,q_{n}\rangle$ is 
\begin{eqnarray}
    \label{eq:rho_offdiag}
    \nonumber
    p_{(q_1\dots q_n)_2}^{(a_1,\dots ,a_{n})} = \frac{1}{2^n} &+& R_{(a_1\dots a_n)_4}(-1)^{q_1}\dots(-1)^{q_n} + \dots \\
    \nonumber
    &+& \sum_{i<j} R_{(0\dots a_i\dots a_j\dots 0)_4}(-1)^{q_i} (-1)^{q_j}\\
    &+& \sum_{i} R_{(0\dots a_i\dots0)_4}(-1)^{q_i},
\end{eqnarray}
which has an identical structure as Eq.~(\ref{eq:rho_diag}), except that now we can access all Pauli coefficients. 
Following similar steps, we convert the above equation into a matrix form $\mathbf{p}^\mathbf{a}=2^{-n} + G\mathbf{r}^\mathbf{a}$, where $G$ is the same matrix given by Eq.~(\ref{eq:tomography_matrix}). 
The $k$-th component of the vector $\mathbf{r}^\mathbf{a}$ is $R_{\mu^\mathbf{a}(b_1\dots b_n)_2}$ when $k=(b_1\dots b_n)_2-1$, where the map $\mu^\mathbf{a}(b_1\dots b_n)_2=(\mu_1\dots\mu_n)_4$ is defined by setting $\mu_i=a_ib_i$.
By measuring $n$ qubits along all $3^n$ axes, we reconstruct $\rho$ by solving for its Pauli coefficients $\mathbf{r}^\mathbf{a}=G^{-1}(\mathbf{p}^\mathbf{a}-2^{-n})$ from measured probabilities.

Finally, let us count the degrees of freedoms. Measuring along each $\mathbf{a}$ requires a single circuit of single-qubit gates. At the end of the gates, we obtain $2^n-1$ probabilities, so after $3^n$ circuits, we obtained $3^n(2^n-1)>4^n-1$ probabilities. The excess is due to the fact that many Pauli coefficients are extractable from multiple circuits. 
For example, $R_{(03)_4}$ can be extracted from both $(Z_1, Z_2)$ and $(X_1, Z_2)$ axes.  
More generally, $R_{(\mu_1\dots\mu_n)_4}$ is insensitive to the measurement axis of the $i$-th qubit if $\mu_i=0$. 
Suppose $k$ out of the $n$ elements of $(\mu_1,\dots,\mu_n)$ are zero, then $R_{(\mu_1\dots\mu_n)_4}$ can be measured in $3^k$ different ways. 
Accounting for this redundancy, the independent d.o.f. each circuit can extract is $\sum_{k=0}^{n-1} \binom{n}{k} \frac{1}{3^k}=(1+\frac{1}{3})^n-\frac{1}{3^n}$. The summation does not reach $k=n$ because no Pauli coefficient has all $\mu$'s equal to zero. Therefore, from $3^n$ circuits, we only obtain $3^n[(1+\frac{1}{3})^n-\frac{1}{3^n}]=4^n-1$ independent d.o.f. as expected.  
On ideal quantum computers with no error, different ways of measuring the same Pauli coefficient should give the same result, up to shot noise. However, current devices are noisy, and the results we observe are different beyond statistical fluctuations. For results shown in Sec.~\ref{sec:quantum}, we simply average the results. One could imagine a more sophisticated scheme to extract unbiased results and estimate error bars.

%%%%%%%%%%%%%%%%%%%%%%%%%%%
\subsection{Pure states\label{sec:pure}}
When the density matrix is known for a pure state, the state vector can be found by diagonalizing the density matrix. All except for one eigenvalue is $0$, and the only nonzero eigenvalue is $1$, whose normalized eigenvector is the state vector, up to a global phase. On the other hand, if the state is a mixed state, $\rho$ has multiple eigenvalues between $0$ and $1$. The eigenvector of the largest eigenvalue corresponds to the dominant pure state. In Sec.~\ref{sec:quantum}, we know the states are mixed, because the purity of $\rho$ is less than $1$. In this case, full state tomography is necessary. However, for future quantum computers, where the state is close to pure, it is a waste to measure along all $3^n$ axes, if all we want is to extract the pure state. In this section, we describe a method for inferring a pure state using only $(2n+1)$ measurement circuits.

Using similar notation as in Sec.~\ref{sec:mixed}, we relate decimal indices to binary numbers $l=(q_1\dots q_{n})_2$, so that a $n$-qubit pure state is spanned by 
\begin{equation}
    \label{eq:pure}
    |\psi\rangle=\sum_{l=0}^{2^n-1} c_l|l\rangle.
\end{equation}
To measure $|\psi\rangle$ means to determine both the amplitude $r_l$ and the phase $\theta_l$ of all nonzero coefficients $c_l=r_l e^{i\theta_l}$, up to a global phase. % 
To determine $r_l=|c_l|$, we measure all qubits along their $Z$ axes. The probability of measuring state $|l\rangle$ is $p_l=r_l^2$. Next, to determine the relative phase between $c_l$ and $c_k$, we mix the two states $|l\rangle$ and $|k\rangle$ and observe their interference. 
Suppose $l=(q_1\dots 0_i\dots q_n)_2$ and $k=(q_1\dots 1_i\dots q_n)_2$ differ only by the $i$-th bit. To mix these two states, we rotate the $i$-th qubit.
Performing $U_X=\exp(-i\frac{\pi}{4}X)$ or $U_{Y}=\exp(+i\frac{\pi}{4}Y)$ rotation on the a qubit gives 
\begin{subequations}
\label{eq:XY_rotations}
\begin{eqnarray}
    U_{X} |q\rangle = \frac{1}{\sqrt{2}}\Big(|q\rangle &-& \,i \;|\bar{q}\rangle\Big),\\
    U_{Y} |q\rangle = \frac{1}{\sqrt{2}}\Big(|q\rangle &-& (-1)^{q} |\bar{q}\rangle\Big),
\end{eqnarray}    
\end{subequations}
where bar denotes spin flips, namely, $\bar{0}=1$ and $\bar{1}=0$. Then, $U_{X_i}|\psi\rangle \!=\!c_l[\dots\otimes\frac{1}{\sqrt{2}}(|0_i\rangle \!-\! i |1_i\rangle)\otimes\dots] + c_k[\dots\otimes\frac{1}{\sqrt{2}}(|1_i\rangle \!-\! i |0_i\rangle)\otimes\dots] + \dots =\frac{1}{\sqrt{2}}(c_l-ic_k) |q_1\dots 0_i\dots q_n\rangle + \frac{1}{\sqrt{2}}(c_k-ic_l) |q_1\dots 1_i\dots q_n\rangle + \dots$. 
We see the probability of finding $U_{X_i}|\psi\rangle$ in $|l\rangle$ or $|k\rangle$
is $\frac{1}{2}|c_l\mp ic_k|^2$. 
Using $|c_l\pm ic_k|^2 = r_l^2+r_k^2\pm 2 r_l r_k\sin(\theta_l-\theta_k)$, we can determine $\sin(\theta_l-\theta_k)$. 
Similarly, $U_{Y_i}|\psi\rangle =\frac{1}{\sqrt{2}}(c_l+c_k) |q_1\dots 0_i\dots q_n\rangle + \frac{1}{\sqrt{2}}(c_k-c_l) |q_1\dots 1_i\dots q_n\rangle + \dots$, so by measuring the probabilities $\frac{1}{2}|c_l\pm c_k|^2$, we can determine $\cos(\theta_l-\theta_k)$ from $|c_l\pm c_k|^2 = r_l^2+r_k^2\pm 2 r_l r_k\cos(\theta_l-\theta_k)$. 
Knowing both $\sin$ and $\cos$, the relative phase angle $\theta_l-\theta_k$ can be determined unambiguously.
%We see to determine the relative phase between two states separated by a single bit requires two circuits.

The above procedure of using single-qubit rotations to determine the relative phase can be generalized to two states separated by any number of qubits.
Suppose two occupied states differ by $m$ qubits. Among these qubits, we choose to act $U_X$ on qubits $i_1,\dots , i_u$ and act $U_Y$ on qubits $j_1, \dots, j_w$, where $u+w=m$. 
To keep the notation compact, we only track qubits that are different, and reorder the qubits to the form $|q_{i_1}\dots q_{i_u} q_{j_1}\dots q_{j_w}\rangle$. Using Eq.~(\ref{eq:XY_rotations}) and $(-1)^q=-(-1)^{\bar{q}}$, we obtain
%\begin{widetext}
\begin{eqnarray}
    \label{eq:general_rotation}
    \nonumber
    &&\bigotimes_{k=1}^{u}\! U_{X_{i_k}}\! \bigotimes_{l=1}^{w}\! U_{Y_{j_l}}\! \big(c|q_{i_1}\!\dots\! q_{i_u} q_{j_1}\!\dots\! q_{j_w}\!\rangle \!+\! \bar{c}|\bar{q}_{i_1}\!\dots\! \bar{q}_{i_u} \bar{q}_{j_1}\!\dots\! \bar{q}_{j_w}\!\rangle\! \big) \\
    &&=\frac{1}{2^{m/2}}\big[c +\! (-i)^u\! \prod_{l=1}^w (\!-1\!)^{q_{j_l}} \bar{c} \big] |q_{i_1}\!\dots\! q_{i_u} q_{j_1}\!\dots\! q_{j_w}\rangle\! +\! \dots,
\end{eqnarray}
%\end{widetext}
where the omitted terms contain mixed $q$ and $\bar{q}$. 
We denote $d_{lk}$ the Hamming distance between $l$ and $k$, which equals the number of bits by which their binary representations differ. 
As long as the $d<m$ Hamming sphere of $|l\rangle = |q_{i_1}\dots q_{i_u} q_{j_1}\dots q_{j_v}\rangle$ is empty, namely, all states with fewer than $m$ tracked qubit flips from $|l\rangle$ are unoccupied, then the coefficient of $|l\rangle$ in Eq.~(\ref{eq:general_rotation}) is unaffected by the presence of other states, which differ from $|l\rangle$ by the other $n-m$ untracked qubits.
After performing single-qubit rotations and measuring the probability of $|l\rangle$, we can observe the interference between $c$ and $\bar{c}$. 
Determining their relative phase only requires knowing $|c\pm\bar{c}|^2$ and $|c\pm i\bar{c}|^2$, so two circuits are sufficient. Suppose the first circuit uses $(u_1, w_1)$ and the second uses $(u_2, w_2)$, then they give different probabilities if and only if $u_1-u_2\equiv 1\, (\textrm{mod}\, 2)$. In practice, we choose $(u_1, w_1)=(m,0)$ and $(u_2, w_2)=(m-1, 1)$.

Finally, we use a tree graph, which we call a phase-inheritance tree, to determine relative phases between all occupied states. 
The Hilbert space of $n$ qubits can be visualized as an $n$-dimensional hyper cube. Vertices of the hyper cube represent quantum states, which can be labeled by $n$-bit binary strings, and parallel edges of the hyper cube represents the same single-qubit flip.
%Let $v_1$ and $v_2$ be two vertices, we denote their Hamming distance by $d(v_1, v_2)$.
%For example, for $v_1=(11)_2$ and $v_2=(01)_2$, we have $d(v_1, v_2)=1$.
%
The pure state Eq.~(\ref{eq:pure}) is represented by a subgraph of the hyper cube. The subgraph is comprised of vertices that the state occupies, namely, $G=\{v_{l}| c_l\ne0 \}$.
We say $G$ is simply connected if all its vertices can be connected by $d=1$ edges. In this case, one can build a tree graph, constituted of a list of directional edges $E=\{e_{lk}=(v_l, v_k)\}$, from a seed vertex, such that all other vertices are connected to the seed vertex by a unique path of $d=1$ edges. The tree graph, which is free of loops, is not unique. 
Because $d=1$ edges correspond to single-qubit flips, we determine the relative phase $u_{lk}=e^{i(\theta_k-\theta_l)}$ associated with $e_{lk}$ using the method described after Eq.~(\ref{eq:XY_rotations}). Notice that flipping one qubit may enable the phase determination of multiple edges. 
Using the phase inheritance tree, phases of all vertices can be traced back to the phase of the seed vertex, and thus uniquely determined up to a global phase. 
When the graph $G$ is not simply connected, we first build simply-connected subgraphs $\{G_l\}$. We define the distance between two subgraphs 
$G_1$ and $G_2$ by $d(G_1, G_2)=\min_{v_1\in G_1, v_2\in G_2} d(v_1, v_2)$, namely, the minimum Hamming distance between the two subgraphs. Suppose the minimum is attained at $v_1$ and $v_2$, then the two subgraphs can be connected by flipping all $m=d(G_1, G_2)$ qubits by which $v_1$ and $v_2$ differ. Using Eq.~(\ref{eq:general_rotation}), the phase of $v_2$ is determined from $v_1$. Using $v_2$ as the seed for $G_2$, the phase of $G_2$ is thus inherited from $G_1$. Repeating this procedure for all simply-connected subgraphs, the phases of all vertices are determined. 
Because the maximum number of qubits that need to be flipped is $n$, we need at most $2n$ circuits of single-qubit gates, where $2$ corresponds to two choices of $(u,w)$ in Eq.~(\ref{eq:general_rotation}).
Adding the trivial circuit where all qubits are measured along their $Z$ axes, pure state tomography then requires at most $(2n+1)$ measurement circuits.

%\nocite{*}
%\bibliography{references}% Produces the bibliography via BibTeX.
\providecommand{\noopsort}[1]{}\providecommand{\singleletter}[1]{#1}%

\end{document}